# Report of 2017 NSF Workshop on Multimedia Challenges, Opportunities and Research Roadmaps

January 2018

**Workshop Participants**

Shih-Fu Chang (co-organizer)
Louis-Philippe Morency (co-organizer)
Alex Hauptmann (co-organizer)
Terry Adams
Sameer Antani
Dick Bulterman
Carlos Busso
Joyce Chai
Julia Hirschberg
Ramesh Jain
Ketan Mayer-Patel
Reuven Meth
Raymond Mooney
Klara Nahrstedt
Shri Narayanan
Prem Natarajan
Sharon Oviatt
Balakrishnan Prabhakaran
Arnold Smeulders
Hari Sundaram
Adam Wolfe
Zhengyou Zhang
Michelle Zhou

**Report Citation:**

Shih-Fu Chang, Alex Hauptmann, Louis-Philippe Morency, Sameer Antani, Dick Bulterman, Carlos Busso, Joyce Chai, Julia Hirschberg, Ramesh Jain, Ketan Mayer-Patel, Reuven Meth, Raymond Mooney, Klara Nahrstedt, Shri Narayanan, Prem Natarajan, Sharon Oviatt, Balakrishnan Prabhakaran, Arnold Smeulders, Hari Sundaram, Zhengyou Zhang, Michelle Zhou, "Report of 2017 NSF Workshop on Multimedia Challenges, Opportunities and Research Directions," arXiv, 2019.

This material is based upon work supported by the National Science Foundation under Grant No. 1735591. Any opinions, findings, and conclusions or recommendations expressed in this material are those of the author(s) and do not necessarily reflect the views of the National Science Foundation.
We also thank Microsoft Research for the generous support for this workshop.

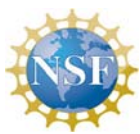

# Report of 2017 NSF Workshop on
# Multimedia Challenges, Opportunities and Research Directions

# Section 1

# Chapter 1.  **Introduction**

The multimedia and multimodal (MM) research community is dedicated to research on the acquisition, communication, analysis, modeling, interface design, and impact of applying multimodal-multimedia data to challenging problems such as semantic information extraction, co-processing and interpretation of multiple heterogeneous modalities, seamless human-machine interaction, and scalable distributive collaboration. This community of researchers also aims to advance  important applications such as education, healthcare, advanced communications and social networking. One of the key objectives for MM research is to model and in some cases exceed humans' ability to process multi-sensory information  during activities like communication and learning. The multidisciplinary MM field is distinct from other research areas  in its  emphasis on deep, meaningful integration of multimodal data to enhance the reliability and depth of information derived, as well as  the quality of multimedia user experience. A major advantage of fusion-based multimodal interaction and systems is greater robustness and coverage of information processing and improved experience of interaction, which often exceed that possible based on the sum of individual modalities. And, the synergy between multimedia-multimodal data sources is often the prime source of generating new knowledge.

The MM community has demonstrated extremely productive activities in the last two to three decades. Starting with the inaugural ACM Conference in Multimedia (ACMMM) in 1993, the International Conference on Multimodal Interaction (ICMI) in 1996, and high-quality publications in other professional societies, the MM community has celebrated major technological breakthroughs with  a major impact on industry and society. Within the past decade, the dominant computer interface paradigm worldwide has become a multimodal-multimedia one on a mobile smartphone. Multimodal-multimedia interaction and communication technologies have been deployed widely in a variety of consumer products, including smartwatches and smartphones, video conferencing and collaborative systems, augmented and virtual reality systems, smart cars, and elsewhere. In addition, multimedia search and recommendation engines have been developed by major IT companies and social media platforms.

With these transformative technologies and the rapidly changing global R&D landscape, the MM community is now faced with many new opportunities and uncertainties. With the open source dissemination platform and pervasive computing resources, new research results are being discovered at an unprecedented pace. In addition, the rapid exchange and influence of ideas across traditional discipline boundaries have made the emphasis on multimedia multimodal research even more important than before. To seize these opportunities and respond to the challenges, we have organized a workshop to specifically address and brainstorm the challenges, opportunities, and research roadmaps for MM research.

The two-day workshop, held on March 30 and 31, 2017 in Washington DC, was sponsored by the Information and Intelligent Systems Division of the National Science Foundation of the United States. Twenty-three (23) invited participants (see list above) were asked to review and identify research areas in the MM field that are most important over the next 10-15 year timeframe. Important topics were selected through discussion and consensus, and then discussed in depth in breakout groups. Breakout groups reported initial discussion results to the whole group, who continued with further extensive deliberation. For each identified topic, a summary was produced after the workshop to describe the main findings, including the state of the art, challenges, and research roadmaps planned for the next 5, 10, and 15 years in the identified area.

**Research Areas Identified:**

Given the breadth and depth of the MM field, we have identified two groups of important topics: (1) cross-cutting MM research and (2) MM applications. The cross-cutting section includes a large number of core areas, whose advances can be integrated to build transformative applications with high impact on industries and society. Each one in the application section exemplifies an important domain that calls for significant advances across multiple MM fundamental areas in the next decade.  Finally, a few topics of great interest to the community were also identified and included as supplemental materials.

*Major Areas in Cross-Cutting MM Research:*

- Multimodal Foundational Methods
- Multimodal Knowledge Discovery
- Joint Knowledge Representation and Processing - Grounding
- Person-Centered Multimodal Interaction
- Multimedia Content Generation
- Multimedia and Multimodal systems
- Data, Challenges and Evaluation

*Major Areas in MM Applications:*

- Education
- Healthcare
- Smart Infrastructure
- Social Good

*Training, Infrastructure and Funding:*

- Training the next researcher generation
- Research Infrastructure
- International Funding and Collaboration

A total of fourteen (14) chapters summarizing the main findings of the above listed areas are included in this document. Due to the limited time of the workshop, a few areas were identified but were not assigned to breakout discussion groups during the workshop. These include, but are not limited to, the following topics. Additional workshops may be organized in the future to discuss these areas and augment the current report with the additional findings.

*Areas for Future Discussion:*

- Privacy/Personalized MM
- Multimodal-multimedia social network and HCI
- Multimodal Internet of things
- Public safety utilizing MM data

# Section 2    **CROSS-CUTTING RESEARCH**

# Chapter 2. Fundamental Methods for Multimodal Learning

**Chapter Editors:**

Carlos Busso, University of Texas at Dallas
Louis-Philippe Morency, Carnegie Mellon University

**Additional Workshop Participants:**

Terrence Adams, Reuven Meth, Raymond Mooney, Balakrishnan Prabhakaran and Maria Zemankova (listener)

## 1. Introduction

An important research endeavor is to develop foundational methods to analyze, model, and represent multimodal information. These methods are characterized as foundational since they are relevant to many applications and research topics. The unique and challenging aspect of multimodal research is the heterogeneity of the multimodal data and the challenges in integrating and interpreting such heterogeneous data coming from multiple modalities. Each modality has its own characteristics that have to be considered [Maragos et al., 2008]. For example, images are 2-dimensional (or sometime even 3-dimensional) while speech is 1-dimensional. The way we describe how a bird or an animal look will be very different from how we describe how it sounds. Creating computers able to understand this multimodal data brings many fundamental problems which can be grouped into five main classes, following the taxonomy of Baltrušaitis et al. [2017]: representation, alignment, fusion, translation and co-learning (Figure 1).

Representation: Learning how a computer can represent numerically the heterogeneous data from multiple modalities. These computational representations should be designed for both efficient modeling and better visualization. For example, a joint representation of how a person looks and sounds when they are happy will allow computers to better recognize human emotions. These joint representations will be the most efficient when they can take advantage of the natural dependencies between modalities. Another objective is to improve the interpretability of multimodal data. By identifying commonalities and differences between multimodal data, a multimodal representation provides an avenue to bridge the gap between continuous versus discrete data, and numerical versus symbolic data.

Alignment: The process of establishing spatial and/or temporal connections between events across modalities. For example, when reading the caption of an image, alignment is the process where words are linked to specific objects or groups of objects in the image. Other examples of alignment include automatic video capturing, and identifying the acoustic source in a video. One challenge in alignment is dealing with data stream with different sampling rates (e.g., continuous signals versus discrete events). Alignment may require defining a similarity metric between modalities to identify the connection points.

The alignment can be temporal, as when we align the audio and images of a video, or it can be spatial alignment, as when we try to morph between two face images.

<u>Fusion:</u> The combination of information coming from two or more sources to uncover or predict a pattern or trait of interest. Examples of multimodal fusion include multimodal emotion recognition, audiovisual speech recognition, and audiovisual speaker verification. The information can be redundant which helps increase robustness, or complementary which often helps increase accuracy. The challenges are multiple since the modalities do not need to be synchronized or even have the same sampling rate. Some modalities may be incomplete with missing information. Some modalities may provide continuous streams of data, while other modalities may be intrinsically discrete providing information about given events.

<u>Translation:</u> The transformation or mapping from one modality into another. Examples include speech-driven animation, and text-based image retrieval. The foundational methods learn the relationship between streams of data, capturing their dependencies. The goal of translation can be generative in nature, creating a new instance in one modality given information from another modality. It can also be descriptive models, where one modality is used to increase the characterization of another modality (e.g., describing the information conveyed in an image).

<u>Co-learning:</u> The transfer of knowledge from one modality to help with the prediction or modeling task in another modality. Co-learning examples are more technical in nature. One of the most popular examples these days is the use of language to help generalization of computer vision algorithms, specifically for object recognition. Co-learning aims to leverage the rich information in one modality, in the learning of another modality, which may have only limited resources (e.g., small number of examples with limited annotations or noisy input). Example of co-learning algorithms include co-training, zero-shot learning and concept learning.

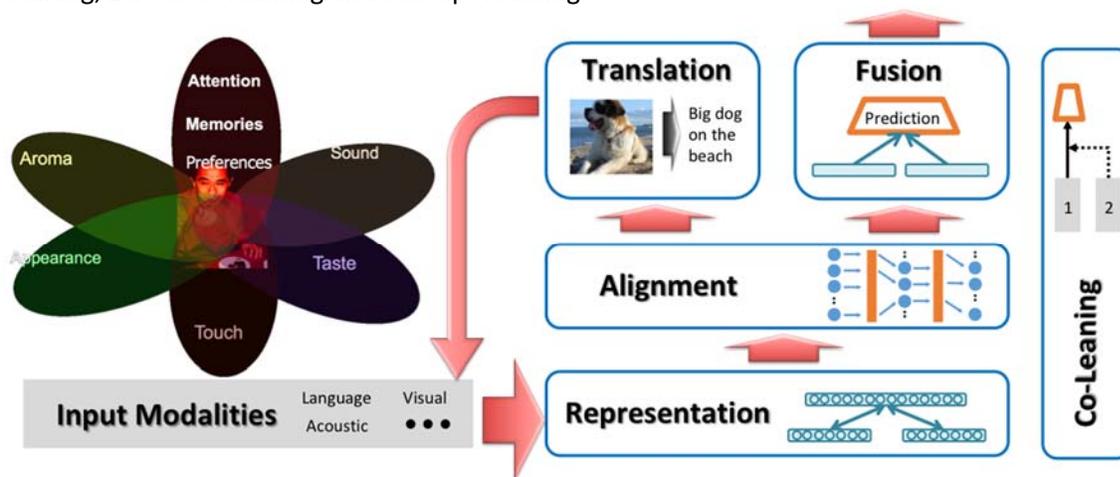

Figure 1: Taxonomy for multimodal learning.

Problems on multimodal processing may involve combination of these categories. Importantly, crosscutting research in these areas will open opportunities to better understand and interpret multimodal data across domains, serving as instrumental tool for the community. These tools can be generic, working across problems. They can also be specific to determined problems or modalities.

The following table summarizes key challenges and outlines a roadmap for future work on this topic.

| State of the art | Key Challenges | Road map | | |
|---|---|---|---|---|
| | | **5 years** | **10 years** | **15 years** |
| - Great achievements for individual modalities, but limited research integrating them.<br>- Promising results integrating vision and language for describing objects and scenes.<br>- Multimodal translation from images to text descriptions (e.g., image captioning)<br>- Multimodal alignment techniques based on attention models. | - Modeling of more than two modalities (e.g., visual, acoustic and language modalities)<br>- Modeling multimodal information present in videos (e.g., event or behavior segmentation)<br>- Learn how to synchronize and coordinate the information from multiple modalities.<br>- Being able to interpret and visualize the representations and interactions from multimodal data<br>- Learning from limited multimodal data | - Learn joint representations of objects and scenes from three modalities<br>- Model asynchrony in audio-visual emotion recognition<br>- Interpretation and visualization of multimodal representations learned from language and vision<br>- Machine Learning frameworks that leverage domains with limited information<br>- Better integration and modeling of understudied modalities in multimodal processing | - Multimodal video representations; Learning how to segment and represent two or more modalities in videos<br>- Alternative multimodal frameworks that address learning in limited data settings<br>- Confidence in multimodal decision<br>- Learning interpretable models of affect in videos | - Interpretable multimodal models of human social interactions in small groups<br>- Multimodal machine translation; translating multimodal intent between languages<br>- Models able to generate human multimodal behaviors |

# 2. State of the Art

Early work on multimodal processing used simple strategies to handle data from multiple modalities. New advances in machine learning have provided appealing frameworks that can be directly applied to multimodal data. This section summarizes some of the advances in foundational methods for multimodal processing. Recent surveys on this area can be found in Atrey et al. [2010], Baltrušaitis et al. [2017], Potamianos et al. [2017], and Katsaggelos et al. [2015].

## 2.1 Representation

A popular framework to represent multimodal data is *canonical correlation analysis* (CCA). CCA transforms two different modalities into a common space where their correlation is maximized.  This is a powerful framework to understand the dependencies between modalities, which do not have to have the same dimension. For example, this approach has been used to explore the relationship between acoustic and facial features [Busso & Narayanan, 2007], to estimate speech articulation from audiovisual data [Katsamanis et al., 2008], to separate audio sources using video assisted techniques [Darrell et al., 2000] and to localize and track speakers [Hershey & Movellan, 1999].

Deep learning using neural networks provides a principled framework to create shared representations between modalities [Ngiam et al., 2011]. Studies have used several alternative deep-learning frameworks to combine multimodal data. An appealing approach is *multi-task learning* (MTL) where different but related problems are simultaneously solved [Caruana, 1997]. One or multiple layers are shared across problems providing a common feature representation. An advantage of MTL is that solving secondary tasks regularizes the deep learning structure, improving the generalization. MTL has been successfully used in emotion recognition  [Ringeval et al., 2015, Parthasarathy & Busso, 2017; Zhang et al., 2016].

## 2.2 Alignment

Dynamic time warping (DTW) [Kruskal, 1983] is a dynamic programming approach that has been extensively used to align multi-view time series. DTW measures the similarity between two sequences and finds an optimal match between them by time warping temporal sequences (inserting frames). It requires the timesteps in the two sequences to be comparable and requires a similarity measure between them. DTW can be used directly for multimodal alignment by hand-crafting similarity metrics between modalities; for example Anguera et al. [2014] use a manually defined similarity between graphemes and phonemes; and Tapaswi et al. [2015] define a similarity between visual scenes and sentences based on appearance of same characters to align TV shows and plot synopses. DTW can be combined with CCA to learn the similarity metric between modalities while aligning them [Shariat, 2011, Zhou, 2009, Zhou, 2012]. While CCA based DTW models are able to find multimodal data alignment under a linear transformation, they are not able to model non-linear relationships. This limitation has been addressed by the deep canonical time warping approach [Trigeorgis, 2016], which can be seen as a generalization of deep CCA and DTW.

As an alternative to explicit modality alignment, alignment can be used as an intermediate step in a downstream task. Multimodal translation is an example of a task that can often be improved if alignment is performed as a latent intermediate step. A very popular way to accomplish this is through neural *attention* models [Bahdanau, 2014], which allows the decoder to focus on sub-components of the source instance. This is in contrast with encoding all source sub-components together, as is performed in a conventional encoder-decoder model. An attention module will tell the decoder to look more at targeted sub-components of the source to be translated – areas of an image [Xu, 2015], words of a sentence [Bahdanau, 2014], segments of an audio sequence [Chan, 2016, Chorowski, 2015], frames and regions in a video [Yao et al., 2015, Yu, 2016], and even parts of an instruction [Mei et al., 2015]. For example, in image captioning instead of encoding an entire image using a Convolutional Neural Network (CNN), an attention mechanism will allow the decoder (typically a Recurrent Neural Network) to focus on particular parts of the image when generating each successive word [Xu and Saenko, 2015]. The attention module which learns what part of the image to focus on is typically a shallow neural network and is trained end-to-end together with a target task (e.g., translation). This approach leads to a latent alignment between the image and a corresponding caption.

## 2.3 Fusion

Multimodal fusion is one of the original topics in multimodal machine learning. In technical terms, multimodal fusion is the concept of integrating information from multiple modalities with the goal of predicting an outcome measure: a class (e.g., happy vs. sad) through classification, or a continuous value (e.g., positivity of sentiment) through regression. It is one of the most researched aspects of multimodal machine learning with work dating to 25 years ago [Yuhas et al., 1989]. There are two types of multimodal fusion *model-agnostic* approaches that are not directly dependent on a specific machine learning method; and *model-based* approaches that explicitly address fusion in their construction.

Early work focused on model-agnostic methods where either the features from different modalities are concatenated (e.g., feature level integration), the decisions from different modality-dependent classifiers are combined (e.g., decision level integration) [Busso et al., 2004, Maragos et al., 2008], or a hybrid approach is taken [Lan et al., 2014]. Model-based approaches, on the other hand, are designed to deal with multimodal data. There are three major categories of such models: kernel-based methods, graphical models, and neural networks. *Multiple kernel learning* (MKL) methods are an extension to kernel support vector machines (SVM) that allow for the use of different kernels for different modalities/views of the data [Chen et al., 2014, Gonen and Alpaydin, 2011, Poria et al., 2016]. As kernels can be seen as similarity functions between data points, modality-specific kernels in MKL allows for better fusion of heterogeneous data. Graphical models are also commonly used to deal with fusion, especially of temporal signals [Ghahramani and Jordan, 1997, Song et al. 2012, Baltrušaitis et al., 2013]. Finally, methods based on *neural networks* and deep learning are facilitating novel formulations resulting in powerful tools for multimodal fusion [Ngiam et al., 2011, Katsaggelos et al. 2015], including work on *recurrent neural models* (RNNs) to capture temporal relationship between modalities [Tao & Busso, 2017].

## 2.4 Translation

Given an entity in one modality the task of multimodal translation is to generate the same entity in a different modality. For example, given an image we might want to generate a sentence describing it or, inversely,  creating an image that represents the text description. It is a long studied problem, with early work in speech synthesis [Hunt and Black, 1996], visual speech generation [Masuko et al., 1998], facial animation [Busso et al., 2007,Mariooryad & Busso, 2012, Sadoughi et al., 2017], and video description [Kojima et al., 2002]. More recently, multimodal translation has seen renewed interest due to combined efforts of the computer vision and natural language processing communities [Bernardi et al., 2016], tackling problems such as image description [Vinyals et al., 2014] and video captioning [Venugopalan et al., 2015]. While the approaches to multimodal translation are very broad and are often modality specific, there are two major categories – *example-based*, and *generative*. Example-based models use a dictionary when translating between the modalities. Generative models, on the other hand, construct a *model* that is able to produce a translation. This distinction is similar to the one between non-parametric and parametric machine learning approaches.

Many of the early multimodal translation systems rely on example-based translation by retrieving the closest matching instance in a stored dictionary, including work on image captioning [Ordonez et al., 2011] and speech generation [Hunt and Black, 1996]. Generative models are arguably more challenging to build as they require the ability to generate signals or sequences of symbols (e.g., sentences). This is difficult for any modality – visual, acoustic, or verbal, especially when temporally and structurally consistent sequences need to be generated. Initial work relied on restricted grammars and rule based systems [Li et al., 2011], but more recent models have relied on deep learning to generate text [Vinyals et al., 2014], images [Reed et al., 2016], sounds [Owens et al., 2016], and facial animations [Sadoughi & Busso, 2017].

## 2.5 Co-learning

Co-learning is defined as aiding the modeling of a (resource poor) modality by exploiting knowledge from another (resource rich) modality. It is particularly relevant when one of the modalities has limited resources – lack of annotated data, noisy input, or unreliable labels. We call this challenge co-learning as most often the helper modality is used only during model training and is not used during test time. We identify three types of co-learning approaches based on their training resources: parallel, non-parallel, and hybrid.

*Parallel-data* approaches require training datasets where the observations from one modality are directly linked to the observations from other modalities. In other words, the multimodal observations are from the same instances, such as in an audio-visual speech dataset where the video and speech samples are from the same speaker. Co-training is such a method that builds weak classifiers in each modality to bootstrap labels in other modalities for the unlabeled data [Blum and Mitchell, 1998]. Co-training has recently been used in audiovisual emotion recognition [Zhang et al., 2016]. Another example is SoundNet, where acoustic models are built by transferring information obtained from object recognition in videos [Aytar et al., 2016]. In contrast, *non-parallel data* approaches do not require direct links between observations from different modalities. These approaches usually achieve co-learning by using overlap in terms of categories. For example, in zero shot learning when the conventional visual object recognition dataset is expanded with a second text-only dataset (e.g. from Wikipedia) to improve the generalization of visual object recognition [Frome et al., 2013, Socher et al., 2013]. Finally, in the *hybrid* data setting the modalities are bridged through a shared modality or a dataset. For example, by using the Bridge Correlational Neural Network [Rajendran et al., 2015], which uses a pivot modality to learn coordinated multimodal representations in presence of non-parallel data. For example, in the case of multilingual image captioning, the image modality would always be paired with at least one caption in any language.

# 3. Challenges Unmet

In this section, we present some significant challenges that are currently unmet or under-studied by the state-of-the-art approaches.

**Verbal, vocal and visual representations** While great progress has been made in language and vision representations, we still need more research to be able to properly represent the human communicative behaviors, expressed primarily through three modalities: verbal, vocal and visual: what is said, how it is said and the visual gestures.

**Multimodal video representations** Language and vision research has shown progress when analyzing images, but videos are still largely an open problem. Videos are particularly challenging when they are unsegmented; when the video contains more than one scene or event. Multimodal video analysis needs to not only integrate multimodal information but also perform segmentation in each modality and over all modalities.

**Multimodal models that learn asynchrony and coordination between modalities** Modeling the alignment from asynchronous multimodal data is still an open problem. For example, in a video recording of a public speaker, the gestures and spoken words may not be synchronized, but there is still a relationship between them that needs to be modeled. Handling this asynchrony is an important challenge in multimodal video analysis.

**Interpretability and visualization of multimodal models** As neural architectures become more prominent in multimodal research, a key challenge will be better interpretability and visualization of the multimodal models. We should design our multimodal models with the goal of being able to interpret the intermediate representations and alignments. New techniques are needed to be able to visualize the multimodal interactions. This is particularly important when we start using these multimodal models in real-world scenarios where users will need to trust and understand, at least at a high level, what the computer is doing.

**Learning from limited multimodal data** Another important challenge is the limited data available for multimodal processing. While we have almost unlimited resources for videos (e.g., YouTube, Vimeo), images (e.g., Flickr, Facebook, Instagram), and audio clips (e.g., iTunes, Soundcloud), these resources do not have the appropriate labels to train multimodal methods. As the cost of annotating datasets is expensive, the research community has to find alternative machine learning approaches to maximize the use of existing labeled and unlabeled resources. For example, multimodal models can be partially trained with different subsets of datasets [Ranjan et al., 2017]. The models can be pre-trained with unsupervised frameworks that do not require labeled data, leveraging an almost unlimited amount of data.

For other multimodal problems involving less conventional modalities, the data is often not publically available, which prevents real advances in this area.

**Real-time, scalability** An important consideration in designing multimodal systems is the complexity of the algorithms. These systems will be impractical for real-time systems if the computation required to process and integrate multiple modalities is too high. For example, early work that modeled the asynchrony between speech and lip motion using hidden Markov models relied on composite audio-visual states, which exponentially affected the Viterbi decoding. The ideal multimodal solutions should also be modular so the system will scale when extra modalities are needed.

# 4. Road Map and Milestones

Advances in fundamental methods for multimodal learning are instrumental for building many of the applications mentioned in this report. Advances in fundamental methods can facilitate solutions addressing common problems across areas in education, healthcare, smart infrastructure and social good. Therefore, it is critical that the research community works on these areas. This section presents milestones in fundamental methods for multimodal learning for the next 5, 10 and 15 years.

## 4.1 5-year Milestones

**Trimodal representations from segmented or static data** As a short-term goal, we should be able to learn multimodal representations from three modalities: language, vision and acoustic. These trimodal representations should be learned from either static data (e.g., visual images) or pre-segmented data. One of the simplest versions of this would be to learn a representation of digits: how they look, how they sound, and they are used linguistically.

**Discovering temporal patterns in audio-visual streams** When a person is speaking, their tone of voice and prosody can be temporarily linked with gestures and facial expressions. Discovering these temporal patterns can help us better understand human communication and affect. As a starting pointing, we propose to discover them from audio and visual streams of information, keeping language modality as a later milestone.

**Interpretation and visualization of multimodal representations learned from language and vision** Building upon recent progress in language and vision, the next big milestone will require to create interpretable versions of these multimodal representations.

**Machine Learning frameworks that leverage domains with limited information** While deep learning is a powerful framework for multimodal processing, it requires enough data for training the models. We expect that alternative learning methods that can create models with limited data will provide solutions to many of the multimodal problems involving unconventional modalities (models trained with sparse optimization with heavy regularization).

## 4.2 10-year Milestones

**Better integration and modeling of understudied modalities in multimodal processing** Most of the progress in multimodal processing includes examples that combines modalities from natural language processing, speech processing and image/video processing. There are many other relevant domains that we have not carefully integrated in multimodal systems. Examples include signals from inertial measurement units (IMUs) and physiological sensors, which are playing an important role in the healthcare domain. With the increase in emphasis on reproducible research, we expect that more multimodal databases will be made available to the community, which will lead to better multimodal solutions to model and integrate these signals.

**Multimodal video representations** The goal of this milestone is to create multimodal techniques that can represent information from multiple streams. For example, how can we identify events or affective behaviors in a long video which has not be pre-segmented. The multimodal algorithms will need to resolve both video segmentation and multimodal representation at the same time.

**Multimodal learning in limited data settings** Many multimodal problems will have a limited amount of annotated data, which makes it important that we can create models that are able to learn multimodal interactions from a limited amount of information.

**Confidence in multimodal decision** As we build these multimodal classifiers, it is important to also include a methodology to evaluate the prediction confidence of these classifiers. A second modality could help boost the confidence.

**Learning interpretable multimodal models from videos** Building interpretable multimodal models is important also for the video domain, where we need to not only integrate multiple modalities, but also handle the potential asynchrony between multimodal streams.

## 4.3 15-year Milestones

**Multimodal models of small group social interactions** This milestone will build not only multimodal representation, but also social representations where the multimodal behaviors of multiple participants are integrated. It is important to model the dyadic and small group interactions.

**Multimodal machine translation** How can we translate multimodal messages between different languages and cultures? The goal is not only to translate the verbal information (i.e., the spoken words), but also to translate the nonverbal information. The cultural aspect of multimodal modeling will be central to this milestone.

**Models able to generate human multimodal behaviors** We need multimodal models that are able to perceive and understand multimodal information, but we also need models that can generate multimodal information. This will be useful for robots and virtual assistants.

# Chapter 3.   **Multimodal Knowledge Discovery**

**Chapter Editors:**

Sharon Oviatt, Monash University
Arnold Smeulders, University of Amsterdam

**Additional Workshop Participants:**

Sameer Antani, Shih-Fu Chang, Alexander Hauptmann, Ramesh Jain, Prem Natarajan and Adam Wolfe

## 1. Introduction

### 1.1   Knowledge Discovery: The Heart of Science

Knowledge discovery is a fundamental scientific process that historically has aimed to establish cause and effect relations among variables. In ancient days, it occurred when Archimedes discovered a method of measuring the purity of gold without demolishing the object, resulting in the first "*eureka*" moment of insight.

Knowledge discovery involves the use of many scientific methodologies to collect and evaluate data in a manner that pursues a process of elimination, or that isolates individual variables, in order to assess their causal impact. This process has centered on hypothesis testing, and the establishment of scientific methods that can support accurate inferences about cause and effect relations among the variables of interest.

Knowledge discovery can culminate in different levels of specificity and power. It can lead to individual findings, more general principles, or to broader and more powerful theories. In order to understand the full scope and importance of a new finding or theory and its applicability, the scientific process also pursues a lengthy process of replicating and testing generality under varied conditions.

Apart from quietly sitting down to think, and using pen and paper tools, digital tools now expedite insightful knowledge discovery. In this chapter we focus on the transformation of knowledge discovery using multimodal-multimedia data, technologies, and analytics. The richness and contextualization of such data have the potential to expand knowledge discovery exponentially, and for many previously unimagined purposes. In particular, it will be invaluable for developing a better understanding of complex systems interactions.

Knowledge discovery that involves multimodal-multimedia data is qualitatively different from past data sources because it is: (1) extremely rich, dense, multi-stream data; (2) analyzable at multiple levels; (3) capable of recording raw information in a highly contextualized form; (4) often human-centered data based on communication and physical activity patterns; (5) processed by computers using probabilistic recognition-based methods, rather than discrete methods of the past. As an illustration of their *richness*, multimedia signals typically involve multiple time-synchronized data streams. Information within these data streams can be analyzed across *multiple levels*, including signal, activity pattern, representational, and transactional. When multimodal-multimedia data are analyzed at multiple levels, a vast multi-dimensional space is created for exploring and discovering new knowledge. *Human-centered* data can include verbal and non-verbal expression, paralinguistic signal information (pitch, volume), movement patterns—which, in combination, are particularly apt for assessing human intention, emotion, cognition, mental health, and broader health status. The sections that follow discuss implications of these characteristics for achieving advances in future knowledge discovery and innovation.

Figure 1 illustrates multi-stream multimodal data collection during collaborative learning, which involved 12 synchronized data streams (speech, images, writing) from three students. These rich data are analyzable at multiple levels (signals, activity patterns, representations, etc.), which results in a vast multi-dimensional space for exploration and knowledge discovery. This example is only exemplary, since input could be expanded to include other modalities, sensors, and levels of analysis (e.g., brain activation patterns). Note that the grounding chapter focuses mainly on discovering patterns between the visual or audio-visual modalities and lexical-level content (i.e., one type of representation). In contrast, this chapter expands the topic of knowledge discovery to multiple modalities and levels of analysis. See the table "Summary Research Roadmap and Timeline" for an overview of challenges and required research progress.

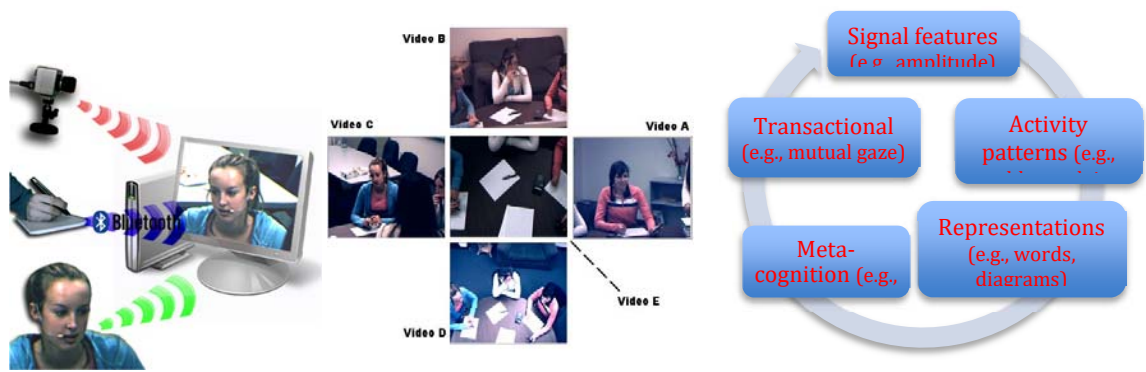

Figure 1: Synchronized multimodal data involving images, writing, and speech (left), collected from a group of students during collaborative learning (middle), yielding data for analysis at multiple levels (right). Multi-level multimodal analytics results in a vast multi-dimensional space for exploration and knowledge discovery.

## 1.2   Exploring Rich Multimedia Knowledge Discovery: A Palette of Strategies

Multimedia data opens the door to different strategies for knowledge discovery, which may be applied for many different purposes. The following illustrate several general strategies and tactics for multimodal-multimedia discovery tools.

### 1.2.1   Multimodal and Cross-Media Knowledge Discovery

Many interesting observations can be discovered at the intersection of modalities. Ambiguity of meaning becomes evident when switching from pictures to language descriptions, and may be resolved when combining them. Depth of meaning can be achieved when using multiple media to convey a message. For example, verbal expressions are clarified by viewing body language. Richly combined multimodal-multimedia information provides a cradle for deeper knowledge discovery, and it will have a broad impact on all areas of the arts and sciences.

Likewise, knowledge discovery benefits from assembling facts on a complex situation from various media and sources, such as audio and video recordings taken by different observers. Although a mobile smartphone versus television camera both collect video, their intended use and therefore content revealed may be very different. Each source may elucidate partial context regarding a complex situation, which together produces a more nuanced and coherent understanding.

### 1.2.2   Multi-level Multimodal Integration for Knowledge Discovery

Observations for knowledge discovery often start with a data-driven hypothesis that needs to be refined and verified.  Initial hypotheses may come from parallel observations noticed in different modalities, which together suggest a more general finding. Similarly, hypothesis verification may occur based on replicated findings across modalities. As such, one powerful paradigm involves discovering general knowledge that is independent of any specific modality data. Computing techniques that support the extrapolation of general knowledge from repeating patterns in multimodal-multimedia data can provide an opportune strategy for discovering new findings, principles, and theory. Similarly, observations noticed across different levels of analysis (signal, representation) can provide cumulative clues that give rise to general knowledge discoveries.

### 1.2.3   Cross-Cultural Knowledge Discovery

Using a similar strategy for abstraction, knowledge discovery can benefit greatly by comparing multimodal-multimedia data on different cultures. In an attempt to understand human behavior, what is hidden in one culture is open in another. Each culture displays its distinct patterns, which may be apparent by observing different modalities. Cross-cultural comparisons provide an opportunity to identify common versus uniquely held intentions, beliefs, and behaviors— including the basis for cultural frictions, and clues for resolving them.

### 1.2.4   Unsupervised Digital and Hybrid Methods for Knowledge Discovery

In contrast with the control over individual "factors" of interest in traditional experimental science, automated machine learning methods instead co-process multiple information sources to identify correlational patterns that jointly predict phenomena of interest. In unsupervised multimodal knowledge discovery, patterns derived from one sensor (e.g., audio) can be used to gauge those in another (e.g., visual), using potentially very large corpora and without hand labeling. With abundant naturally occurring unstructured data, unsupervised machine learning has the potential to explore and model complex systems-level phenomena in multimodal-multimedia datasets (e.g., global climate change). Understanding such phenomena can require analyzing large amounts of data during system functioning at multiple levels, for long time periods, and within a natural context. However, the benefits can include discovering nonlinear patterns and surprising emergent phenomena. To optimize knowledge discovery, new hybrid methods could be developed that combine unsupervised learning with empirical/statistical techniques discussed in Section 2.2.

## 1.3   Significance

The process of knowledge discovery has supported civilization and its advances over thousands of years. It is at the heart of science. We depend critically on it to advance all areas of science, and to fuel the innovation pipeline that leads to new business concepts, applications, and job creation. As such, knowledge discovery is a fundamental cross-cutting activity and driver of new commercial markets.

## 1.4   Focal Challenges and Benefits

In this chapter, we discuss examples of major limitations with society's current approach to knowledge discovery, and how next-generation multimodal-multimedia computational tools could be developed to overcome them.

First, there is a need for (1) ***new methods to evaluate the validity and quality of the growing body of digitally transmitted data***, which currently is jeopardized by dissemination of explicitly false data, biased data, and propaganda that is predatory rather than prosocial in nature. Among the application areas and societal institutions that would be most heavily impacted are education, journalism, advertising, politics, and international relations. Without verifiably accurate data, knowledge discovery cannot occur.

Secondly, there is a need for (2) ***computational tools that correct for systematic biases in human cognition and decision-making.*** For example, human reasoning is prone to rampant overgeneralization, a cognitive bias that undermines objective knowledge discovery.  New multimodal-multimedia computational tools could be developed to overcome these biases by designing synergistic human-computer systems in which computer functioning targets tasks where humans are weakest. New systems could be applied for humanity's benefit in medical practice, schools, preservation of environmental resources, and other areas;

Third, there is always a need for (3) ***new methods to extract emergent semantic knowledge from combined information in multiple modalities***. When the same event is recorded by two or more modalities, for example an image and eye movements, there is an opportunity to learn otherwise hidden semantics by following the salient path of gaze locations. All multimodal combinations (e.g., body language and speech, facial expressions and tone of voice) provide greater semantic depth that can reveal beliefs, attitudes, truthfulness, stress, and other information for interpreting a situation correctly.

In addition, a major long-term goal is to develop (4) ***systems-level theories*** in all areas including, but not limited to, human health and medicine, learning, cross-cultural behavior and intentions, and the environment. System theories provide a coherent account of how multiple levels of functioning interact, for example to produce the global climate we experience, which is complex and involves emergent phenomena that cannot be attributed to singular factors, simplistic models (e.g., linear), reductionist scientific beliefs, or traditional cause-and-effect inferential reasoning. The development of systems-level theory requires applying multidisciplinary perspectives, observing system functioning within its natural context, analyzing multiple levels of functioning, and documenting change in system functioning over long time periods. The aim is to discover system dynamics, principles of emergent phenomena that can be unpredictable or even catastrophic, and to derive theories of complex phenomena. Such theories become more tractable when there are opportunities for longitudinal analysis of interconnections in multilevel-multimodal data.

## 2. Summary Table and Research Roadmap

| State of the Art | Key Research Challenges | Summary Research Roadmap & Timeline | | |
|---|---|---|---|---|
| | | **5 years** | **10 years** | **15 years** |
| Digital sensors for collecting wide range of communication & physiological data; | Automated tools for large scale multimedia data collection, alignment, annotation, modeling, analysis & for testing the generality of new findings; | New tools for analysis of modalities other than A-V (writing, haptics, gaze, etc.); | Formalizing data description and workflow for "data-to-discovery" process; | Machine learning tools for analyzing heterogeneous data sources across multiple levels in a multi-dimensional space; |
| Algorithms for interpreting context, activity patterns, and user state; | New computational tools for discovery using data varying in granularity, dimensionality, reliability; | New analytic tools beyond visualizations;  New tools to support collection & analysis of longitudinal sensor & multimedia data; | New tools to automate testing the generality of scientific findings;  Fact verification and means of crossmodal plausibility checks; | Development of systems-level theories in multiple domains (education, medicine, cross-cultural conflict resolution); |
| Collection, retrieval & analysis of key data in multimodal datasets;  New platforms for collecting large corpora; | Computational tools to analyze multidimensional data involving multiple modalities & levels of analysis; | New tools to facilitate hypothesis generation & verification; | Demonstrations of trans-domain & trans-modal learning; | Continuous proactive knowledge discovery and prediction from streaming data, with ultra-reliable accuracy based on strategic fusion of data sources; |
| Automatic classification & labeling of objects & events in A-V data; | Computational tools to support cause & effect reasoning, interpretation of data & new theory; | New tools to evaluate data validity & authoritativeness; | Machine learning tools for analyzing extreme heterogeneity in sensor & multimedia data; | |
| New machine learning tools (DNNs, CNNs), and application to boosting accuracy across many domains; | Computational tools to verify that multimedia evidence is factual, and to rate its authoritativeness; | New machine learning tools that support analysis, modeling & prediction beyond bimodal A-V data; | Demonstrations of ultra-reliable multimodal-multisensor systems for identifying cognition and health status; | |
| Deployment of deep learning within modular hybrid architectures (e.g., AlphaGo); | Computer interfaces that deploy techniques to correct for human decision-making biases; | New hybrid system architectures that incorporate deep learning tools within modules for plan-based reasoning ; | Demonstrations of synergistic human-computer interfaces that correct systematic biases in human cognition in different domains; | |
| Emergence of multilevel multimodal analytic paradigms for knowledge discovery & theory generation; | Ultra-reliable models & predictors of human intention, emotion, cognition & health; | | | |
| Identification of major themes in multilevel multimodal analytic data; | Formation of systems-level theories with broad applicability that can account for unpredictable emergent phenomena; | | | |
| Initial multimodal analytics for predicting human cognition and health status; | | | | |

# 3. State of the Art

## 3.1   The multimodal revolution

During the last decade, mobile devices with multimodal user interfaces based on combined new media have eclipsed keyboard-based graphical interfaces as the dominant computer interface worldwide. The proliferation of multimodal interfaces based on natural input modes (e.g., speech, multi-touch, gestures) has been largely driven by the trend toward mobility. Many converging factors have supported this paradigm shift, including the emergence of digital sensors for recording human images, audio, activity patterns, and physiological state— and new algorithms for deriving users' physical context, activities, social interactions, emotional and mental state, and health status. The availability of platforms for collecting very large corpora, including informal social media like Facebook and Google translate, as well as formal corpora involving medical diseases, has fueled the emergence of new algorithms and machine learning techniques. Recently, spectacular progress has occurred in the automatic classification of objects and events in these databases using audio-visual analyses.

## 3.2 Empirical-Statistical Techniques

Empirical and statistical inference techniques are beginning to be applied to unique multimodal datasets, for example including time-synchronized multi-stream data involving speech, handwriting, images, and linguistic content. Based on multi-level multimodal analyses conducted during data-driven grand challenge events, the following main themes have been determined by research findings:

**Valuable predictive information is present in all modalities and at multiple levels of analysis, and does not necessarily require any linguistic content analysis.** For example, signal- and activity-level patterns alone can be highly predictive of emotional and mental states. These findings are due to new computational tools, which now support the collection of extensive longitudinal data on signal- and activity-level patterns during interpersonal activities like learning. These findings imply that automation of some predictors may be achievable in the near term.

**Analysis of combined multimodal data can yield substantially higher reliabilities than unimodal information sources**, in some cases achieving 98-100% accuracies. Demonstrations frequently have shown that as additional information sources are modeled (i.e., unimodal, bimodal, trimodal), prediction accuracy increases. This implies that ultra-reliable fusion-based multimodal-multimedia systems could be developed not just for security, but also medicine, education, and similar market sectors.

**Major findings based on multi-level multimodal analyses frequently are evident across several modalities and levels of analysis,** providing convergent evidence for a finding. This indicates that multi-level multimodal analyses can yield a multiplicity of cues, which could increase the likelihood of (1) rapidly extracting a scientific insight about causality, and (2) formulating causal inferences that are broader in scope—leading to more powerful theory.

Multi-level multimodal analytics based on empirical and statistical techniques represent a recently emergent methodology for scientific exploration and discovery of the vast multi-dimensional space described in Section 1.1. This new methodology currently is in its infancy, but already has begun to produce systems-level theory.

**As one example:** Oviatt and colleagues discovered that systematic changes in signal-level dynamic handwriting patterns are predictive of domain expertise in mathematics. Modeling of the convergent pattern of individual findings led to the causal inference that math domain experts *expend less total energy* forming written strokes when working on math problem content, compared with novices. Based on this work, a limited-resource systems theory was introduced on adaptive conservation of energy during the process of learning, its function, and its emergent outcomes [Oviatt et al. in press].

## 3.3 Computational techniques

Existing methods of digital knowledge discovery include machine learning involving multimodal and multimedia signals, which is supported by the abundant digital data. Neural networks and deep learning constitute powerful tools for knowledge discovery. Deep learning is a computational technique whereby raw signal data and actions are provided as labels to a system for the purpose of learning observations hidden in new data never seen before, with the aim of predicting certain information— such as making medical diagnoses. Deep learning currently is being applied to the analysis of both unimodal and bimodal data—such as the audio-visual data discussed in Section 2.1. The application of deep learning has led to rapid advances on many problems, including classification of image content, spoken language processing, machine translation, mastery of complex games, and simulation of human audio-visual sensory perception.

**As one example:** Ngiam and colleagues [2011] applied deep networks to cross modality audio-visual feature learning for visible speech. Using a bimodal deep auto-encoder, they learned correlations in multimodal representations that produced predictions corresponding with the McGurk speech perception effect. While this particular result replicated a known scientific finding, it nonetheless demonstrates that deep learning techniques can lead to knowledge discovery.

**A second example:** Most recently, deep learning software called AlphaGo, developed by Google, defeated the world's top human GO master during a competitive game. The system was not preprogrammed to play Go, but rather learned using a general-purpose approach that allowed it to interpret the game's patterns. Deep learning tools were deployed within modules of a hybrid system architecture that included planning. Future challenges involve creating algorithms that can transfer learned patterns from one task, such as GO, to other different tasks [Gibney, 2017].

Given the rich data sources described in Section 1.1, machine learning techniques also are currently being used to analyze human speech, writing, facial expressions, movement patterns, and physiological indices to predict deception, emotion, cognitive load, domain expertise, and mental health status.

When properly designed, deep neural networks for signals spell out the invariants in the signal, versus invariants of the categories of interest in the signals. The ability to do this using a homogeneous architecture of processing elements makes deep networks very suited for hardware acceleration and software libraries—both of which have occurred with the advent of GPU and Caffe-type libraries. Although deep networks have accelerated many fields of study, they nonetheless have significant limitations. For example, they require that many design parameters and architectural features be tailored to a given problem. Secondly, however successful deep learning's black-box solutions may be, they cannot elucidate cause and effect relations without pursuing further scientific analysis.

As mentioned in Section 1.1, multimodal-multimedia data are beginning to be analyzed at multiple levels (e.g., signal, activity pattern, representational, transactional), which presents a vast multi-dimensional space for exploration. New machine learning techniques will be needed to extract novel insights and knowledge from this type of multidimensional data. In particular, new methods will be required to handle bimodal data beyond audio-visual, data involving three or more modalities, more heterogeneous data sources, and to extract meaningful patterns from multidimensional datasets containing multi-level multimodal information. In addition, new hybrid methods will be needed to extract a causal understanding of the basis for prediction success, and to successfully generalize the prediction results to new datasets.

# 4. Open Challenges

Numerous challenges remain to advance this area, including the need for:

- New computational techniques to discover information in data varying in granularity, dimensionality, reliability, and continuity
- New computational techniques to extract and analyze correlational patterns in multidimensional data involving heterogeneous fused media and sensors, each of which can be assessed at multiple levels
- New computational tools to accelerate human interpretation of the patterns in domain-specific data, which could precipitate related theory
- New computational techniques to verify that multimodal-multimedia evidence is authentic and factual, and to rate its authoritativeness
- Development of synergistic human-computer interfaces that incorporate new computational techniques that correct for biases in human decision making
- New tools to automate the process of multimodal-multimedia data collection, analysis, modeling, and also accelerate testing the generality of new findings in order to clarify the power and applications of related theory
- Ultra-reliable human-centered models of intention, emotion, cognition, health, and mental health (including cross-cultural differences), based on strategic fusion in multimodal-multisensor systems
- More coherent systems-level theories, which have broader applicability and can account for unpredictable emergent phenomena

# 5. Achievable Milestones

**5 years:** New visualization tools; New tools to support collection and analysis of longitudinal data; New tools to accelerate analysis of modalities other then speech and vision (handwriting, gaze, haptics); Tools to facilitate hypothesis generation and hypothesis validation; Initial tools to evaluate the validity and authoritativeness of data; New machine learning techniques for bimodal prediction beyond audio-visual, and prediction involving three or more modalities.

**10 years:** New tools to automate the process of testing the generality of scientific findings; Formalizing data description and workflow for "data-to-discovery" process; New machine learning techniques for handling extreme heterogeneity in multimedia-multisensor data;

Demonstrations of ultra-reliable multimodal-multisensor systems for identifying cognition and health status; Demonstrations of synergistic human-computer interfaces that provide targeted correction of systematic biases in human cognition in different domains.

**15 years:** New multi-dimensional machine learning techniques for simultaneously analyzing multiple heterogeneous information sources (media, sensors), each at multiple levels; Systems-level theories in multiple domains (education, medicine); Multimodal-multisensor analytics for modeling cross-cultural conflicts of beliefs and intentions, and related systems for promoting conflict resolution

# 6. Summary of Benefits

The advancement of analytic techniques to process multidimensional data, specifically multi-level analysis of rich multimodal-multimedia information, is a cross-cutting activity that will be essential for progressing in all areas of science. New methods for mining this type of data is expected to yield a multiplicity of clues due to convergence of findings across modalities and levels of analysis, which will increase the likelihood and speed of extracting scientific insights. In addition, these methods are expected to result in formulating causal inferences that are broader in generality—leading to more powerful theories with greater applicability, as well as more systems-level theories that society currently lacks. This progress in knowledge discovery will fuel the innovation pipeline, leading to new business concepts, products, technologies, and market directions.

# Chapter 4.  **Multimodal Representation, Processing and Grounding**

**Chapter Editors:**

Joyce Chai, Michigan State University
Shih-Fu Chang, Columbia University
Raymond Mooney, University of Texas at Austin
Arnold Smeulders, University of Amsterdam

**Additional Workshop Participants:**

Louis-Philippe Morency and Jie Yang (listener)

## 1. Introduction

How can language describe the world around us? We use language almost every day, sometimes well-conceived at other times carelessly. But the amazing thing is that we can almost always find the right words to express what we mean, and to be understood by others. It is nothing less than a miracle that we can condense the infinitely complex world around us into just a few words. And, from the fact that we are able to describe almost situations, including the ones we have never seen before, lingual expressions must be powerful both in their expressive capacity as well as in their grounding of reality. In this chapter, we focus on *grounding, the connecting of lingual expressions with the continuous reality in all its complexity*. In grounding, we aim to understand how this language-miracle of coding efficiency works. How does language describe the audio or visual world? By studying grounding, we aim for a more natural and at the same time richer and deeper representations of scenes, actions, and events. This helps to make communication less ambivalent (disambiguation).

Current solutions in modern AI are capable of recognizing the concept of the audio, visual and tactile world. A computer can recognize a cow when it sees one. It can do the same for things (from forks to refrigerators), scenes (from weddings to mountains), and subject types (from Dalmatians to elephants). It has learned to do only the recent years from many labeled examples. The most important driving factor behind the recognition of all things and all concepts has been the open-innovation competitions (e.g., TRECvid by NIST). *Since recent, a machine can perform a category labelling of the world.*

A decade earlier AI - machines became capable of reasoning with symbolic statements in a logical or probabilistic manner: If the building carries a sign for "Hotel", and the door is a glass door, then that is likely to be the entrance door. In these systems, also applied to dermatology diagnosis, to quality control, and other complex reasoning, the human was performing the sensory observation to feed into the reasoning. Since longer, a machine can do symbolic inference.

A machine could reason about what the user saw and specified to the machine, or it could produce the most primitive of all lingual expressions about the world: labels. In grounding, we aim to fuse the reasoning capacity with the world-labelling capacities such that intelligent systems can generate coherent sentences and can communicate with humans in their natural way.

These enabled capabilities through grounding will benefit a variety of applications that involve:
(1) communication with autonomous systems (e.g., human-robot communication);
(2) knowledge discovery and construction (e.g., common sense knowledge or knowledge population from live sources).

In effect that is much of the long-term aims of current systems, for example in training and education, in health care, surveillance, driving, visual diagnosis, and all applications where eye-hand coordination is of relevance.

Grounding implies the discovery what words actually mean. Through categorization we now know how to find an elephant, a mountain or a bus in a picture. But we cannot solve the question yet: "find the car behind the white car". As simple as it seems, behind is not uniquely defined and in the common way of expression can easily be to the right or even before the white car. In addition, when we say "look out for the old lady", "old" is a qualifier which depends on intersubjective standards. Can the visual or audio imprint of "old" be learned independently or is old always connected to the noun it connects to? Can we learn the sentiment adjectives convey? Can we learn a cynical intonation (which implies we should do the opposite)? And, by the way, when the query phrase contains a negation, is it clear what we mean, or is negation in general ambiguously grounded in reality?

The following table summarizes key challenges and outlines a roadmap for future work on this topic.

| State of the art | Key Challenges | Road map | | |
|---|---|---|---|---|
| | | **5 years** | **10 years** | **15 years** |
| - Grounding words, phrases, verb frames and language instructions to simple environment with limited success | - Data collection, alignment, and annotation (granularity of data)<br>- Learning from limited data - > one shot or zero shot learning | - Theoretical contributions to grounding by understanding expressions of sentiment.<br>- Ground language to a | - Theoretical contributions to grounding by understanding the expressive power of adjectives.<br>- Ground language to unconstrained | - Theoretical contributions to grounding by understanding metaphoric description of new scenes. |
| -<br>- Factual image caption via deep learning<br>- Visual Relation detection and Visual QA on a limited domain<br>- Multimedia event extraction of small predefined ontologies | - Data bias, learning from subjective labels<br>- The disparity of abstraction of the different media<br>- Domain adaptation and generalization<br>- Automatic discovery of concepts in new domains<br>- Extend to other modalities beyond vision<br>- Transparency and interpretability<br>- Common sense knowledge | constrained physical environment with high success (e.g., for collaboration, task learning)<br>- generate descriptions from general videos in structured sentences with some success<br>- Visual Question Answering on videos<br>- Limited queries to ego-centric videos (e.g., police body cameras)<br>- Knowledge construction from parallel text and image data<br>- Detection of video event classes in limited domains | environment for health care, training/education applications<br>- Visual Question Answer beyond *what* questions (e.g., why and how questions)<br>- Knowledge construction and summarization from non-parallel text and multimedia data<br>- Predictive multimedia (future event prediction using video, audio, text, and contextual information)<br>- Detection of large numbers of event classes in new domains. | - Natural multimodal communication with robots (including understanding of human emotions)<br>- Continuous knowledge discovery, question answering, summarization, and prediction with high success using large multimedia multilingual streaming sources<br>- Detection of new event classes in new domains. |

# 2. State of the art

## 2.1 Grounding Language to Physical Environment

Grounding language to perception is to connect linguistic constituents (e.g., words, phrases, sentences) to the perceived physical environment (e.g., visual attributes, objects, and actions). It is related to symbol grounding (Harnad 1990) in cognitive science which argues the meanings of symbols or words should be grounded to sensorimotor interactions with the world. rounding language to physical world allows computer systems (e.g., artificial agents such as robots) to understand human language with respect to their internal representations of perception and action, to learn and reason about the physical world, and to communicate with humans in language.

Multiple levels of grounding have been explored. Words for describing color names, spatial relations, and object types and attributes are often grounded to the sensory feature space of the perceived scene (Roy 2002). For example, words can be represented by classifiers trained on visual features (Matuszek et al., 2012, Kennington and Schlangen., 2015)and compositional semantics can be applied to derive meanings for noun phrases or sentences based on the words. More recent works have applied multimodal embedding to link words with visual features (Socher et al., 2013, Frome et al., 2013, Lazaridou et al., 2014, Gella et al., 2016), where words are represented by distributional semantic vectors. The mapping between word embedding vectors and visual representation vectors can be learned from training data, for example, data with large vocabularies and sparse labels (Zhang et al. 2016). Since the embedding captures semantic similarities between words, the trained models can be used for zero-shot prediction, where image labels are not provided during training.

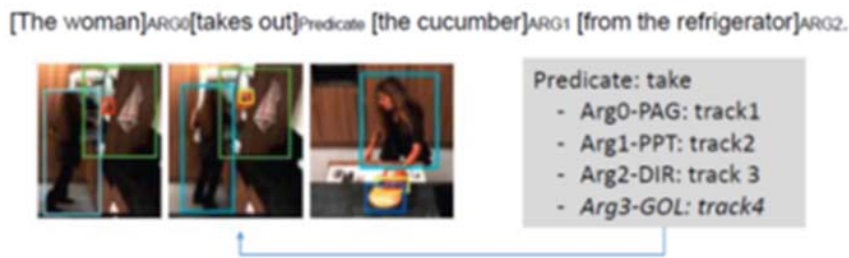

Figure 1: An example of grounding verb frames from a sentence to the video

Recent years have seen an increasing amount of work on grounding referring expressions to the perceived environment (Liu and Chai 2015) and grounding sentences (e.g., language instructions) to perceived activities in images or videos or action execution in the environment (Chen and Mooney 2011, Tellex et al., 2011; Artzi and Zottlemoyer 2013; Krishnamurthy and Kollar, 2013; Yu and Siskind, 2013, Tellex et al., 2014, Tu et al., 2014, Naim et al., 2015, Yang et al., 2016). Most works treat this process as two steps. In the first step, language is parsed into some meaning representation (e.g., frame-based semantic role representation, attributed relational graphs, first-order logic, etc.) and videos/images are processed by computer vision algorithms to identify bounding boxes of objects or tracks of objects. The second step involves mapping the linguistic meaning representation to the vision representation (e.g., through graph matching) or joint learning of such mapping using graphical models such as factorial HMMs and conditional random fields. For example, as shown in Figure 1, the semantic roles (e.g., agent, patient) of the sentence ``*the woman takes out a cucumber*'' are grounded to object tracks from the video (Yang et al., 2016). More recent work has applied deep learning, for example, to map language instructions and visual observations to actions in a single end-to-end model (Misra et al., 2017).

## 2.2 Image and Video Caption Generation

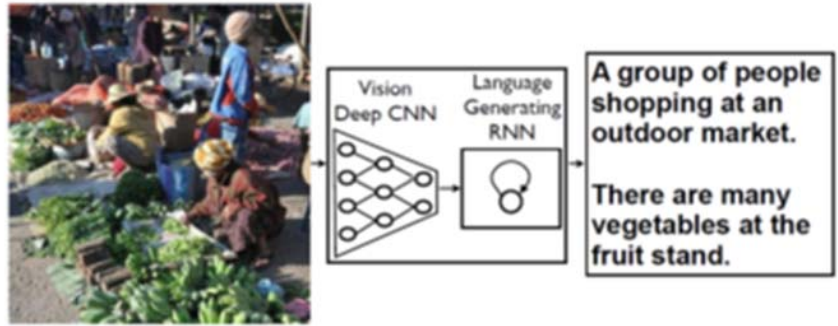

Figure 2: Sample Image Caption Generation from Vinyals et al. (2015)

One of the most well investigated problems in language grounding is the automatic generation of natural language descriptions (i.e. captions) of images and videos. There has been significant progress on this task in recent years. Most of this work has focused on image captioning, but there has also been reasonable progress on video description. A sample generated caption is shown in Figure 2.

The earliest work on image (Farhadi et al. 2010, Kulkarni et al., 2011, Yang et al., 2011, Kuznetsova et al., 2012) and video (Barbu et al, 2012, Krishnamoorthy et al. 2013) description used traditional computer vision object and activity recognizers that employ manually engineered features to determine nouns and verbs relevant to the visual content. They would then use manually engineered grammars or statistical language models learned from raw text to determine a fluent natural language sentence that incorporates these words.Recent work has focused on deep learning approaches that integrate a Convolutional Neural Networks (CNN) to extract a learned semantic visual representation with a Recurrent Neural Network (RNN) to generate a natural language sentence from this representation. The current standard architecture for image captioning (Donahue, et al. 2015, Vinyals et al., 2015) fine-tunes a CNN trained on ImageNet for object recognition (Russakovsky, 2014) to produce a "deep" vector representation and then generates a sentence from this representation using a Long Short Term Memory (LSTM) (Hochreiter and Schmidhuber, 1997) recurrent network. Similarly, a typical architecture for video description (Venugopalan, et al., 2015) uses a "sequence to sequence encoder/decoder framework" (Sutskever et al. 2014) to train a series of two LSTMS (an encoder and a decoder) to map a sequence of "deep" CNN video-frame representations to a sequence of words.

Captioning systems are evaluated by comparing automatically generated captions to human written ones using machine translation metrics such BLEU (Papineni et al., 2002) or through human quality judgements. The standard corpus used to evaluate image captioning is MS-COCO (Lin et al., 2014). State-of-the-art results on this corpus show that machine generated captions match human descriptions as equally well as human captions match each other (Fang et al. 2015); however, this is based on automatic scoring not human judgments. In human evaluation, only 27% of machine captions are as good as or better than human generated captions (Vinyals et al., 2017). Additionally, MS-COCO only includes images containing one or more objects from a set of 91 pre-specified categories. Video description systems are typically trained and tested on a corpus of YouTube videos with crowdsourced human descriptions (Chen and Dolan, 2011) and movie clips with human captions from commercial Descriptive Video Service (DVS) annotations for the visually impaired (Rohrbach et al., 2015; Torabi et al.2015). Although current captioning systems generate reasonable descriptions for many images and videos, they still perform poorly on objects and activities that are not well represented in the training data (Hendricks, et al., 2016).

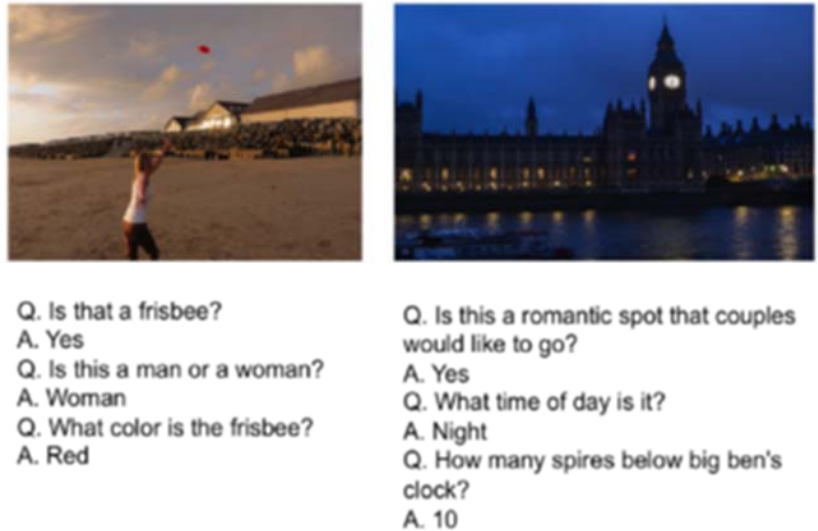

**Figure 3:  Sample VQA Images and Questions/Answers**

## 2.3 Visual Question Answering

Another important task that clearly requires connecting linguistic symbols to continuous perceptual data is answering questions about images and video, i.e. visual question answering (VQA) as shown in Figure 3.  In the last couple of years, there been significant activity and progress on this task thanks to a recently released dataset and challenge problem (Antol et al., 2015). This dataset contains over 200K images (from MS COCO) and 600K crowdsourced questions (with answers) about these images, a sample of which is shown below. To enrich the types of the questions, recently other datasets (Krishna et al., 2016) have been developed to provide dense annotations of objects, attributes, and more importantly, relations between objects.

The typical approach to VQA is a deep learning approach that uses an LSTM to encode the question, a CNN to encode the image, and a deep neural network to predict and answer from these two representations (Antol et al. 2015, Noh et al., 2016, Fukui et al., 2016). To detect the rich visual relations (e.g., spatial, preposition, comparative relations), recently models have been developed to learn the representation of visual relations in the common embedding space described earlier (Zhang et al., 2017). As shown in Figure 4, visual relations extracted from images are then used to answer questions involving both relations and entities (e.g., "what is the color of the jacket worn by the man under the clock?) Current state-of-the-art performance on the standard VQA test set is in the low 60's% correct. Therefore, current systems are able to answer interesting questions about images but are clearly not yet at human-level performance.

Many of the questions can be answered without even looking at the image (Agrawal et al., 2016); for example, if a person thinks to ask a question "Is there an X in the image?" the answer is very likely "Yes". Therefore, a new version of VQA challenge problems has recently been released that are more challenging and require seeing the image (http://www.visualqa.org/, v2.0). There has also been interesting recent work on "visual dialog" in which a system can engage a human in a conversation about an image, for example, asking a *series* of questions to identify a particular entity in the image.

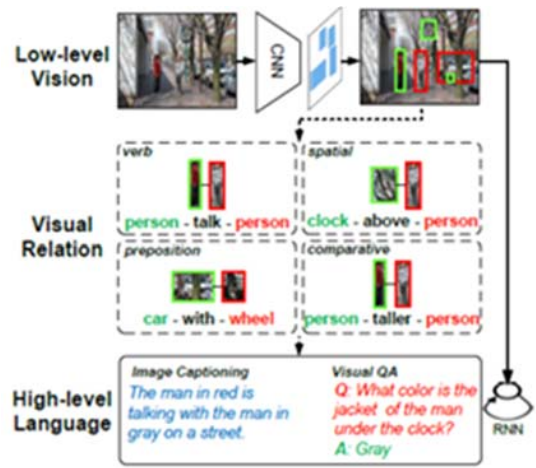

Figure 4. Visual Relation Detection and Applications in Visual QA

## 2.4 Multimodal event extraction

Recent efforts have shown interest and advances in detecting multimedia events in unconstrained videos. Such events may include "how-to" procedural events like "making a sandwich" and social events like "wedding" or "demonstration" (Jiang et al., 2013). Community-wide evaluations over predefined events or open-vocabulary classes (Smeaton et al., 2006 and Soomro et al., 2012) have attracted broad participation. Early systems relied on kernel methods like support vector machines (SVM) over a diverse set of multimodal features (visual, audio, as well as OCR text). Recent systems use deep learning models (CNN, 3D CNN, and recurrent networks) with significantly improved results. Recent efforts also showed expansion of the event ontologies to a richer pool like EventNet (Ye, et al., 2015) or ActivityNet (Caba Heilbron et al., 2015), extension of evaluation tasks from classification to precise temporal localization (Gorban et al., 2015, Shou et al. 2016, Shou et al. 2017), and application of models like GAN (Generative Adversarial Network) in predicting future events in videos (Vondrick et al., 2016).

# 3. Challenges

## 3.1 Data for Cross-Media Grounding

Deep learning methods typically require large amounts of supervised training data; however, in many applications of grounding, labeled data is limited and expensive to obtain. Many learning techniques attempt to reduce the demand for labeled data, such as unsupervised and semi-supervised learning, active learning, and zero-shot learning. It is unclear to what extent these methods can be successfully applied to multi-modal data. Grounding typically tries to exploit "found data" that naturally contain both linguistic and sensory information such as captioned or tagged images or videos, or videos with DVS. Finding additional sources of "naturally supervised" data showing natural grounding of linguistic descriptions to sensory perceptions for various applications is a key challenge (e.g., grounding of mentions of semantic instances in long news articles to objects in image tweets about the same event). The alternative is to use crowdsourcing to collect supervised data at scale. Methods for improving such crowdsourcing, such as active learning and "gamification" is another challenge. For example, Thomason et al. (2016) use an "I Spy" game as a more engaging way for human users to train a robot to learn multi-modal grounded word meanings.

## 3.2 Learning

Most current language grounding systems utilize deep learning with CNNs and RNNs. Such methods allow for complex multimedia processing while supporting end-to-end training. However, these methods require large amounts of training data and significant computational resources utilizing specialized hardware (e.g. GPUs, TPUs). They also require that complex images, video, audio, and language be represented as fixed-length vectors of real values that are opaque and therefore hard to interpret or explain.

Many complex scenes, events, and situations involve multiple entities and activities with a rich relational structure. Traditionally, semantic networks (a.k.a. knowledge graphs) or logical formulae are used to represent such rich semantic structures. Trying to "cram" such complex representations into fixed-length vectors seems highly suboptimal. However, automatically learning such rich structures is challenging, particularly in a framework that allows end-to-end training of a complete, complex processing pipeline in order to optimize some ultimate performance metric.

There has been some recent progress developing "hybrid" systems that integrate traditional structured relational representations with deep learning, such as the VQA system of Andreas et al. (2016) that composes "modular" neural networks from independently learned components. However, such systems generally require manually specified structured representations (such as a syntactic parser trained on supervised parse trees). A challenging problem for future research is the development of a hybrid system that learns its own structured representations for a complex processing pipeline, while still supporting end-to-end training that optimizes overall performance.

## 3.3 Representation (i.e., disparity in representation across different modalities)

Despite the recent advances in recognizing semantic classes in sensory data like images and videos, there are still great disparities in representations in language and vision. For example, entities detected in visual content are typically generic (e.g., "a car", "fire") except for the named entities like databases of named faces or named landmarks. Linking such generic entities to specific instances mentioned in language (e.g., "my Ferrari car", "2016 California forest fire") will be very useful in fusing information extracted from sources of different modalities. In addition, information such as organizational, transactional, and causal attributes extracted from textual documents is at a much higher level than those extracted from sensory data. Reconciliation of such disparities and fusion of the complementary information from multimodal sources will be important.

## 3.4 Modality (i.e., beyond language and vision)

Almost all work in grounding has worked on connecting words and symbols to visual data. Grounding in other modalities such as audio, olfactory, gustatory, and haptic (i.e. touch and proprioceptive) sensory data has not been well investigated. Gathering robotic sensory data at the scale required for deep learning is particularly challenging. It is also unclear whether the techniques used for visual grounding (e.g. CNNs) will work well for these other sensory modalities.

## 3.5 Transparency and Interpretability

When grounding language to perception , many earlier approaches   first treat language processing and vision processing separately and then combine their partial results together. This pipelined  integration relies on feature engineering that identifies informative features to train the model for joint inference or grounding. It often addresses how to effectively combine multiple modalities together to support mutual disambiguation. In general, the pipelined architecture is more transparent in terms of its internal representations and relatively easier to interpret system behaviors. More recent approaches to joint language and vision processing apply neural networks and deep learning. These approaches have the advantage of learning representation without feature engineering and empirically have shown better performance in various tasks (e.g., image description generation and visual question answering). However, these approaches also have problems with transparency and interpretability, which need to be addressed in the future.

## 3.6 Common Sense Knowledge

Lack of common sense knowledge is one of the major obstacles in developing AI systems, which also applies to grounding language to perception and action. In human-human communication, common sense knowledge is often presupposed. For example, in the sentence ``*the woman takes out a cucumber*" (as shown in Figure 1), the agent role and the patient role of ''take out" are explicitly specified, but not the destination role. However, in order for computer systems (e.g., a robot) to fully understand this action,  the destination role will also need to be grounded to the perceived environment (Yang et al., 2016). Recent work has explored physical causality knowledge related to action verbs (i.e., "slice a cucumber'' implies the state of the cucumber after the action will change from one piece to several smaller pieces) (Gao et al., 2016) and other types of physics implied by action verbs (e.g., "he threw a ball" implies that "he" is bigger, heavier, and faster than "the ball") (Forbes and Choi 2017). The idea is that this type of knowledge can potentially provide guidance and additional cues for AI agents to ground language to perception and action. However, despite the availability of many different types of knowledge bases, the very basic principles or knowledge about how the physical world works are still limited for AI systems.

## 3.7 Domain adaptation

In most applications, the data in the final test application is drawn from a somewhat different distribution than the available supervised training data. There are a wide variety of machine learning techniques for domain adaptation or transfer learning that attempt to address this issue (Pan & Yang, 2010, Jiang, 2008). However, it is unclear how these techniques apply in a multimodal setting, in particular, should each modality be adapted independently, or should adaptation should be done jointly across all modalities?

Another major challenge incurred when changing application domains is the lack of a predefined set of classes (objects, scenes, events, and relations) over which annotated data can be collected and detection models can be learned with certain optimization metrics. Given a new multimedia corpus without such predefined ontologies, though certain tasks can still be attempted such as open-vocabulary caption generation, QA, or zero-shot semantic retrieval, construction of knowledge bases reflecting salient semantic entities in the new domains will be difficult. Thus, automatic techniques for discovering new concepts, relations, and their representation structures in new domains will be highly desirable. For example, without prior manual specification, can machines automatically discover new entities, events, and associated syntactic/semantic rules by analyzing and mining through streams of images, videos and related documents in a new domain (e.g., soccer, baseball, etc.).

Solutions in this direction will greatly contribute to the ambitious goal of automatic knowledge construction. Recent work such as (Li et al., 2016) developed a joint deep learning and data mining model to automatically discover event-specific visual concepts and automatically name such discovered concepts based on a cross-media joint representation. The method discovered new event types, subtypes, event-specific arguments, and their roles. However, extending such methods to discover rich knowledge structures and handle diverse open-source multimodal data (beyond image-caption or video-transcript pairs) remain major challenges and opportunities.

# 4. Milestones

## 5 years:

**Theoretical contributions to grounding by understanding expressions of sentiment.** One of the key components of grounding is to understand the expression of sentiment as it will color all the objective part of the communication. There is promising work on fixed adjective-noun pairs (like beautiful sunset and happy couple) (Borth et al, 2013). The target here is to enhance the scope to non-fixed adjective noun pairs.

**Grounding language to constrained physical environment with high success.** Efficient algorithms and tools for constrained physical environments (e.g., home, hotel, etc.) with well-trained computer vision models. These algorithms could be applied to offline data to support search and analytics. They can also be applied to applications where real-time grounding is essential (e.g., human-agent language-based communication about the shared physical environment).

**Grounding of structured sentences** with limited degrees of freedom to describe new scenes with a limited repertoire of states. Structured sentences help to solve the unbounded scope of language to zoom in and solve one of the dimensions of contextual ambiguity.

**Generation of captions for images and video in a wide variety of domains that are similar in quality to human descriptions**. In particular, a useful milestone would be generating human-quality descriptions of first-person video (from a wearable camera or police body camera). This could be useful as an aid for visually impaired individuals, or for producing searchable logs of police activity.

**High accuracy visual question answering (VQA) for video.** Again, this might be particularly useful for first-person video to support enquires from the visually impaired, video surveillance data, police body camera data, or recorded "life log" video data from ordinary users' daily experience.

**Knowledge construction.** Automatic discovery of entities, relations, and event types or subtypes from textual documents as well as closely coupled multimodal data (e.g., news images plus captions, social media videos plus speech transcripts or titles) in new domains. These can be used to construct or update ontologies or schemas in an automatic or semi-automatic fashion.

**Video event detection in limited domains.** Extend the current event detection methods to handle a large number of event classes in new domains with few training samples (such as videos from body worn cameras and life-log videos).

## 10 years:

**Theoretical contributions to grounding by understanding the expressive power of adjectives.** Adjective are one of the hardest components to grasp in language. The visual component of "old" is very different for "cars" as they are for "cheese" or "man". Yet we perceive such immediately even for unknown and never seen before categories. The target here is to understand adjectives the visual grounding thereof among different categories.

**Grounding language to unconstrained physical environment.** Systems that can ground language to unconstrained physical environments where computer vision models for the environment can be updated through incremental learning or zero-shot learning.

**Causal Visual Question Answering.** Systems that can answer more difficult "why" and "how" questions about images and video, such as "Why did the man go to the kitchen?" (Answer: To get a cup of coffee) and "Why did students wear caps and gowns? (Answer: It was the university commencement day.).

**Multimodal question answering (MQA).** Systems that reliably answer questions from diverse multimodal input such as multiple video streams with audio and associated social media streams. May be particularly useful for smart city applications where there are many surveillance cameras and microphones, social media post streams, and crowdsourced streams.

**Multimodal knowledge discovery, construction, and summarization.** Automatic discovery and construction of knowledge from loosely coupled multimedia, multilingual, and multi-source data sources (beyond image-caption pairs or video-speech transcripts in limited domains).

**Multimodal future event prediction.** Prediction of event occurrence at different time scales in the future in specific domains (e.g., self-driving cars, life-logs) using video, sound, text, as well as contextual information.

**Large numbers of video event detection in new domains.** Extend the 5 year goal into the detection of new, never seen before domains. Demonstrate the detection capabilities and transferability of models into existing ontologies.

## 15 years:

**Theoretical contributions to grounding by understanding metaphorical descriptions of new scenes.** Metaphoric expressions are needed to describe a not yet seen before scene, such a microscopic image or the scene after a disaster. In fact, all description will include some form of metaphor as the lingual system is very different from numerical features of reality. Some form of metaphor is needed as a foundation to grounding.

**Natural multimodal communication with robots.** Systems that can ground human language not only to robots' perception system (e.g., ego-centric vision), but also the action system (so that robots can perform actions to change the perceived physical world). These systems will allow humans to communicate with robots in language, teach robots new tasks, and collaborate with robots to achieve joint tasks. The systems will also have the abilities to understand high-level concepts such as human goals and emotions.

**Continuous knowledge discovery, summarization, and prediction.** Systems that are able to automatically discover, summarize, and verify knowledge (including entity, action, as well as emotion) from large distributed multimodal multilingual sources in a continuous fashion. They can be used to construct and update knowledge based and to reason, inference, and predict outcomes of future events.

**Detection of new video events in new domains.** Extend the 10 year goal into the detection of new, never seen before domains and new, never seen before events therein. Demonstrate real-time live detection capabilities and transferability of models learned over existing ontologies.

# Chapter 5.   **Person-Centered Multimodal Interaction**

**Chapter Editors:**

Michelle Zhou, Juji Inc.
Hari Sundaram, University of Illinois Urbana-Champaign

**Additional Workshop Participants:**

Terry Adams, Dick Bulterman, Carlos Busso, Joyce Chai, Julia Hirschberg, Tatiana Korelsky (listener), Ketan Mayer-Patel, Shri Narayanan, Sharon Oviatt, Arnold Smeulders and Zhengyou Zhang

## 1. Introduction

In recent years, hundreds of millions of people have left their multimodal digital footprints online—Facebook posts, YouTube uploads, and myriad of mobile health and e-commerce application data. These footprints are a veritable cornucopia of data.

In this chapter, we use the word "*multimodal*" to refer to multiple forms of user-generated data, including text, images, and videos. Such data captures diverse user behavior through different interaction modalities, including spoken utterances, facial expressions, and gestures.

These unprecedented amounts of multimodal digital footprints enable a better understanding of individuals and their unique characteristics, such as personality, motivations, and interests, at scale.  In turn, such understanding of individuals enables person-centered, multimodal interaction between human beings as well as between an individual and an Artificial Intelligence (AI) agent.

A sophisticated understanding of individuals and their associated behavior, including inferring their psychological characteristics and decision making-strategies, will have a profound impact on our science directions as well as on our society. Let us illustrate the impact at different scales: billions of individuals and millions of communities.

At an individual level, for example, a physician can better decide on a treatment plan for a particular patient based on a fine-grained understanding of the patient's motivations and her abilities to cope with stress. A virtual interview agent based on artificial intelligence (AI) technologies may engage with a person in a multimodal conversation to collect information about the person's interpersonal skills, her abilities to deal with stress, and then recommend career opportunities or development plans that suit her best.

At a community level, government agencies as well as corporations are interested in promoting certain population behavior changes in several areas. Consider the alarming rise of child obesity in developed countries including the U.S. Obesity can cause irreparable harms to the physical and mental development of these children. In addition, it harms the productivity of the future workforce and strains the demands on the nation's healthcare system. Imagine the use of multimodal communications tailored to each individual as well as families to "nudge" them to adopt more healthy eating and exercise habits can positively impact the well-being of a nation's population. Governments, healthcare providers, and insurance companies are stakeholders in improving the nation's health at scale. Corporations are motivated since the ability to understand and engage with each of their customers as a unique individual will optimize their business outcomes and improve customer satisfaction.

## 2. Research Topics

To understand the distinguishing characteristics of individuals from their multimodal digital footprints, we need to develop a suite of intelligent multimodal technologies that broadly fall in the following three research areas:

- **Multimodal Human Behavior Analysis**. In both human-human interaction and human-machine interaction, humans often use verbal and non-verbal, multimodal cues to communicate information and express intent. Multimodal human behavior analysis is to understand the complex relationships across the modalities and infer the underlying communication meaning and intent. One example of such analysis is to interpret human affective states, while the other example is to infer human agreement or disagreement intent.
- **Computational Psychology from Multimodal Human Behavior**. Beyond interpreting human multimodal behavior, another related research topic is to infer the inherent and developed psychological characteristics of humans from such behavior. For example, one goal is to infer one's personality traits and motivation based on her multimodal communication behavior, expressed through text and/or videos.
- **Person-Centered Multimodal Persuasion**. Based on the interpretation and inference of human multimodal behavior and psychological characteristics, another research topic is to develop multimodal persuasion technologies tailored to individuals. This touches upon how to tailor multimodal persuasions (e.g., words and video) based on one's current behavior (e.g., expressing disagreement and disgust), inherent psychological traits (e.g., impulsive temperament), and developed characteristics (e.g., changing psychological needs or evolving cognitive skills).

## 3. Drivers and relevant applications

The technologies introduced above are applicable across a wide variety of domains and industries, from healthcare to education to marketing, to enable person-centered decision-making and optimize business outcomes. While Table 1 summarizes a list of applications, we mention a few example applications below:

**Healthcare**. Technological advances in the areas mentioned above will enable healthcare providers to better understand patients, including their current state and psychological characteristics, and use such insights to develop person-centered care plans. Governments will be keen to shape healthcare policies and develop interventions to fight unhealthy behaviors (e.g. smoking, poor exercise, and eating habits) and pandemics that affect the welfare of an entire nation.

**Marketing**. Similarly, technological advances in the areas mentioned above will enable the understanding of a target audience as unique individuals at scale and the creation of tailored persuasions based on individual differences.

**Education**. Educational institutions and beyond can better understand students as unique individuals and then offer person-centered guidance and learning materials.

Since human communication is inherently multimodal, multimodal research not only will bring unique perspectives in interpreting and inferring human characteristics and predict human behavior, but also facilitate the development of effective persuasion technologies that will facilitate better human-human and human-computer engagements. In particular, multimodal research makes unique contributions in the following three aspects:

**More accurate interpretation of human behavior**. The whole is bigger than the sum of parts. Analyzing human multimodal input—verbal and non-verbal—based on the relationships among the modalities enables a more accurate interpretation of human behavior than unimodal analysis.

**Deeper inference of human characteristics**. Human multimodal behavior encodes richer information and captures human characteristics more comprehensively than unimodal behavior. Analyzing human multimodal behavior enables a deeper, and more accurate inference of inherent and evolving human characteristics, e.g., inferring one's personality from both verbal and non-verbal communications.

**More effective persuasion**. Persuasion technologies that integrate multimodal communication channels are more effective for two main reasons. First, persuasions may be delivered in different modalities to communicate different types of information effectively (e.g., text for communicating abstract information while images for conveying visual information) [Zhou & Pan 2005]. Second, persuasions may be delivered to different people using different modalities or different combinations of modalities based on their characteristics (e.g., people with different cognitive style may prefer different types of persuasive information) or context (e.g., primarily visual in a noisy and crowded environment) [Cialdini 2006].

| Applications | State-of-Art | Key Challenges | Road Map | |
|---|---|---|---|---|
| | | | 5 Years | 10-15 Years |
| • Healthcare<br>• Marketing<br>• Education<br>• Human Capital Management<br>• Security | • Limited multimodal human behavior analysis<br>• Computational psychology from human unimodal behavior<br>• Manual composition of persuasion technologies<br>• Limited methodologies for sampling content on graphs; field experiments<br>• Data-driven view of persuasion<br>• Information Asymmetry | • Person-centered data sparsity including a lack of longitudinal data<br>• Robust and extensible human multimodal behavior analysis<br>• Cross-disciplinary methods for interpreting human multimodal behavior and inferring human characteristics<br>• Cross-cultural understanding and interpretation of human multimodal behavior<br>• Automated synthesis of multimodal persuasions<br>• Methodological: dealing with data skew, data sparsity, effective sampling from graphs; ability to conduct field experiments at scale<br>• Cross-disciplinary: integration of data analysis with behavioral economics, advertising and algorithmic game theory<br>• Protocols to ensure privacy; User interfaces | • Improved algorithms for analyzing human multimodal behavior from diverse data<br>• Detection of 50 basic human multimodal activities<br>• Limited understanding of human psychological characteristics and qualities for specific tasks and context<br>• Semi-automated synthesis of multimodal persuasions; Preliminary results from integrating ideas from behavioral economics, game theory with persuasion technologies at a small scale<br>• Methods to effectively sample social graphs with content, dealing with sparsity<br>• A privacy preserving protocol; UI for easy specification of differential privacy constraints | • Individualized multimodal engagement for person-centered decision making<br>• Ubiquitous hyper-personalized multimodal experience in real and virtual worlds<br>• Multimodal personal pet and companion<br>• Methodologies to influence behavior at large scale.<br>• Universally implemented protocol that grants individuals ability to not be tracked; interfaces that allow them to easily specify what can be shared. |

Table 1. Summary

# 4. State of the art

Here we highlight the state-of-the-art technologies in the three research areas mentioned above. This is not intended to highlight individuals' research. Instead, it is the collective assessment of the field.

- **Limited multimodal human behavior analysis**. Currently there are algorithms that can recognize specific human multimodal behavior [Bousmalis et al. 2013], although such recognition is often limited to particular human activities or in a particular context.
- **Computational psychology from human unimodal behavior**. There are various algorithms that can automatically infer human psychological characteristics from human unimodal behavior. For example, certain algorithms can infer personality, motivations, and basic values from one's text communication [Gou et al., 2014, Li et al., 2017], while others can infer personality from one's voice (e.g., [Lepri et al. 2012]), images (e.g., [Al Moubayed et al. 2014, Liu et al., 2016]), or video (e.g., [Batrinca et al. 2011, Lepri et al. 2012]). So far such inferences are limited to the use of human unimodal behavior and to the inference of basic human characteristics, such as the most well-known Big 5 Personality Traits.
- **Limited composition of multimodal communications**. There are algorithms and methodologies that can automatically synthesize multimodal communications during a human-computer interaction. For example, there are algorithms that synthesize multimodal information presentations, including both verbal and visual presentations for a target audience in a specific context (e.g., [Zhou et al., 2006]), or compose multimodal expressions, including verbal and non-verbal expressions in an Avatar for its interaction with a user (e.g., [Cafaro et al. 2016]). However, such multimodal synthesis algorithms or methods are rather limited to a particular task and context.
- **Limited methodological frameworks.** We can sample from large social networks, but the design of current sampling schemes preserves network structure not structure of the content. We can conduct large scale experiments online, but not in the physical world.
- **Data-driven persuasion.** Today, computational persuasion relies on data analysis at scale; but it ignores research in behavioral economics on human decision making.
- **Information asymmetry.** Businesses (e.g., advertisers) know about individuals' online behavior, but tracked individuals have little say in restricting the data collected and its use.

# 5. Challenges

To facilitate person-centered multimodal interaction, no matter in human-human or human-computer interaction, there are several key challenges.

- **Person-centered data sparsity and a lack of longitudinal data**. While there is abundant of data generated by people online, it is often difficult to find sufficient data centered around an individual, which can then be used to infer a reliable understanding of the individual let alone understanding the development of that individual. This is because majority of people online are consumers of data instead of producers of the data, while collecting longitudinal data about a person requires a long-lasting platform.
- **Robust and extensible human multimodal behavior analysis**. So far existing algorithms for human multimodal behavior analysis are trained for recognizing specific human activities in a particular context. It is often difficult to reuse not alone extend such algorithms for the analysis of human multimodal behavior for different activities for in a different context. Research advances are required to make such analysis algorithms more robust and extensible in analyzing human multimodal behavior especially across tasks and contexts.
- **Cross-disciplinary methods for interpreting human multimodal behavior and inferring human characteristics**. Inferring human psychological characteristics and qualities, such as personality, cognitive, and social styles, from one's multimodal behavior requires the collaboration of multiple disciplines, including Computer Science, Psychology, and Social Science. Establishing collaborations across multiple disciplines is never easy and often requires multi-level support, including both financial and organizational support.
- **Cross-cultural understanding and interpretation of human multimodal behavior**. People under different cultures may use multimodal interactions differently (e.g., some are more verbal-oriented while others use more gestures or visual cues). In addition, people under different cultures may even use similar multimodal interactions to signal totally different information or imply different intent. However, existing multimodal behavior analysis algorithms and methods hardly take into account cultural influences to accurately interpret human multimodal behavior and characteristics across the globe.
- **Automated synthesis of multimodal persuasions**. While there are many theories on synthesizing multimodal persuasions, few computational approaches are developed to synthesize multimodal persuasions and tailor such persuasions to different audiences for different tasks or under different contexts or cultures. The multimodal community can contribute to developing such computational persuasion approaches uniquely because the community has already been focusing on understanding the effects of different modalities and their integrated effects on people.
- **Limited methodological frameworks.** A large number of people use social networks, such as Twitter or Facebook. While there are efficient samplers for sampling graphs to preserve network structure (Leskovec and Faloutsos, 2006, Hubler et al., 2008), uniform sampling, the gold standard for content, is infeasible on graphs due to a lack of random access. Second, uniform sampling is sub-optimal for data mining content, since online data exhibits data skew (Kumar and Sundaram, 2016). While performing large scale experiments on Amazon Mechanical Turk is straightforward, performing in-situ field experiments in the physical world at scale is hard. For example, how will today's busy individuals with limited time and attention respond to persuasive messages while shopping at a grocery store, tending to their baby, or walking and talking on their mobile? Today we can conduct these experiments in the laboratory, but not in the field.

- **Data-driven persuasion.** Much of the focus in the internet industry is on extensive data collection of individuals' behavior, including understanding of their demographic and behavioral characteristics. While data is essential for persuasion, we need to supplement data analysis with ideas from human psychology [Cialdini 2006] and behavioral economics (Tversky and Kahneman, 1981, Mullainathan et al., 2008), such as information framing, prospect theory, and scarcity. More importantly, it is a challenge to "implement" such theories to guide the generation of persuasions.

- **Privacy protection.** As we learn more about each individual, we need to empower them to make decisions about their privacy (Kravets et al., 2015). For example, one may be comfortable letting her friends know that she is visiting a specific shopping website, but she may want advertisers tracking her to know that she shopped without specifics. Or as she prefers, advertisers cannot track her shopping behavior. Today, there is information asymmetry between tracked individuals and advertisers who track them (Turow, 2012). Individuals know little about who are tracking them, the data collected, and the use of such collected data.

# 6. Milestones

With the continuing accumulation of personal data, we envision major progresses or milestones to be achieved in the following time frame.

## 5-year milestones

- **Improved algorithms for analyzing human multimodal behavior from diverse data**. Within the past several years, researchers have been making progress on coming up with different algorithms to tackle the challenges of interpreting diverse human multimodal behavior. We would expect to have continuing advances in this area with more effective and efficient algorithms for human multimodal behavior analysis including dealing with multi-source, heterogeneous data across domains.

- **Detection of 50 basic human multimodal activities**. As the analysis algorithms improve, the community as a whole should be able to identify at least 50 basic human multimodal activities, which can then be used for various applications, e.g., doctor-patient engagement or teacher-student interaction.

- **Limited understanding of human psychological characteristics and qualities for specific tasks and context**. Since we are able to understand human psychological characteristics from certain modalities, we expect this area of work to advance and infer human psychological characteristics in certain tasks or context (e.g., a person's openness to experience in the context of healthcare vs. commerce) from human multimodal behavior.
- **Semi-automated synthesis of multimodal persuasions**. With the better understanding of individuals' behavior and characteristics, synthesis algorithms or systems can then be built to guide the generation of multimodal persuasions with a human in the loop. For example, a marketer may synthesize multimodal persuasions to a target audience based on the guidance given by a system on the characteristics of the target audience as well as effective persuasion techniques (e.g., using verbal vs. visual communication cues). Integration of large scale data analysis techniques for creating persuasive messages with ideas from behavioral economics (information framing, decision making under scarcity), and algorithmic game theory (mechanism design).
- **Methodological**. Content-aware task-specific samplers from graphs that deal with data skew and sparsity. Design of mobile application frameworks that allow researchers to deploy field experiments at scale.
- **Privacy.** Development of protocols that ensure differential privacy settings of an individual. User Interfaces that allow individuals to specify policies that regulate any data collection agencies and the use of such collected data.

## 10-15 year milestones

- **Individualized multimodal engagements for person-centered decision making**. With technology advances in the areas mentioned above, we envision a completely new individualized engagement paradigm that utilizes multimodal interactions to facilitate person-centered decision making. For example, in remote medicine or distance learning where doctors/teachers will engage patients/students in multimodal interactions and such interactions are automatically analyzed in real time to infer the patients' or students' characteristics and needs, which will enable doctors and teachers to devise individualized treatment or coaching plans.
- **Ubiquitous hyper-personalized, multimodal experience in real and virtual worlds**. Every individual desires different experiences in their life based on their own needs and wants as well as the context. For example, when visiting Paris, some people prefer a more refined, culture-oriented experience while others desire a rugged, adventurous experience. Based on the given context (e.g., time of the year and tourist volume) and individual desires, hyper-personalized experiences will be created and delivered to different individuals in both real (e.g., physical places to visit in Paris) and virtual worlds (e.g., shared experiences with others).

- **Multimodal personal pets and companions**. With the advances of intelligent multimodal interaction technologies, we also envision a new generation of intelligent systems, with or without a physical form, which will become people's new companions or pets. Such a companion will fully understand what a person wants and helps a person fulfill his/her psychological or physical needs as desired.
- **Methodologies to influence behavior at scale**. Be able to set declarative goals at scale (e.g. specify a 25% decrease in a city's carbon footprint through behavior change over the next five years)
- **Privacy**. Universally implemented protocols that all companies will follow will grant individuals ability to opt out of tracking and to be able to specify differential outcomes for different agents. Interfaces that allow individuals to set policies that govern sharing of their data interactions with dense IoT environments.

# Chapter 6. **Multimedia Content Generation**

**Chapter Editor:**

Dick Bulterman, Centrum Wiskunde & Informatica

**Additional Workshop Participants:**

Carlos Busso, Joyce Chai, Shih-Fu Chang, Raymond Mooney, and Michelle Zhou

# 1. Introduction

This chapter considers the various means in which people and information processing agents generate multimedia content. While content generation is a broad concept, we are particularly interested in situations where content is generated on-demand for a particular user (or groups of users) based on merging a flexible content model with a user profile. The goal of such content generation is to provide a customized, personalized and tractable user experience for simulation, synthesis or entertainment purposes. The content may be generated in one atomic operation or iteratively in a series of successive-refinement steps. In this chapter, our narrow definition of content generation is **applying personal or situational preferences to a content model in order to generate a coherent (multimedia) experience**.

Content generation has historically been a handcrafted activity that is strongly dependent on the relatively wide distribution of a single content projection. A good example is a feature-length movie production, in which a script is interpreted by a production crew (including directors, actors, audio/visual/lighting/scenic/costume and other area-specific professionals), then refined in targeted editing post-production activity, and then finally packaged for distribution via a collection of distribution channels (such as movie houses, video-on-demand/DVD distribution, or adaptation of hybrid media types). In this process, there is no fundamental restriction that each step results in a single content projection.

A formal framework for content generation that will allow for flexible composition or reuse will provide the following benefits to both producers and consumers of content:

(1) A single story line may include multiple content paths that would allow for the reuse and reapplication of content object;
(2) The interests/backgrounds/abilities/contexts of the use could be taken into account in order to generate tailored experiences that meet the needs of particular groups;
(3) Information can be enriched or localized by applying model refinements or by integrating third-party experiences;
(4) By providing tractable adaptation choices, users can gain additional insights into why particular content was generated for them and help them to better understand the information flow they receive.

Advances in content generation technology with support the following industrial segments:

(1) Content production, content storage, content archiving;
(2) Information sharing and enriching in across a wide spectrum, from education to entertainment, from tourism to training, and from professional to peer-level communication.

These applications will have societal/global impact in all areas when complex information can be scaled and modified to meet the contextual needs of users. A key element in this process is not only supporting a mapping from an abstract content model to a particular user's instance, but also in providing a mechanism to allow a user (if desired) to understand how information was tailored to meet his/her needs based on some model of transformation provenance.

The following table summarizes key challenges and outlines a roadmap for future work on this topic.

| State of the art | Key Challenges | Road map | | |
|---|---|---|---|---|
| | | **5 years** | **10 years** | **15  years** |
| - Repurposing existing content<br>- Generating simple content descriptions from existing content<br>- Generating simple interactive and dynamic content sequences from narratives<br>- Creating content from context in games<br>- Synthesizing new content from collections of user-generated content. | - Labeling content for reuse and integration.<br>- Moving beyond visual data for interpretation and understanding<br>- Supporting automated modality transformation for accessibility and understanding.<br>- Supporting ultra-personalization of content.<br>- Support context modeling in addition to content modeling.<br>- Support information provenance. | - Support for common labeling of content within domains<br>- Adding support for sensor data as media.<br>- Allow ultra-personalized identification of content<br>- Support for the explicit modeling of context<br>- Support for understanding the authenticity of content fragments. | - Support for multi-view and time-variant labeling of content within domains<br>- Process rich media in the compressed domain.<br>- Allow personal control over reuse of personal media in non-local environments.<br>- Support for multi-view and time-variant context models<br>- Support for understanding the authenticity of content collections. | - Support for multi-view and time-variant labeling of legacy content within domains<br>- Process legacy rich media in the compressed domain.<br>- Allow tracking and verification of personal media in non-local environments.<br>- Support for multi-view and time-variant context models in legacy content<br>- Support for understanding the authenticity of legacy content collections. |

# 2. State of the art

Since the advent of directed multimedia content generation research, progress has been made across a wide range of component disciplines. In this chapter, we focus on the state of the art in five areas: traditional media experiences, games and entertainment, news and information sharing, and integrated professional/user-generated content.

## 2.1 Generating content from existing media experiences

One of the first issues addressed under a broad definition of content generation was the repurposing of existing content (often in non-digital formats) to an increasing diverse set of end user devices. One of the earliest uses of multimedia data was the repurposing of content that was generated for the film and broadcast industries. Content generation in this context typically consisted of adapting existing media for the spatial and bandwidth constraints of (for the time) non-traditional delivery platforms, such as PCs and mobile devices. Rather than speaking of content generation, such uses can be more correctly considered to be content adaptation for the specific task of content repurposing (El Saddik and Hossain, 2006). Such work depended on a series of seminal results in content coding and compression that, in many ways, formed the foundation for the digital content revolution (Poynton, 2012).

Targeted content generation in early multimedia systems was often constructed as an authoring activity (Bulterman and Hardman, 2008). While rich authoring systems were developed using informal or formal models, the bulk of multimedia content shared across the Internet remains the distribution of lightly-edited content saved directly from media capture interfaces, such as smartphone cameras and microphones.

## 2.2 Generating content from content

Once content was available for wide-scale sharing, a significant challenge became making content accessible to this wide audience. As with conventional repurposing, this required content to be reformatted for a wide range of devices and encoded in transportable formats. From a content perspective, new problems included summarizing content, either for general purpose content (Shipman et al., 2003, Lienhart et al., 1997, Zhu et al. 2004), or for specific domains (Kopf et al., 2004, Zhang and Chang, 2002), and the (automated) development of alternative content representations, often to support access within the assisted technology community (Hersch and Johnson, 2008). These summaries, which include reducing the temporal or semantic complexity of content, or making them available in different modalities, have enable content to be made available for specific tasks in education and training, and for supporting the personalized delivery of content.

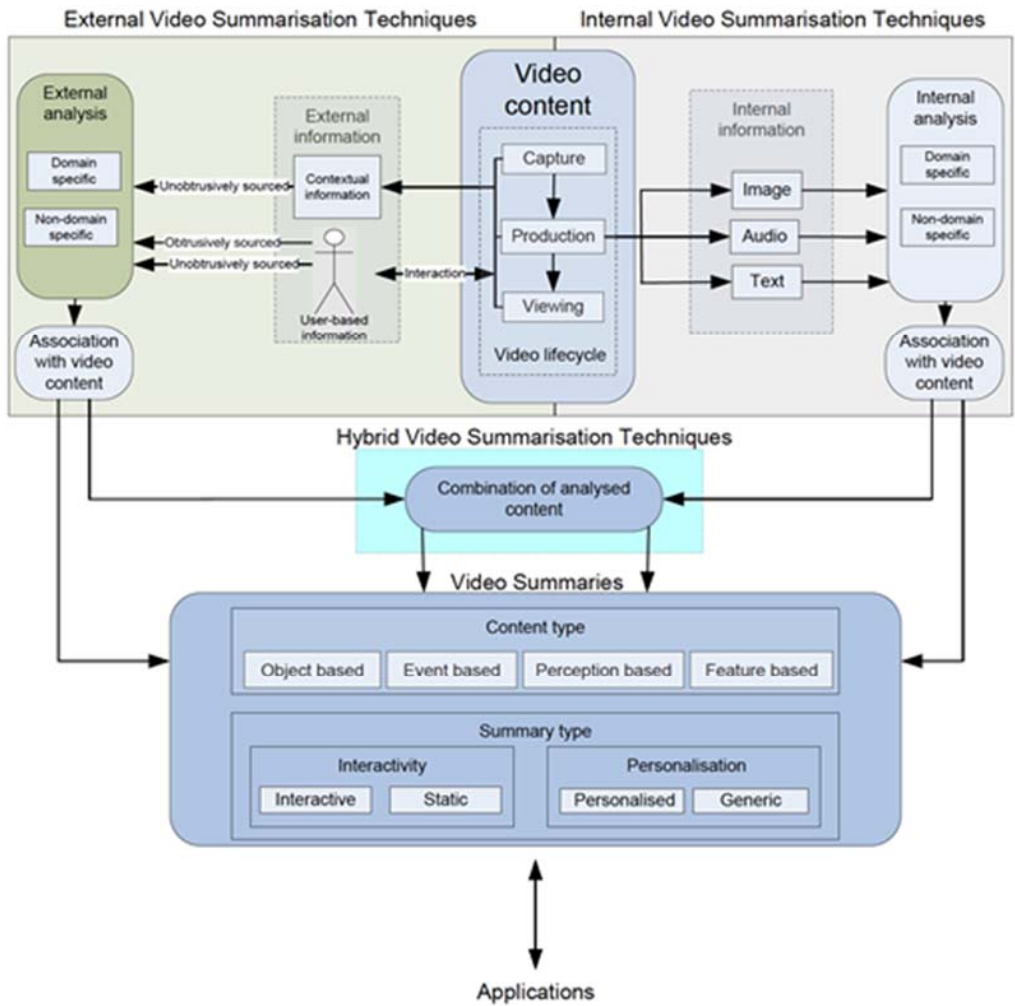

Figure 1: Generic models for generating content summaries (Money and Hargis, 2008).

## 2.3 Generating content from narrative structures

The production of any form of content is explicitly or implicitly based on a narrative structure of that content (Aristotle, 350 BCE; Freytag, 1863). While the narrative structure in conventional content is typically linear (that is, the story starts, the story unfolds, the story ends), several attempts have been made to support interactive narratives, in which the unfolding of the plot is influenced by user interaction. An early experiment in such non-linear narratives was Činčera's *One man in his house*, a production made for the Czechoslovak pavilion at Expo 67 in Montreal. Using a device called a Kinoautomaat (or, automatic cinema), a film was shown to a live audience. At nine times during the production, the audience was able to vote on which of several pre-defined branching paths was to be followed. All paths we crafted to provide a coherent story line. A more recent version of the same principle was followed in the Finnish production Accidental Lovers, with the modernization that users could vote via text messages (Zsombori, et al. 2008).

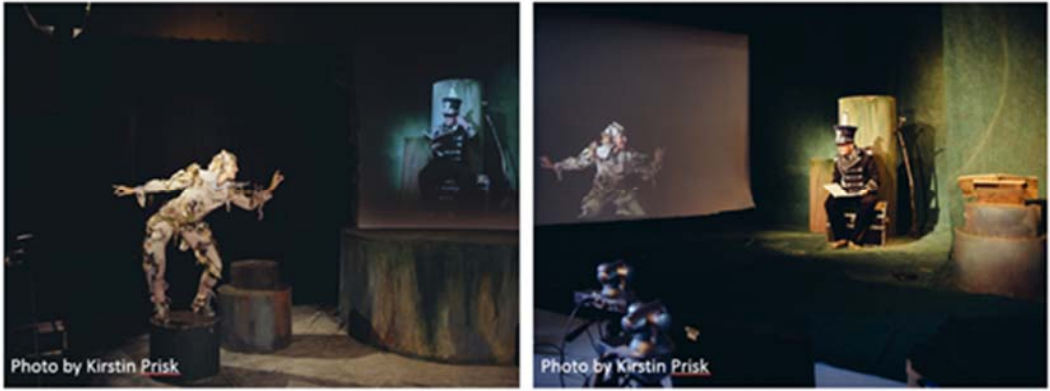

Figure 2: Generic multi-view projections based on narrative structure (Williams et al, 2015).

The high cost of production that accompanies the branching model of narrative generation has stimulated work in semantic modeling of stories and content, with the end user being an explicit or implicit agent in selecting content. Several choices guide such dynamic content generation: does every version, given the same control inputs, deliver the same experience, or is each content experience unique? That is, is a new narrative projection created each time (which can inhibit the development of shared social experiences in favor of highly personal ones), or is a branching tree developed – and perhaps exposed to the end-user – that makes the user an explicit co-creator?

Narrative-based content generation shares many concepts with hypermedia models. While successful hypermedia is often limited to hypertext, this is only true because text has efficient transfer and loading properties (Gao et al, 2008). The use of narrative structures as a basis for content generation is not limited to entertainment applications. Increasingly, narrative structures are being used to define navigation models for education content and medical diagnoses. In many ways, pre-defined and dynamically generated narratives, in spite of early enthusiasm, remain plagued by the high costs of producing compelling content.

## 2.4 Generating content based on context

By far the most sophisticated application of dynamic content generation has not been in the development of content based on narratives, but in generating content based on the context of use in electronic games (Iosup et al 2010). In most of these situations, content generation is managed as part of the computer graphics interface provided to the user. Depending on the power of the local rendering environment, new content is dynamically added to enrich the user experience.

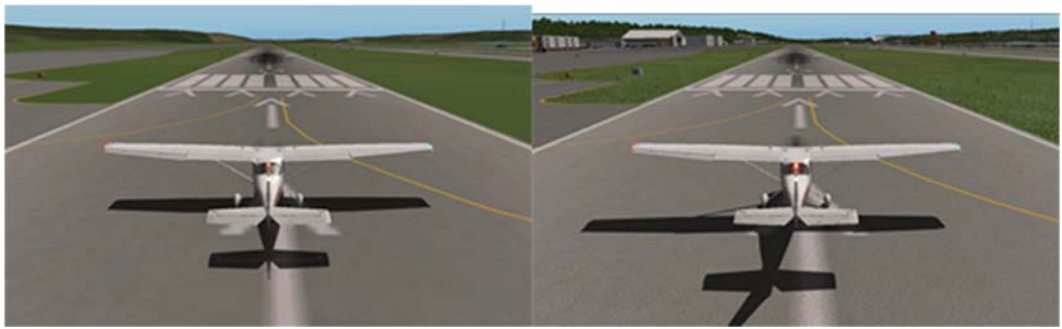

Figure 3: Two content projection that are generated based on local facilities.

The environment shown in games is often artificial and under control of the game creator. This is also true for content generated in virtual reality worlds. Here again, domain constraints help reduce the scope of the content models required to support virtual worlds: much like games, medical applications can benefit from providing coarse models for background content and can provide detailed models for medically-relevant aspects of the application (Van Krevelen and Poelman, 2010).

A hybrid form of context-based content generation is in AR: augmented reality. Here, content in the real world (such as devices, artifacts or even people) can form the basis for generating additional new content that can be used to provide extra or related information to a user. Augmented reality systems can benefit from the fact that most of the background content already exists in the real world. At the same time, the situational awareness and richness of a content model that must be supported for effective AR systems may limit AR's practical deployment.

## 2.5 Integrating professional and user-generated content

A relatively recent phenomenon in wide scale media distribution is the sharing and tailoring of user-generated content (UGC) in social and professional situations. As a base for content generation, UGC holds great promise: it is vastly abundant, often quite timely and is created from a highly personal perspective. Many recent projects have tried to harness the power of UGC and to integrate this content into a broader distribution model (Zsombori et al, 2011).

One use of UGC in content generation is via media mash-ups. Here, content from a common event is gathered and temporally aligned and then automatically reduced to a single 'community' presentation of that event (Shrestha et al 2010). Mash-ups benefit from the large collections of visual angles provided by having multiple sources at one event, allowing a composite presentation to be constructed using well-understood cinematic rules (Ursu et al, 2009). The lack of metadata with the media, however, often makes it difficult to provide highly personalized experiences of these concerts: mash-ups work because the various pieces of content can be rather easily partitioned into content equivalence classes, from which content can be drawn nearly at random.

An alternative to using community content that are then combined using a concert narrative is to let the content itself define the narrative motif. By mining a user's own photographs and video content, a narrative can be created from the collection of media assets available. These can be built around popular persons in the media collection, temporal events (such as vacations), or yearly media summaries. These generated presentations provide a partial solution to a well-understood 21$^{st}$ Century problem: that people spend large amounts of time capturing media but almost no time actually viewing their own media.

One of the criticisms of the social consequences that accompany automatically generated content is that there is no longer a common base for shared experiences when viewing personal media. One interesting project to overcome this problem is a product being developed by Smule, a California start-up. Using the Smule capture and ingestion tools, groups of users can collaborate over time to generate common audio/video productions. The company's tools allow content to be aligned and synchronized. If video is provided, the system provides automatic shot selection to highlight individual contributors.

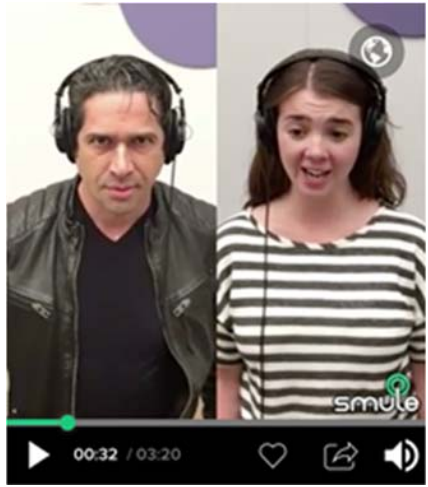

Figure 4: Generating a collaborative version of *Là ci darem la mano* from Don Giovanni, by two artists who have never met, using Smule.

# 3. Challenges

## 3.1 Labeling multimedia content that can serve as the basis for generation

The proliferation of low-cost, high-quality media capture devices, coupled with widely available network-based storage for media assets has help flood the Internet with content. Finding objects of interest (either in a smaller scale, in terms of a particular fragment, or in a larger, in terms of a situational or event capture) is no small task. Traditional approaches to labeling media (using metadata to identify people, places and events) have always held more promise than that it a reliable basis for search. Automated approaches to content classification still operate at a very coarse level, making selection of content fragments a hit-or-miss process. New technology is needed to help users locate content of interest within their own social networks, with adequate protections and pull-back options that would protect people inside content fragments. This includes locating and identifying individual objects within content sequences that can be verified.

## 3.2 Moving beyond video and images

Nearly all of the progress that has been made in the development of content generation tools has concentrated on the visual domain. This has obvious advantages, but also inhibits the reuse and regeneration of content with different modalities. It is very limiting to think of multimedia media information as containing only text, images, audio and video. In an environment where not only people but also sensors and device provide a time sequence of data, new tools and techniques will be required to not only extract meaning from data, but allow that data to be repurposed and recombined to make content more understandable and compelling.

## 3.3 Supporting automatic modality recoding

The focus of content generation (and, to a large extend, understanding) has been to classify and delivery content in a particular modality. In a diverse world, where multiple languages are spoken, many device formats must be supported and people with cognitive abilities at multiple semantic levels need to be accommodated. There is a tremendous need to be able to retarget content for its broadest possible impact. Automatic captioning from recorded *and* visual content (providing not only content but also context) is a compelling example. Another is making content understandable by users from diverse backgrounds and with diverse abilities. This goes far beyond compression and introduces the need to explicitly support current and future context. Especially the integration of a 'future context' is important, since media of all types is typically interpreted at time of consumption rather than at time of production or coding.

## 3.4 Supporting ultra-personalization during generation

The tailoring of content to specific user groups remains an intensive area of study. Examples discussed in this chapter include the development of, for example, concert mash-ups. Such mash-ups can provide generic compelling summaries of large events, but they fail when participants have ultra-personal affinity with other participants. If the 'concert' being shared is a school concert, then shot selection to meet the needs of parents of different participants may vary widely. The same is true for the building of 'media' archives for users in educational, medical, training and information environments. Part of this process is supporting the fine-grained identification of content (including personal content), while also providing adequate privacy guarantees and controls on the further distribution of information.

## 3.5 Supporting complex domain modeling of content

For applications such as augmented reality, the need to provide complex (and time-variant) domain models is essential. Creating such models need to support in the same way that geo- and topological information is considered national priorities. While processing this data (and the addition of local value) may offer protected commercial opportunities, the baseline data should be created and curated as a national resource. New initiatives also need to be developed that support the time-varying modeling of content. All content changes its meaning as it is used, with often differing characterizations based on the context of consumption. Adding context as well as content to information increases its potential to be reused and its potential to be better understood.

## 3.6 Supporting information provenance

As automatic content generation becomes increasingly feasible, the problem of information provenance takes on increased urgency. Understanding the heritage of information, understanding the underlying narratives used to structure content, understanding the authenticity of individual content sequences and explicitly recording the context of capture and use becomes essential in maintaining an open information society. With the current state of the art barely able to support synchronized timestamps on co-recorded content, there is a strong need to develop methods that – if needed – provide information users with the confidence that the message embedded in the media is verifiable and accurate. To date, there has been limited action to protect even basic rights of content ownership but nearly no action to protect the rights of content consumers in the same way that the rights of users of household appliances or consumer products are protected. In the information society, understanding the basic qualities of that information is essential.

# 4. Milestones

## 5 years:

**Content labeling.**  Providing uniform definitions for semantic tagging of content in a number of broad domains. Once these definitions have been standardized, new tools can be developed to label new and legacy content.

**Moving beyond video and images**.  As content encodings become richer and more diverse, new standards need to be developed that allow multiple forms of rich media to be encoded, searched and shared. For a content generation perspective, this means supporting sensor data, text data and graphics data as baseline content formats.

**Supporting automatic content recoding.**  In order to meet the needs of diverse users, content should be able to be automatically recoded from one modality to another, across language and social barriers. This can include automatic captioning and summarization.

**Supporting ultra-personalization.** New tools and techniques need to be developed that allow users to control the labeling, storage and migration of information on their private repositories. Just as we recognize that our telephone numbers are not the property of a telecom provider, we should recognize that our private and collective media annotations belong to the content owner, not the storage service provider.

**Supporting complex domain modeling of context.** Tools should be developed that support the ability to define a context model within a general domain associated with content. This model should allow intent to be encoded in addition to objects and persons.

**Supporting content provenance.** Basic facilities should be developed to authenticate the heritage of individual content fragments: this should include content creation verification in terms of content and not only encoding, as well as information on content alteration over time.

## 10 years:

**Providing layered and extensible models for new content labeling.** Moving beyond the baseline labeling established early, layered and extensible models for content descriptions need to be developed that can encode multiple view and contexts of content over time.

**Moving beyond video and images**.  Beyond collecting and storing rich collections of time-sequenced data, analysis tools will need to be developed to process this information efficiently. Given the expected volumes of information, this will include searching, understanding and reusing information in the compressed domain.

**Supporting automatic content recoding.**  In addition to converting basic content from one modality to another, tools should be provided that add visual context cues and emotional mappings within several targeted application domains. It is clear that this can be important for entertainment, news and educational content.

**Supporting ultra-personalization.** Content generation and selection should support the notion of personal *exclusion* rather than only personal *inclusion*. Owners of personal content should be able to have control over how their personal representations are used and shared, including the ability to recall or withdraw their content's use on demand.

**Supporting complex domain modeling of context.** Domain models for context should be extended to provide multi-view mappings based on time-varying associations. This context should be externalized to allow separate contextual analysis.

**Supporting content provenance.** Extending facilities to authenticate the heritage of a collection content fragments: this should include collection creation verification in terms of content and not only encoding, as well as information on content alteration over time.

## 15 years:

**Supporting multi-context labeling of legacy content.** Assuming support has been defined for new content analysis and labeling, facilities need to be provided to analyze and catalogue legacy content. Such content would exist both in the public and the private domain. In short, someone should be able to ingest their great-grandparent's content and 'understand' what they were seeing, both in terms of the view of the great-grandparent and the modern day offspring.

**Moving beyond video and images**. In addition to understand the objects contained in information, the context of that information will need to be made explicit and external. This context should support multiple interpretations and multiple contexts.

**Supporting automatic content recoding.** The ability to generate new content forms based on existing content will be extended to include explicitly manipulating narratives and tracing the narrative development of legacy content. Understanding the narratives is essential in understanding the underlying information.

**Supporting complex domain modeling of context.** Tools should be developed that support the extension of context analysis to legacy content. Standardized context models should be developed for key domains (entertainment, medical, educational, security, etc.) that allow for the multi-view generation of content summaries and extracts.

**Supporting content provenance.** Extending facilities to authenticate the heritage of a collection of legacy content fragments: this should include collection creation verification in terms of content and not only encoding, as well as information on content alteration over time. The verification can include the integration of personalization and contextual information.

# Chapter 7. **Multimedia and Multimodal Systems**

**Chapter Editors:**

Wu-Chi Feng, Portland State University
Ketan Mayer-Patel, University of North Carolina at Chapel Hill
Balakrishnan Prabhakaran, University of Texas at Dallas

## 1. Introduction

Addressing systems-level challenges faced by multimedia and multimodal (MM) applications is a key research area in order to fully realize and enable the potential broader impacts of these applications on society. These kinds of systems often have complex intermedia relationships that greatly affect how system resources and tradeoffs are best managed. Furthermore, MM applications must adapt to dynamic operating conditions such as varying available bandwidth, computing resources, and power. Often application requirements and priorities can not be determined a priori and may be highly dependent on context including the computational resources available to the user at the time, the ways in which the user is interacting with the system, and the semantics of the media content with respect to the higher-level goals of the user.

These are challenges that have been continually addressed over the past twenty years and the multimedia systems research community has made steady progress in developing guiding principles of systems design and demonstrating their effectiveness in real applications. The current state of the art in MM systems research, however, has effectively reached the end of existing lines of research with diminishing returns. The time is ripe to leap forward to new paradigms of system design and develop mechanisms that more fully incorporate higher-level semantics and a contextual understanding of operating conditions.

What exactly makes MM systems (and thus MM systems research) different from systems in general? Fundamentally, this difference is rooted in the complex and non-linear relationship between the overall quality of experience (QoE) achieved by MM applications and the underlying performance of system components. Every aspect of systems research is informed by this relationship and is in some form or another the root of every systems-level challenge that needs to be addressed. General measures of system performance uninformed by the QoE function are inadequate as are system-level mechanisms that blindly optimize with regard to them.

For example, networking mechanisms that generally "improve" performance by lowering packet loss may be counterproductive for an interactive multi-participant video conferencing application if that improvement comes at the cost of additional latency. Similarly, a 1% packet loss rate indiscriminately applied to a compressed media stream could result in a significantly inferior QoE than a 5% packet loss directed toward stream data that are known to be less important to the reconstruction of future application data units.

Multimedia Systems research lies between multimedia applications and the underlying systems that either compute or deliver the expected result. Multimedia systems research focuses on being able to provide (i) mechanisms to run the application in real-time or near real-time, (ii) mechanisms to allow the application to scale to large numbers, or (iii) provide the best quality-of-experience to the user.

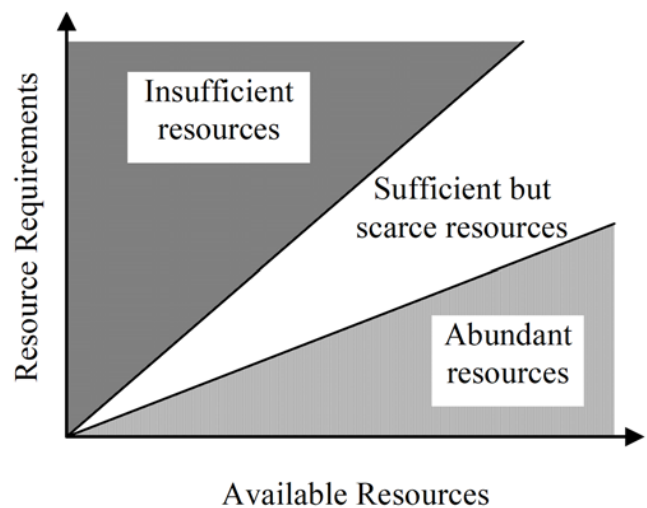

Fig 1. The Window of Scarcity Framework [Jeffay 1986]

One way to reason about multimedia systems research is within a framework often expressed as "The Window of Scarcity" illustrated in the figure above [Jeffay1986]. This conceptual framework describes three main development phases resource intensive multimedia applications go through including: insufficient resources, sufficient but scarce, and abundant.

Many advanced multimedia applications start out in the "insufficient resources" window (e.g., DVD-quality playback on an Intel 486 PC in the late 80's). When a class of applications is in the insufficient resource window, little can be done to realize practical, scalable, real-time implementations. However, for some applications, multimedia systems research can provide techniques to lower resource requirements or at least adapt to insufficient resources available, making the application plausible (e.g., dropping video frames to allow for the appearance of real-time playback).

Once in the "sufficient but scarce" phase, techniques that focus on providing adaptability, scalability, and/or efficiency play an important role in the further development of these applications. These techniques often require deep integration with the underlying software system, the underlying hardware, and/or the network that supports the application. As advances in computing and networking capabilities make the relative resource requirement less onerous, the multimedia systems techniques that were employed to make the multimedia system feasible may be called upon to provide additional scalability or make use of richer media types. In essence the resource requirements of the class of multimedia systems within the window expands to remain within the window (e.g., greater resolution, faster frame rates, larger bitrates, etc.).

Eventually, a class of multimedia systems may no longer need to rely on negotiating resource tradeoffs in sophisticated ways at all. Indeed, straightforward and less complex implementations may easily achieve desired performance given the computational and and networking infrastructure available. At this point, these multimedia systems fall out of the window of scarcity, enter the window of abundance, and are generally no longer of significant interest to multimedia systems researchers. However, in the best of worlds, the techniques and insights gained by researchers while development was within the window of scarcity can be generalized or reapplied to a new class of multimedia systems which may now be entering the window of scarcity.

Thus, in summary, multimedia systems research focuses on adaptability, scalability, efficiency, (and more recently security) mechanisms necessary to enable the current state-of-the-art generation multimedia systems within the window of scarcity to maximize the quality of experience delivered to the user. Given this perspective, our purpose in this section is threefold:

- Identify the key principles that this research community has distilled so far in developing the current state-of-the-art in multimedia and multimodal systems approaching the end of their time in the window of scarcity.
- Characterize the classes of multimedia and multimodal systems that are just now entering the window of scarcity and the systems-level challenges faced by them.
- Suggest a roadmap for key systems-level challenges that will need to be addressed over the next five, ten, and fifteen years.

## 2. Key Principles of Multimedia Systems

Early multimedia systems research generally addressed challenges of streaming media in interactive and broadcast applications. Adapting media streaming to available bandwidth, supporting heterogeneous users, and scalable media delivery were central concerns. These applications included video conference, video-on-demand, computer-supported collaborative work, collaborative authoring, and composable media processing and services. Over the course of the first two decades of multimedia systems research, several important key principles of multimedia systems design emerged including:

- Application-level framing
- Client-driven adaptation
- Graceful degradation and eventual consistency
- QoE coherence

*Application-Level Framing* (ALF) is a foundational principle of multimedia systems first articulated by Clark and Tennenhouse in the context of network protocol design [Clark1990]. The core idea behind ALF is to, as much as possible, respect the fact that application data should be manipulated in accordance to the highest-level goals and requirements of the application. In other words, the application should be able to define and organize data into Application Data Units (ADUs) which are atomically useful (i.e., independent and idempotent). The goal is to minimize the effect of data loss and latency by aligning adaptation and loss mitigation mechanisms with representation structure (e.g., compression) and application-level semantics.

Ideally, the ALF principle mandates that underlying computation and networking resources are configured and tuned to respect ADU organization. In practice, however, the ALF principle often works in reverse. Hidden costs and limits within lower layers of the networking stack and operating system are exposed in order to allow the application to best form the most appropriate ADUs for communication and processing tasks. For example, the Maximum Transmission Unit (MTU) size associated with a network path is exposed (or discovered) in order to allow an application to appropriately configure a video codec to produce independently decodable ADUs comprised of a subset (i.e. slice) of macroblocks of a video frame that can fit within a single packet minimizing the effect of packet loss.

*Client-driven adaptation* is another commonly applied principle of multimedia systems design that serves to simplify server mechanisms and support scalability. The idea behind client-driven adaptation is to encode and transmit media data in a way that allows clients to appropriately adapt to dynamic network conditions (e.g., available best effort bandwidth and loss), computational resource constraints (e.g., power and processor capabilities), and overall context (e.g., user-level goals and related media). Ideally, this can be accomplished without maintaining per-client state at the originating media source (i.e., streaming camera or server). Per-client state and computation is a bottleneck to scalability and stateless server design is a well-understood principle in distributed systems which is reflected in the principle of client-driven adaptation within multimedia systems. Exemplary applications of the principle of client-driven adaptation in multimedia systems include the early seminal work of receiver-driven layered multicast by McCanne [McCanne1996] and the more recent development of Dynamic Adaptive Streaming over HTTP (DASH) [ISO/IEC23009-1] now in widespread use to support adaptive large-scale video-on-demand streaming.

Multimedia systems characteristically involve many different streams of media data integrated with application-level state and context resulting in complex multimodal relationships between how media is represented, communicated, manipulated, and used. Often participants and agents within a distributed multimedia system are heterogeneous with respect to the types of displays and computational resources available and may need to transition between high-resource environments and low-resource environments dynamically. For example, a teleimmersion system may need to support users within a high-resource CAVE environment that includes large-scale 3D displays that completely surround participants interacting with remote users participating with a Google Cardboard VR headset using a mobile phone. Even in a relatively simple multimedia system such as a shared desktop video conferencing system, dynamic network conditions may require negotiating tradeoffs between media streams.

*Graceful degradation and eventual consistency* are principles of multimedia system design that can be brought to bear to tackle the challenges encountered when available resources are insufficient to sustain a multimedia system at peak operating conditions. In large part, however, multimedia systems incorporate these principles in ad hoc, application-specific ways using relatively simple heuristic policies and mechanisms that do not scale to more complex and contextual systems. For example, sacrificing video frame rate and quality to preserve continuous quality audio is well-understood and easily realized policy born from a widely shared understanding of how these media types contribute to the QoE within a video conferencing system.

As the number of media streams involved increases and as the heterogeneity of user capabilities and user roles grows, the ability to meaningfully apply the principles of graceful degradation and eventual consistency via ad hoc application-specific policies explicitly specified by the system designer essentially vanishes. Consider the same media types as before in the video conferencing example, but now within a remote body camera system used by law enforcement. Context, in this case, matters a great deal and the same policy as before (i.e., audio preserved at the expense of video) may not be so easily recognized as beneficial in all or even most cases. If the officer is engaged in a traffic stop with a citizen, audio may in fact be the most important media type to preserve at the expense of video quality or frame rate. If the officer is engaged in a foot chase with a fleeing suspect, audio may be pointless and maintaining a relatively high video capture frame rate may now be paramount. Alternatively, such a situation may best be served by abandoning video altogether and resorting to capturing a sequence of fast-motion, high quality still images.

The principle of *QoE coherence* informs system design by recognizing that maintaining a stable and expected quality of experience in a multimedia system is often as important and sometimes even more important than maximizing peak quality. In other words, variance in quality should be minimized at the expense of average quality achieved. The canonical example of this principle is maintaining a steady perceived visual quality throughout the playback of a video stream even if this means forgoing short-term periods of increased available bandwidth. The same principle applies for latency in interactive multimedia systems (e.g., multi-user virtual environments, games, etc.). While there are certainly limits to latency with regard to usability, users often adjust expectations to latency within allowable ranges and changes in latency (i.e., jitter) can be disruptive and detrimental to QoE.

# 3. The Window of Scarcity Today and Tomorrow

The challenges associated with interactive streaming, large-scale video distribution, and video-on-demand have been key components of applications and systems within the window of scarcity over the past twenty years and have driven the development of modern compression codecs and QoE-adaptive streaming protocols and techniques. These applications continue to remain relevant to multimedia systems research, especially as display and camera resolutions continue to increase, phone-based mobile systems add power constraints, and multi-view stereoscopic streaming becomes possible.

Looking forward, however, several new application domains have more recently entered the window of scarcity and we can anticipate new systems-level challenges that will need to be addressed in the coming five, ten, and fifteen years. These new application areas will require closer collaboration with other communities of researchers including computer vision, machine learning, graphics, and mobile systems.

**Computer Vision Systems**

As we move beyond simply delivering bits and more toward multimedia systems that are actually content-aware (as opposed to being data-compression-aware in current video streaming technologies), integrating computer vision-based algorithms into the multimedia systems are going to be extremely important. This includes a broad range of applications such as (i) employing computer vision into compression, particularly for stereoscopic or multi-view / multi-camera video [Merkle2007], (ii) prioritizing transmission of video data based upon image content [Hadizadeh2013], and (iii) scalable distributed video command and control [Batstone2015].

To leverage computer vision algorithms in multimedia systems requires bringing adaptation and real-time into traditionally black-box computer vision algorithms. This will involve altering the algorithms to have "exit" points where partial results can be returned to the underlying multimedia system in the event that there is insufficient time. Techniques to integrate the partial results into the multimedia system and to balance between computer vision and systems work (e.g., compression) will need to be investigated.

Finally, content-aware control and adaptation techniques that balance both content and systems will be required. One approach to providing content-aware adaptation is to close the loop with the underlying computer vision algorithm and provide per-pixel saliency feedback to the media source to guide compression tradeoffs and transmission priorities. Current state-of-the-art compression and streaming techniques which have been so highly optimized for human visual system perception over the years have actually become a roadblock for developing effective computer vision-based multimedia systems.

**3D Teleimmersion**

3D Tele-Immersion (3DTI) environments are a new medium for human interactions and collaborations in the areas of education, sports training, rehabilitation, etc. Typically it is a distributed, multi user, multi-site environment that depends on the stochastic network like the Internet for communication.

A traditional 3DTI system immerses a person in a virtual environment through the use of 3D stereovision cameras and audio sensors. 3DTI systems have been used in various fields such as entertainment, meetings, gaming, tele-rehabilitation etc. Multiple persons can interact with virtual objects and among themselves in the shared virtual space. Adding a tactile element to these systems makes them even more engaging and opens up the possibility of new and exciting applications where the collaborators can not only see each other in the virtual environment but also "touch".

One example of such a 3DTI system is a tele-rehabilitation virtual therapy session between therapists evaluating a patient with upper limb disorders from a remote location. The basic modalities that we considered for this scenario are data streams from a depth camera (i.e., Microsoft's Kinect), haptics and wearables. Depth cameras facilitate face-to-face interactions and capture details about body parts, such as muscles that are involved in the disorder. Haptic devices provide measurements of patient muscle strength to the therapist. Wearable sensors on the patient's body provide details of body movement. The end result of such a 3DTI system is a tele-rehabilitation monitoring application where a therapist can evaluate a patient thoroughly despite being at a remote location. A 3DTI application such as this one has to be highly interactive and contextually adaptive , collecting and coordinating related information from the physical and logical data sources  that are related to high-level therapeutic goals via a complex and difficult to express QoE function.

**Augmented Reality Systems**

Augmented reality systems have been proposed for video gaming and informational overlays. Some significant systems research issues will need to be solved including creating adaptive computer vision algorithm to run on small mobile devices (e.g., glasses) with limited power and computation. Computer vision algorithms will need to be downscaled to run on more resource impoverished devices. Alternatively, having separable computer vision computation that can do some processing locally and leverage cloud computing simultaneously (and adaptively in real-time) will be necessary.  Finally, integrating the different systems such as video localization and identification, networking, and the augmented reality system will be required.

**Scalable Video-based Sensor Systems**

Video streaming technologies have been around for over two decades. While they have provided significant entertainment value for users, they fall short on applications that require analysis or understanding of the actual content in the stream. Video-based sensor systems that can collect a large amount of data, still require significant manual (human) intervention to extract meaningful results for all but the simplest of tasks. Mechanisms that support the "programming" of such large systems in a way that can be done by the domain (e.g., civil engineer studying bicycle traffic) need to be developed and scaled to large numbers of sensors. For many video sensor systems, the ability to use non-visual input (e.g., inductance loops or other scalar and binary sensor) may greatly improve the efficiency, scale, and accuracy of the system. Efficient mechanisms to quickly and efficiently integrate such multi-modal information will be required. Time-adaptive computer vision algorithm and resource cognizant computer vision algorithms that will be deployed near the edges of the network will also be necessary. Finally, mechanisms to more easily integrate disparate systems such as video codecs, computer vision algorithms, and other systems in pipelined fashion to process and massage data more quickly will be necessary.

# 4. The Road Forward

The road forward for the multimedia systems research community must build upon the principles that we have developed and applied successfully over the past twenty years as we address the challenges of applications and systems just now becoming feasible within the window of scarcity. Common characteristics of these applications include increased complexity with multiple inter-related media streams and multimodal inputs, application-level goals driven by content understanding and semantics, highly contextual QoE functions, and extremely wide ranges of operating conditions. We suggest that these challenges will require addressing three specific roadblocks as a community, namely:

- Creating a framework to bridge the concerns of multimedia systems with the advancement and integration of computer vision algorithms.
- Developing flexible, modular compressed media representations amenable for deeply exploiting application-level constraints, goals, and semantics.
- Advancing beyond ad hoc explicit policy-driven adaptation and control mechanisms toward implicit contextual solutions derived with machine learning techniques.

**Bridging Multimedia Systems With Computer Vision**

Until recently, there has been little overlap between the computer vision research community and the multimedia systems research community. Generally the computer vision research community has treated video sources and video transport as black box components. Frame rate and quality are provisioned to provide image data that is relatively free of compression noise. Resource tradeoffs between video quality and bitrate and/or frame rate are generally not integrated into vision algorithm design. Tradeoffs between the number of viewpoints used and the quality of those viewpoints on average or allocation of bitrate amongst available viewpoints are similarly not exploited or explored. Although there has been some research into compressed-domain vision algorithms [Babu2016], in large part, this black box view of video as a datatype is a result of a convenient separation of concerns and abstraction when major advances in both multimedia streaming and vision could be made independently.

At the same time, decades of multimedia systems research and development has optimized media representations and streaming protocols for human perception alone. Advances in widely used video and audio codec standards have been designed from the ground up with human perception as the target application. Streaming protocols assume little to no feedback about regions of interest with an implicit assumption that human attention to error is spatially agnostic. Quality control for media representations is usually reduced to a single degree of freedom that controls an overall rate allocation and error control mechanism specifically developed to reduce human perception of noise. Intermediate representational structures such as motion compensation vectors are not conditioned to be consistent or to reflect actual optical flow because these elements are assumed to be discarded after the frame decoding process is complete. These assumptions of human perception are inappropriate, however, when media is being used in service of a computer vision algorithm or system.

In order to bridge this gap, as a research community, we need to invest in developing a new canonical interface to video resources that allows rate/distortion tradeoffs, exposes control latency, and accepts vision-directed error control. A significant first step can be made by developing rate and quality adaptive interfaces to streaming real-time video sources in the OpenCV framework. OpenCV is the framework of choice for a large majority of computer vision research and by creating an API to video sources that exposes rate/quality tradeoffs and latency in a low-effort manner, we can begin to engage the vision community to view system tradeoffs as a relevant factor in the development and advancement of vision algorithms. Such an API should provide a feedback interface for per pixel saliency information that can be used by a vision algorithm to direct and inform error control and adaptation.

**Unbraiding Compressed Domain Media**

The development of increasingly optimized media representation standards has led to increasingly complex compressed domain representations. In particular, the underlying prediction model (i.e., motion compensation, block-structure, global motion modeling, intraframe prediction, etc.) has become increasingly complex with each subsequent coding standard. This trend is driven by the relative increase in computational resources (i.e., Moore's law) and the efficacy of complex prediction models as video resolutions increase. These ever more complex prediction models serve to further entrench the underlying assumptions about the visual input and its consumer. Furthermore, the resulting encoded syntax in modern standards such as H.264 and HEVC has become a brittle intertwined entropy encoding of compressed-domain syntax elements that are difficult to tease apart and process separately.

Over-optimized media representations make it difficult to experiment with new ways of integrating and aligning media representations with domain-specific knowledge and constraints. The gains in encoding efficiency across a wider range of bitrates, while useful, work against the principle of application-level framing. Thus, researchers may sometimes find themselves struggling to take advantage of opportunities for cross-stream, cross-media, or multimodal relationships that may be relevant in complex multimedia systems. An unintended consequence of having highly-optimized media representations seems to be that multimedia systems researchers are beginning to view compressed media data sources as black box components in much the same way as the computer vision research community has, thwarting innovative research in which media representation and streaming reflects and is aligned with application-level goals. Part of this difficulty simply stems from the fact that it is difficult to train students with the intricacies of modern media standards such that they can develop projects to a point of being able to do experiments and get results.

What is required is a more "research-friendly" media encoding standard with functional components that can be mixed and matched in a high-level manner. Model-based prediction encoding should be completely decoupled from residual encoding. Each prediction model (i.e., motion compensation, coding mode signaling, etc.) should be defined independently. By unbraiding the resulting entropy encoded compressed-domain syntax, flexibility is emphasized at the expense of hardware-based optimizations, implementation efficiency, and optimized compression rates. This will allow greater flexibility to incorporate domain-specific content models and constraints and experiment with application-specific and contextual approaches for residual encoding where rate and error control has the most impact. MPEG-RVC is a coding standard framework that is aligned with many of these goals and may represent a good starting point toward developing the tools needed to unlock the next generation of multimedia systems research in which tighter integration with vision algorithms and greater consideration for domain-specific concerns are better supported.

**Learned Contextual Adaptation and Control**

As multimedia systems become increasingly complex, ad hoc explicit system-level adaptation and control becomes increasingly more difficult to develop and less effective. Contextual and multimodal concerns further complicate our ability to identify appropriate control actions as the number of possible operating regimes and the possible tradeoffs available explodes combinatorially. In essence, as we begin to address the challenges associated with the next generation of multimedia systems in the window of scarcity like computer vision driven applications, VR/AR, 3DTI, and large scale multimedia sensing, there are simply too many moving parts in the system. The QoE function associated with these complex systems is difficult to model and highly non-linear.

Applying machine learning techniques to the problem of systemic adaptation and control may provide a fruitful approach for addressing these challenges. By capturing quantifiable measurements of QoE and how it changes in real-time as a system is being used and relating those measures back to operating conditions at the time (i.e., learning the QoE function), it may be possible to implicitly learn the appropriate adaptation and control responses as those operating conditions change. Additionally, developing frameworks for learned control and adaptation will allow us to leverage unbraided compressed media representations in which each compressed-domain intermediate structure can independently negotiate quality/rate tradeoffs in contrast to the single degree of freedom quality tradeoffs we currently rely on that are specifically optimized for human perception without application-level knowledge.

# 5. Milestones

The future of multimedia systems is exciting and more relevant than ever. The incorporation and reliance on multimedia data sources and multimodal relationships is fundamental to building complex real-time systems such as autonomous vehicles, teleimmersion, AR/VR, and large-scale sensing. Although forecasting progress across the next twenty years of research and development is at best speculative, we believe that identifying possible development milestone targets can be a useful way of creating a common vision for progress and guiding support for new research initiatives and programs. Below we anticipate progress in the avenues of research we have identified and highlighted across the next five, ten, and fifteen years. It should be understood, however, that the specific topics of emphasis we have chosen to concentrate on here does not reflect the true breadth and depth of multimedia systems research overall and new and unexpected avenues of multimedia systems research are likely to have a significant impact as well.

**5 years:**

- Develop and release community-supported rate and quality adaptive APIs for streaming video sources with saliency feedback into commonly used frameworks for vision research (i.e., OpenCV).
- Develop and release free open source research-oriented unbraided compressed video streaming implementations of widely used compressed media standards.
- Define modular specifications for compressed media separating prediction/model streaming from residual streaming.
- Develop generalized QoE management frameworks with appropriate interfaces for applying machine learning techniques.

**10 years:**

- Ratify new vision-aware and vision-specific video compression standards through appropriate standards making bodies based on research experience.
- Foster research community focussed on machine learned real-time QoE-based compression and system control.

**15 years:**

- Disseminate off-the-shelf mechanisms for machine learned contextual compression and streaming.
- Deploy open publicly available self-configuring large scale distributed environment capture as a platform for innovative application development.

- Move toward contextual QoE management derived from high-level application semantics.

# Chapter 8.  Data, Competitions/Challenges, Evaluation

**Chapter Editors:**

Sameer Antani, U.S. National Library of Medicine
Alex Hauptmann, Carnegie Mellon University

**Additional Workshop Participants:**

Terry Adams, Julia Hirschberg, Reuven Meth, and Adam Wolfe

## 1. Introduction and State of the Art

Data is the primary nugget of computable information. As such, its collection has become increasingly important, particularly when it is used for extracting knowledge, understanding patterns, predicting outcomes, and analyzing value. It has been reported that Jeff Bezos, CEO of Amazon, Inc., has anecdotally said that Amazon does not destroy any data. This also means, that all the data that is being acquired – be it through searching, shopping, commenting, watching movies/shows, requesting songs, uploading photos, using compute services – is now a treasure trove for not only advancing solutions, but advancing it where they would be most valuable to the company. Similar efforts have been reported for Google, Apple, Facebook, etc. And, when it comes to healthcare, IBM Watson, NIH, and hospital systems are all aware of the importance of data, and enable their use toward advancing science. This commentary can be summarized in a simple sentence – "Without data I do not exist!". For data to be valuable and useful, it must be accessible.

Definition: *Useful Data* is information that provides value possibly through a variety of ways, including training of algorithms, testing, or data used in an operational setting.

For the data to be useful, it must be accessible – as stated before – but, data that is acquired privately, is rarely available due to its competitive nature and high monetary value. Scientific data, on the other hand, must have mechanisms to enable *replication,* maintain *redundancy*, and enable *reuse*. This idea is not particularly novel and has been expressed many times before. However, the idea is challenging for it brings to the forefront challenges in metadata definition, alignment, error in values, holes, semantics of the value, and purpose of acquisition.

There are several types of data formats: text; signal: audio, sensor; image; and video. There are metadata, and other convolutions of data – it could be raw or processed; structured or unstructured; constrained or unconstrained range; synthetic, simulated, or real; It could be naturally multimodal or be augmented through data fusion. Its purpose could be for training, test (as ground truth), or an outcome of operations. All these divisions are not mutually exclusive, but are intended to provide insight into various ways one can think about data.

Users of these data include data scientists, machine learning researchers and developers, developers of artificial intelligence (AI) applications, data analytics professionals, and companies that provide inference using data in areas such as health care, education, business intelligence, law.

Sources of data are wide ranging. There is social media data (Twitter, Facebook, Instagram, etc.), internet usage data; health data – both in hospitals and clinical care providers, as well as personal iOT sensor data; commercial data – logistics, manufacturing, utilities, transportation, defense, cybersecurity, etc. The problem with using such data outside of the immediate scope is lack of annotation, access control, availability for use, and granularity of the sensor data. And, last but not the least, is the buzzword Big Data which challenges us with velocity with which the data is being acquired, validity of the data, and veracity of the content.

Due to lack of access, the use of synthetic or simulated data is becoming more prominent in the development of data analytics and machine learning algorithms. Data hungry algorithms are being fed augmented data, and simulated noise is being injected into the data streams so that the algorithms can be made more robust. While this is appropriate and a useful strategy, there are concerns that the augmentation and the injected noise may not represent all of the variety in the *real-world data* that the algorithms may encounter.

Data gathering and use policies may determine the extent to which data is collectable in an automated fashion, and how it may be used.

Even a casual search of annotated multimedia/multimodal datasets that are currently used in challenges and competitions yields quite a large set, with ever-increasing numbers of labeled examples and classes.

Bob Fisher at Edinburgh University has indexed over 670 image and video data sets [3] and categorized them into 24 main areas:
- Action Databases
- Attribute recognition
- Autonomous Driving
- Biological/Medical
- Camera calibration
- Face and Eye/Iris Databases
- Fingerprints
- General Images
- General RGBD and depth datasets
- General Videos
- Hand, Hand Grasp, Hand Action and Gesture Databases
- Image, Video and Shape Database Retrieval

- Object Databases
- People (static), human body pose
- People Detection and Tracking Databases (See also Surveillance)
- Remote Sensing
- Scene or Place Segmentation or Classification
- Segmentation
- Simultaneous Localization and Mapping
- Surveillance (See also People)
- Textures
- Urban Datasets
- Other Collection Pages
- Miscellaneous Topics

While the number of available datasets is large, it should be noted that the quality of the annotation (or "ground truth") can be quite variable. Some, such as the Youtube8Million [4] video dataset do not provide any accurate truth at all, but only an algorithmically determined approximation based on multiple other text sources. At the other end of the labeling spectrum are meticulously annotated but much smaller datasets such as VIRAT [5], where each person has a bounding box, all activities are documented with begin and end times, and any motion is tracked in detail.

Dataset size is also measurable in different ways: The most critical number for a dataset are the number of labeled positive and negative examples as well as the total number of classes or types. The worst way to measure size is by disk space. TeraBytes of disk to store the data, which is generally larger for videos, but small for audio and minimal for text. High definition video at high frame rate requires more than low resolution images at reduced frame rates. A reasonable concise description of a dataset could be: X number of videos of Y number of different types of objects each annotated in Z videos with average length of S seconds.

To further illustrate the diversity in datasets, we should point out that some merely provide a word for the caption of an image ("dog") [6], others describe the scene in text (a man walking a dog on field in the sun). A number of datasets have multiple labels for one image [7]; some have multiple shape-labeled images [8]. In the video world, the solution is sometimes to just provide a single label for a whole clip [9, 10], to provide time segmented labels of object visibility or actions [11]. [12], and Kinetics [1]: recognize activities in trimmed or longer video, localize where the actions take place in time or within the frame, with another task of describing all the events in a video. Some of the most challenging datasets deal with human emotion or sentiment in multimodal datasets [2] such as spoken reviews, images, video blogs, human–machine and human–human interactions.

**ImageCLEF [13] and LifeCLEF [14]**: The CLEF organises a number of tasks with a global objective of benchmarking performance in multimedia domains such as  lifelogging summarization and retrieval, bio-medical image concept detection and caption prediction and tuberculosis severity score prediction from CT images; and medical visual question answering; Recent benchmarks have included dataset on:

- Image Annotation: A task aimed at the development of systems for automatic multi-concept image annotation, localization and subsequent sentence description generation.
- Medical Classification: Addresses the problem of labeling and separation of compound figures from biomedical literature.
- Medical Clustering: Addresses the problem of clustering body parts x-rays.
- Liver CT Annotation A study towards automated structured reporting, the task is computer aided automatic annotation of liver CT volumes by filling in a pre-prepared form.
- BirdCLEF: an audio record-based bird identification task
- PlantCLEF: an image-based plant identification task
- FishCLEF: a video-based fish identification task
- SeaCLEF: a visual-based sea-related organisms monitoring task

**VideoNet Benchmark [15, 16, 17, 18, 19]**: VideoNet is a video and video object detection and segmentation benchmark with multiple human annotations; multiple object tracking, and activity forecasting tasks,

**ACM Multimedia Challenges.** The Multimedia Grand Challenges present a set of problems and issues from industry and academia to engage the Multimedia research community in solving challenging questions for multimedia. The first Multimedia Grand Challenge was presented at ACM Multimedia 2009 and has established itself as a prestigious competition in the multimedia community. Among the past challenges were :

- MSR Video to Language Challenge
- Social Media Prediction (2017). Making predictions about the future
- Lane Level Localization on a 3D Map (2017) to localize the vehicle in correct lane and location
- YFCC100m: Yahoo-Flickr Grand Challenge on Tag and Caption Prediction; Yahoo-Flickr Event Summarization Challenge (2015) separate a collection of photos and videos into a set of possibly overlapping events

**MediaEval[20] "**is a benchmarking initiative dedicated to evaluating new algorithms for multimedia access and retrieval**."** The evaluated tasks include: Emotional Impact of Movies Task; Predicting Media Interestingness; . Multimedia Satellite matching; Medical Multimedia aims to predict diseases based on videos of the gastrointestinal (GI) tract.

**TRECVID[21]**: TVNews search, story segmentation, camera motion classification, Documentary search, IACC semantic indexing, copy detection, known item search, localization; BBC Rushes summarization, MED, MER (HAVIC), Multi-camera SED event detection, BBC Video Hyperlinking, Vine Video-to-text

**Other notable datasets due to their size**: Youtube8M [4], IBM/MIT Moments [24], HAVIC [10], YFCC [23], FCVID [22].

The tasks supported by the various dataset also scan a wide range of applications. Event detection, object detection, action detection, activity detection (where an activity is fundamentally just a more complex action), facial recognition, lip-reading, emotion detection, video transcription, video question answering,

Given all these datasets, and new ones emerging virtually monthly, there is clearly a need for sharing repositories, a *datahub* that is similar to *github.com* for source code. Such a data hub may need to be protected and managed by a unbiased third-party. An example in the medical domain is the *Harmoni* project under the auspices of GRNet, the Internet2 data node for Greece. It is charged with collecting, maintaining, and making available for research all radiology imaging data from all clinical sources (private and public, small size or institutional) and archive it – and make it available for research. The discussion of data, therefore, is closely entwined with acquisition, storage, management, access, retrieval, and transport infrastructure, as well as policy that would permit these. Generic rules can be applied to any application domain – smart cities, health systems, health research, logistics, utilities, etc.

Finally, it is important to have data visualization research that supports discovery of what exists, as well as analytics to extract patterns.

| Topic | State of the Art | Key Challenges, Unmet Needs | Road Map | | |
|---|---|---|---|---|---|
| | | | **5 Years** | **10 Years** | **15 Years** |
| Data and Challenges | Data exists in varying sizes.<br><br>Data is mostly under private control with limited to no general access.<br><br>Data acquisition, sharing, and use policies are obscure. Also, the data provider has little or no control over post-acquisition data use.<br><br>Data de-identification strategies are in their infancy.<br><br>Data hubs for data access are limited in availability. Those that are available are inconsistently managed.<br><br>Data interoperability is in its infancy.<br><br>Cross-domain learning is in its infancy.<br><br>Data visualization is in its early years. | There is a need for a data hub. Data hubs may be domain dependent.<br><br>There is a need for developing data acquisition, management, access, sharing, and retrieval policies that are transparent to the data provider, and the consumer alike.<br><br>Ability to separate proprietary data from public data.<br><br>There is a need for cross-domain learning.<br><br>There is a need for developing algorithms for annotating vast amounts of data, or extending pilot annotations to Big Data.<br><br>There is a need for improved data de-identification techniques.<br><br>There is a need for data fusion algorithms that would permit combining data sets from different modalities or domains.<br><br>There is a need for advancing data visualization techniques. | Data management, sharing, and use policies become clearer.<br><br>Community resources: Data hubs start becoming formalized.<br><br>Content and context summarization techniques; data labeling (metadata generation) techniques evolve. Efficient tools are developed.<br><br>Machine learning over big data sets evolves.<br><br>Cross-domain learning evolves.<br><br>Data de-identification techniques issues are meaningfully addressed.<br><br>Data correlation techniques are studied. | Data hubs become common place.<br><br>Data use and sharing policies become transparent.<br><br>Data fusion and alignment techniques.<br><br>Causality is studied.<br><br>Knowledge discovery and representation (beyond metadata) – static model development.<br><br>Hypothesis generation | Hypothesis verification<br><br>Knowledge extraction – dynamic model development. |

# Challenges Unmet

In spite of the recognition that without data there can be no progress, we are at a stage where there is a lack of uniformity in data resources that can further the R&D in multimodal analysis. One suggestion toward correcting that is the need for developing a data hub, similar to github.com. The site provides a project oriented repository for source code and the ability to contribute to the development in teams. There is a need to save details about the dataset along with the ability to cite the data set, and provide attribution to the data provider. Data-hubs may be domain dependent, and require particular metadata to enable appropriate management, and access.

With the creation of data-hubs, there would also be a need for developing data acquisition, management, access, sharing, and retrieval policies that are transparent to the data provider, and the consumer alike. There should be an ability to separate proprietary data from public data. Further research is needed in data de-identification techniques so that PHI and PII elements can be sequestered or removed.

With an eye toward machine learning, knowledge extraction, knowledge representation, and model development, there is a need for developing algorithms for annotating vast amounts of data, or extending pilot annotations to use with Big Data. Can we trust the crowd when annotating data via crowd sourcing methods? Can we get the crowd to give a broad stroke annotation, and then using a more refined method for getting measurable gold standard annotations? Can we annotate data using a cascaded effect (i.e., composing binary tags). There are tools that assist annotation by making intelligent estimates of what the annotator intends, and then allows annotator to easily fix. However, these tools would need advances for large multimodal (possibly aligned or fused) data sets.

Research is needed in machine learning algorithms that operate on large collections of disparate, but related data. There is a need for cross-domain learning. There is a need for data fusion and alignment algorithms that would permit combining data sets from different acquisition purposes, modalities or domains. It would also be worth exploring if machine learning and knowledge extraction algorithms developed for one domain can be extended to another with relatively minimal modifications.

Without visualization techniques, it would be very difficult to guide the development of knowledge discovery and extraction. There have been recent advances in visualization of single domain data, answering a single specific question. There is a need for advancing data visualization techniques for extremely large, multidimensional data sets.

## 2. **Milestones**

**5 years:** Big multimodal data is already under study, however, there are challenges with data management, sharing, and use policies. These will become clearer over the next 5 years with induced changes to enable greater use. This use will be enabled through the formalization of community resources such as data hubs. Efficient tools for annotating data are developed. In addition, content and context summarization techniques and data labeling (metadata generation) techniques evolve. Machine learning over big data sets evolves, and knowledge discovery by through correlation is widely reported. Beginnings of cross-domain learning are observed. Data de-identification techniques issues are meaningfully addressed.

**10 years:** Making the most out of available data. Community utilities: data hubs, data annotation techniques, knowledge discovery and representation methods become common place. Data use and sharing policies become transparent. Data fusion and alignment techniques are studied. While the first 5 years studied correlation, these 5 years will study causality. Research is conducted on knowledge discovery and representation (beyond metadata) – static model development. Hypothesis generation from existing data is studied.

**15 years:** Hypothesis verification research is conducted. Knowledge extraction – application of the static knowledge model and research is conducted into dynamic model development.

# Section 3    APPLICATIONS

# Chapter 9.  **Education**

**Chapter Editors:**

Sharon Oviatt, Monash University

Zhengyou Zhang, Microsoft Research

**Additional Workshop Participants:**

Louis-Philippe Morency, Arnold Smeulders, and Maria Zemankova (listener)

# 1. Introduction

This chapter focuses on major multimodal-multimedia technology advances that are expected to improve educational outcomes by substantially stimulating learning, and to assess the outcomes more accurately. It does not address the broader spectrum of educational technology topics or specific application areas.

## 1.1 Definition

Education needs to support lifelong learning by a diverse community of citizens. During the last decade, multimodal-multimedia interfaces based on combined new media have eclipsed keyboard-based graphical interfaces as the dominant worldwide computer interface, including for education. In addition to supporting mobile and informal learning, this technology enables more expressively powerful and flexible human input to computers (writing, speaking), which can more substantially stimulate cognition and learning compared with keyboard interfaces [Oviatt et al. 2012; Oviatt, 2013]. This cognitive facilitation occurs because modalities like speech and writing are used for human communication, and language is a direct carrier of thought [Vygotsky 1962].

Input modalities like writing/sketching and manual gesturing are capable of conveying precise spatial representations, which are believed to provide a foundation for human thinking and reasoning [Johnson-Laird 1999]. Unlike keyboard interfaces, multimodal-multimedia also represents the emergence of more flexible *multi-component tools* for learning. This multiplicity of input and output alternatives, for example, enables students to select the input that is best matched for a learning activity [Oviatt, 2013]— such as pen input to draw a diagram or to write symbols for math formulas. In addition, multimodal-multimedia technology is the preferred direction for supporting individual differences and universal access to computing.

As a parallel development, new learning analytic techniques now are needed that match these data sources in order to evaluate learning progress. First-generation learning analytic techniques have focused on assessing click-stream data from keyboard input, which has been used for educational system management, tracking student engagement, and advising students on activities leading to graduation. In contrast, second-generation multimodal-multimedia learning analytics are modeling richer data sources like speech, writing, images, and their combination. These methods are beginning to predict learners' actual motivation, cognitive state, and their development of expertise [Oviatt et al in press].

## 1.2 Significance

Modern civilizations all provide educational opportunities for children and young adults, so they can become integrated into adult communities as informed citizens, employed contributors, and self-actualized human beings. Advanced societies also support lifelong learning so their citizens can adapt to unexpected changes and remain informed contributors throughout their lives. Education therefore is a critical foundation for personnel training and job creation that supports the economy. It also supports the mastery of important communication and interpersonal skills, an understanding of societal values, and the ability to succeed at everyday practical tasks like following product or service instructions.

## 1.3 Focal Challenges

Computer technologies have become increasingly influential tools in students' classroom experience, including laptops, interactive whiteboards, clicker systems, online classes, smartphones and other handhelds. However, "technology-enriched learning" as implemented in school systems, including laptop initiatives based on commercial computers, too often has failed to significantly improve student performance on standardized achievement tests or cognitive measures [Campuzano et al. 2009; Shapley et al. 2009; Zucker et al. 2009; Oviatt, 2013]. In some cases, introducing technology has led to a decline in student achievement and an expansion of the gap between student groups [Oviatt et al. 2006; Vigdor et al. 2010]. Today's commercial technology too often emphasizes extraneous and addictive interface features that are designed to sell more computers, even though they undermine students' ability to focus attention and learn [Oviatt, 2013]. Educational technology in our schools also has been too heavily influenced by politics, rather than empirical evidence demonstrating efficacy for learning.

These evaluations underscore that commercially available computer interfaces need to be *completely redesigned* to deliver on the promise of supporting extended thinking and reasoning tasks and student learning. This will require: (1) **multimodal-multimedia interfaces based on input tools for rich "content creation"** (e.g., writing, speech), and that include **complementary and well integrated modalities**; (2) **elimination of addictive and extraneous interface design features,** so students' attention and activity is focused on exerting effort required to master new concepts and skills; (3) **evidence-based design principles** to guide the design of educational applications (e.g., science simulations, second language learning) based on scientific findings and theory [Oviatt, 2013]. It also will require: (4) **learning analytics** that goes beyond organizational management and advising on graduation likelihoods—to assessing students' cognitive state and actual learning achievement. Empirical evidence of learning at this deeper level is needed to personalize educational applications, optimize individual learning outcomes, and also form requirements for the purchase of new educational technologies in our schools.

The following table summarizes key challenges and outlines a roadmap for future work on this topic.

| State of the Art | Key Research Challenges | Summary Research Roadmap & Timeline | | |
|---|---|---|---|---|
| | | 5 years | 10 years | 15 years |
| Multimedia output widely applied to educational technology (science animations, simulated VR worlds like Second Life)

Keyboard-based input shown to limit effective learning, compared with expressively powerful speech, writing and multimodal input

Studies have shown multimodal input improves brain activation, ideation, problem-solving, inferential reasoning, composition & learning

Commercial computers now support speech & pen input to create and retrieve content (e.g., Microsoft Surface) and related applications (e.g., OneNote)

New multi-component input tools now can be matched more aptly to specific learning tasks

Click-stream learning analytics used to manage educational organizations, create student profiles & advise them

Newer multimodal-multimedia learning analytics starting to be developed to identify cognitive state & learning progress

Multimodal learning analytic predictors shown | Multimodal-multimedia technologies that support rich content creation during problem solving, including during distance education (e.g., MOOCs)

Multimodal-multimedia interfaces that accommodate individual differences in physical and cognitive abilities

Multimodal-multimedia interfaces that increase students' engagement and physical or communicative activity

Multimodal-multimedia technologies that support native speakers of non-Roman languages (Hindi, Mandarin, Japanese)

Multimodal-multimedia modeling of students' emotion and cognition

Multimodal-multimedia interfaces that eliminate distracting and addictive features so students can focus on learning

Automated multimodal learning analytics to support evidence-based educational practices and technology adoption

Privacy controls for students and teachers who provide data

Multisensory-multimodal neuroscience findings and learning theory to guide | Privacy controls for student & teacher multimedia data

Reliable recognition of ink representations (symbols, diagrams, numbers, words)

Accurate interpretation of extended student-system spoken dialogues (e.g. question asking)

Reliable recognition of basic emotions for adapting educational applications

Personalized tools for multimodal-multimedia learning

Multimodal-multimedia technologies that minimize distraction

Participatory design with learners & teachers to develop multimodal learning analytics functionality for end-users & classrooms

Wearable multimodal –multimedia activity tracking applied to student learning assessment

Accurate multimodal learning analytics for identifying learning | Active multimodal input for content creation in MOOC environments

Reliable recognition of naturalistic & complex emotions for adapting educational applications

Personalized & adaptive multimodal-multimedia learning tools based on recognizing cognitive state & learning progress

Advanced wearable & tangible multimodal-multimedia learning tools (with highly accurate multimodal activity tracking)

Extensive collaborative ecosystems for multimodal-multimedia educational technologies

Multimodal-multimedia technologies that support learning gains for diverse students (low- vs. high-performers), effectively minimizing the achievement gap

Accurate multimodal learning analytics for identifying learning progress automatically & continuously within a domain at 0-100 levels, enabling elimination of dedicated test taking in schools

Multimodal-multimedia technologies that support substantial learning | Multimodal-multimedia technologies that support substantial learning gains for native speakers of non-Roman languages (Hindi, Mandarin, Japanese)

Implantable multimedia learning tools for physically and cognitively disabled students

Ultra-reliable multimodal learning analytics ( >98% accuracy) for monitoring learning progress automatically and at large scale

Systems-level theories elucidating what occurs during the learning process, based on multi-level multimodal analytics |

| to outperform unimodal ones | design of educational technology, including systems-level theory (see knowledge discovery chap) | progress at 3-4 levels within a domain | improvements across all types of different content | |

## 2. Opportunities and Current State of the Art

To date, commercially available educational technology based on keyboard input has largely been a failed promise. Students use keyboard-based computers for relatively mechanical tasks like text editing, information retrieval from the Web, and email—but not for extended thinking and problem-solving tasks that constitute meaningful educational activities. Compared with keyboard tools, when students use more expressively powerful interfaces (e.g., pen-centric, multimodal), they produce more domain-appropriate ideas, solve more problems correctly, and make more accurate inferences about information [Oviatt, 2013; Oviatt et al. 2012]. They also experience lower cognitive load, communicate and build more on complex ideas, learn more during note taking and knowledge creation tasks, and produce better compositions [Hayes et al. 2010; Mueller et al. 2014; Oviatt et al. 2004]. For the *same students completing the same tasks*, the magnitude of facilitation in different studies has been a substantial 10-60%.

Parallel cognitive neuroscience research has shown that using multisensory-multimodal input tools creates a long-term sensori-motor memory. For example, fMRI scans of children and adults who actively construct complex letter shapes, compared with simply viewing, naming, or typing them, reveal increased neural activation in the brain's Visual Word Form Area, improving visual letter discrimination, word comprehension, and reading [James, 2010]. That is, the motor act of writing leads to corresponding improvement in letter recognition and reading comprehension. In summary, neuroscience research has demonstrated that multisensory-multimodal brain circuits develop to support cognitive processes associated with learning, such as reading [James 2010].

The above findings demonstrate progress in understanding how and to what extent digital tools can mediate thinking and reasoning, and they elucidate the multisensory-multimodal neuroscience foundations of human learning. They emphasize that learning involves multisensory-multimodal brain processes, and that human learning can be optimized when active multimodal input occurs. Multimodal-multimedia technology that elicits complex human actions, both spatially and multimodally, can stimulate deeper encoding and retention of learned information than tapping input using keyboards. Among the most promising future directions for educational technologies are pen-centric, conversational, tangible/manipulable, and multimodal-multimedia interfaces that combine them [Oviatt, 2013]. In recent years, these areas all have been emerging rapidly at the commercial level.

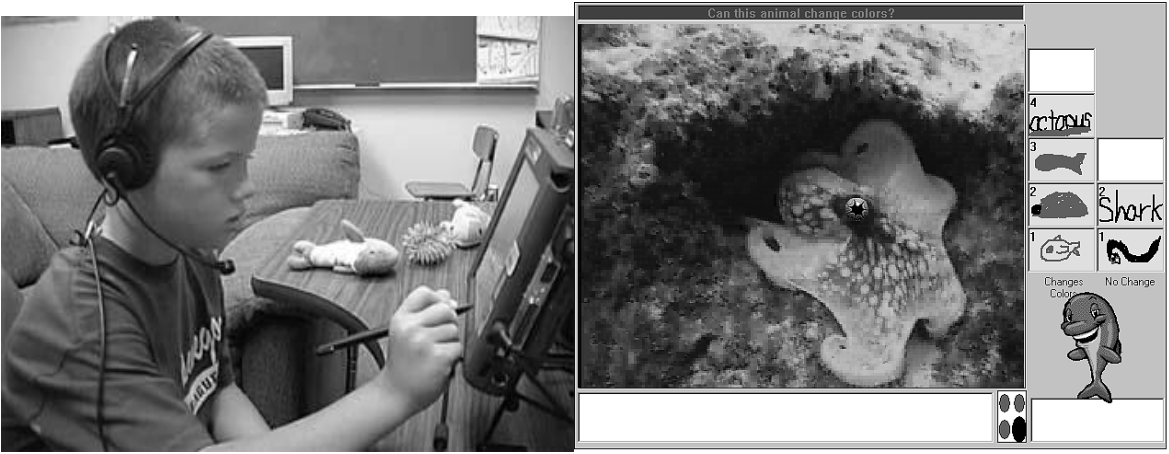

Figure 1. Eight-year-old conversing with animated characters, first to ask the digital octopus questions about itself, and then to talk with the dolphin character while using pen input to make a summary graph about what they learned [Oviatt et al. 2004b]

During the past five years, multimodal learning analytics has begun to emerge for analyzing rich human-centered communication data like speech, writing, facial expressions and other non-verbal movements [for review, see Oviatt et al. in press]. These data sources are especially apt for evaluating motivation and emotion, cognitive state, and the acquisition of domain expertise. Recent work has shown that multimodal learning analytics is a fertile method for accurately predicting students' learning. When mining these rich data sources, the analysis of combined multimodal data also typically supports higher predictive accuracies than relying on unimodal data [Oviatt et al. in press]. For further discussion on the current state of multi-level multimodal analytics, see the chapter on knowledge discovery [Oviatt et al., this volume].

Education of students with physical and cognitive disabilities, especially in the context of their inclusion in regular classrooms, provides significant opportunities to develop and apply multimodal-multimedia technologies. For example, to meet the language and communication needs of a hard-of-hearing individual, we can develop tailored technologies such as real-time speech recognition, captioning, synthesis, and sign language recognition and animation.

Machine learning can also make learning fun and engaging. For example, machine learning algorithms can guess what a person wants to draw, suggest where to draw the next step, and replace a doodle with polished clip art. Another example is when one is composing a document. Machine learning can suggest sentences with better writing style, or find relevant multimedia materials online based on content understanding. However, it is important that machine learning aids such as this be used selectively and, in particular, not in a way that reduces students' effort during active learning.

As the above discussion emphasizes, learning depends on students' engagement in physical and communicative activity. As a result, the development of educational technologies that offer an apt collection of input modalities is critical. Unfortunately, the development of expressively powerful input tools has lagged substantially behind multimedia output, which already has been widely applied to educational technology, including in immersive and animated forms (e.g., explanatory science animations, simulated virtual worlds such as Second Life).

# 3. Open Challenges

Numerous challenges remain to advance this area, including the need for:

- Multimodal-multimedia technologies that support rich and flexible content creation during extended problem solving, distance education (e.g., MOOCs), and for native speakers of non-Roman languages
- Multimodal-multimedia interfaces for accommodating extreme individual differences, including physical and cognitive disabilities
- Multimodal-multimedia interfaces that increase students' engagement, and physical or communicative activity level
- Adaptation of educational technology to complex and naturalistic student emotions that motivate learning
- Multimodal-multimedia interfaces that eliminate distracting and addictive features so students can focus their attention on effortful learning
- Multimodal-multimedia interfaces designed to enhance complementarities among input modes to improve system functionality, usability & robustness
- Multimodal-multimedia ecosystems that optimize educational functionality and usage in collaborative contexts
- Automated multimodal-multimedia learning analytics to support evidence-based technology adoption by schools
- Privacy controls for students and teachers who provide multimedia data
- Multimodal learning theories to guide system design, including systems-level theories that account for emergent learning phenomena

# 4. Achievable Milestones

## 5 Years:

Privacy controls for students' and teachers' multimedia data; Reliable recognition of written representations in major languages (e.g., symbols, diagrams, numbers, words);

Reliable multi-turn student-system dialogue interpretation in major languages (e.g. question asking);

Reliable basic emotion recognition for adapting educational applications; Personalized multimodal-multimedia learning tools;

Multimodal-multimedia educational technologies that minimize distraction;

Multimodal-multimedia educational technologies that stimulate learning gains involving different content;

Participatory design with learners and teachers to develop multimodal learning analytics for end users and classrooms;

Multimodal learning analytics for estimating students' learning progress at 3-4 levels, rather than bimodally.

## 10 Years:

Reliable naturalistic and complex emotion recognition for educational applications;

Advanced personalized and adaptive multimodal-multimedia learning tools (e.g., based on recognizing emotional and cognitive states);

Wearable, tangible and embedded multimodal-multimedia learning tools (with advanced activity tracking algorithms combining physiological and communications data multimodally for improved accuracy);

Collaborative ecosystems for deploying multimodal-multimedia educational tools for greater impact;

Active multimodal content creation and interpretation for MOOC environments;

Multimodal-multimedia educational technologies that produce substantial learning gains for diverse students (e.g., low- vs. high-performers), eliminating the achievement gap;

Accurate multimodal learning analytics for monitoring learning progress automatically, enabling the elimination of test taking;

Evidence-based technology adoption by schools.

## 15 Years:

Multimodal-multimedia educational technologies that yield substantial learning gains for native speakers of non-Roman languages (i.e., 80% of world);

Implantable multimedia learning tools for physically and cognitively disabled students;

Ultra-reliable multimodal learning analytics (i.e., >98% accuracy) for monitoring learning progress automatically at large scale;

Systems-level theories of the learning process.

# 4. Summary of Benefits

Multimodal-multimedia technologies can leverage our multisensory-multimodal brain processes for learning. They have the potential to support substantially deeper encoding and retention of learned information, compared with keyboard interfaces. Multimodal-multimedia technologies that emphasize content creation will be best suited for supporting complex, generative, and extended learning episodes. Over the next ten to fifteen years, this technology and applications based on it are expected to revolutionize worldwide education, especially in countries with languages that are not Roman alphabetic. It will enable mastering new domain knowledge, communication and interpersonal skills, and personnel training related to job creation. Effective multimodal-multimedia technologies will become essential tools for supporting lifelong learning and adjustment to change.

# Chapter 10. **Multimodal for Healthcare**

**Chapter Editors:**

Sameer Antani, U.S. National Library of Medicine
Ramesh Jain, University of California Irvine
Balakrishnan Prabhakaran, University of Texas at Dallas

**Additional Workshop Participants:**

Reuven Meth and Ketan Mayer-Patel

## 1. Introduction

There is an increased interest in evidence-based medicine and data-driven healthcare. There are several efforts underway that aim to gather and use vast troves of health data with an aim to eradicate (or minimize the effects of) disease and improve overall population health. The U.S. National Institutes of Health (NIH) has been charged with the *All of Us* Research Program, which is *a historic effort to gather data from one million or more people living in the United States to accelerate research and improve health[1]*. The effort aims to create opportunities for researchers to uncover paths toward delivering precision medicine by taking into account individual differences in lifestyle, environment, and biology. The NIH also has the Big Data to Knowledge (BD2K) program that aims to *facilitate broad use of biomedical big data, develop and disseminate analysis methods and software, enhance training relevant for large-scale data analysis, and establish centers of excellence for biomedical big data. The BD2K Program also supported initial efforts toward making data sets "FAIR" – Findable, Accessible, Interoperable, and Reusable[2]*.

In the United Kingdom, the Burroughs-Wellcome Trust has set up a website, called Understanding Patient Data, that explains to the lay public how patient data can be used to further biomedical research, the challenges in using such data, and how they could volunteer their information for research[3]. These examples are but a tiny sample of the vast troves of biomedical data that are being gathered and used for research. By no means are these ideas new. In previous years studies/collections such as the Framingham Heart Study[4], which began in 1948 to identify risk factors for heart disease, continues today for cutting-edge heart, brain, bone, and sleep research. Also, the MIMIC collection (Johnson et al., 2016), an openly available dataset developed by the MIT Lab for Computational Physiology, comprising de-identified health data associated with approximately 40,000 critical care patients. It includes demographics, vital signs, laboratory tests, medications, and more.

---

[1] NIH All of Us: https://allofus.nih.gov

[2] NIH BD2K: https://commonfund.nih.gov/bd2k

[3] Burroughs-Wellcome Understanding Patients: https://understandingpatientdata.org.uk

[4] Framingham Heart Study: https://www.framinghamheartstudy.org

These data aren't just tabulated columns of numeric data, nor are they just textual reports or case studies, but include a mix of clinical notes, case reports, billing data, other coded data, genotype data, phenotype data, lifestyle data, biomedical imagery (e.g., radiology images such as X-Ray, MRI, CT Scans; or, digitized microscopy or histology images, etc.), audio (e.g heart beats), video (e.g. observational data for gait, falls, taking medications), and signal data (e.g., ECG, fMRI, IoT data). Analytics applications that are based on all types of such multimodal biomedical data are critical for advancing healthcare research and advising the clinician, informing the patient toward a healthy lifestyle, and preparing the biomedical researcher. Computational science can play a significant role in devising novel algorithms and techniques, and developing applications that can aid in discovering unknown relations, identifying patterns, learning from the complex biomedical data. There are research opportunities within each modality, e.g. image analytics, and machine learning for computer-aided diagnosis (CADx), text analytics for clinical question-answering and medical informatics, etc., and using multiple modalities.

Advances in multimedia analytics for healthcare can positively impact areas of general healthcare applications, and personal health, as a result of applications that can analyze personal health data and other factors such as rapid and continued growth in PHR-EMR use; clinical protocols on smaller datasets; opportunity for population scale research; popularity of personalized wearable sensors and smart devices; and emerging solutions addressing data challenges: silos, ownership, and privacy.

The US NIH defines Precision Medicine as follows: "Precision medicine is an emerging approach for disease treatment and prevention that takes into account individual variability in genes, environment, and lifestyle for each person"[5]. In other words, accounting for individual variability is a different and important challenge from the challenges faced for "general or public" health. A simple example might be the difference between addressing a flu epidemic and handling an inherited heart condition[6]. While the flu epidemic affects a large population, an inherited heart condition might affect a very small population. Tracking, educating the affected persons, and treating (if possible) might involve totally different strategies. For a typical flu epidemic, large amounts of data are available for analysis: historical data, seasonal data, flu strain data, etc. Whereas for a genetically inherited disease, the affected population might be very small. And the data, if available, might be incomplete. It is also possible that other comorbidities might complicate the diagnosis, possible treatment/intervention strategies.

---

[5] NIH Precision Medicine Initiative:
https://medlineplus.gov/magazine/issues/fall15/articles/fall15pg19-21.html
[6] Inherited Heart Diseases:
https://www.hopkinsmedicine.org/heart_vascular_institute/clinical_services/specialty_areas/center_inherited_heart_diseases.html

Hence, in carrying out the discovery of patterns leading to a poor outcome, patient specific physiological conditions would have to be kept in mind. For instance, a patient may have known atrial fibrillation and the physicians may consider this as an accepted baseline for cardiovascular functions. Or, a patient may have known gait impairment and physicians may consider this as the accepted baseline for biomechanical functions. The data acquired through on-body sensors might have to be analyzed keeping these baselines as well as the specific physical therapies prescribed for each patient in mind.

**Definition:** *The tasks within multimodal computational analysis of healthcare data includes but is not limited to:*

- (i)     *multimodal multi-granular multi-temporal health data alignment and fusion;*
- (ii)    *personalized-dynamic health data model development from data acquired through population-wide streaming (iOT) health sensors;*
- (iii)   *data acquisition challenges: (a) maintaining data quality, (b) normalizing data resolution over time and across devices during any functional era;*
- (iv)    *analytics necessary for labeling multimodal data, derivation of semantics, and techniques for multimodal question-answering and cross-modal information retrieval; and,*
- (v)     *data analysis leading to disease prevention (lifestyle), and cure (diagnostics, medications) of disease.*

| Topic | State of the Art | Key Challenges, Unmet Needs | Road Map | | |
|---|---|---|---|---|---|
| | | | **5 Years** | **10 Years** | **15 Years** |
| Multimodal for Health | Computer-vision and Machine Learning for specific disease patterns in small to moderate sized data. | There is a lack of annotated data which is key for training machine learning algorithms. | Discovering and representing semantics in multimedia data | Surge in health focused personalized data acquisition (from private clinics, iOT sensors, and hospital systems.) | Start incorporating live data from sensors (hospital, personal, smart homes or smart surroundings) |
| | Biomedical and clinical informatics applications. | Despite efforts in gathering large data sets, there is limited access to PHR/EMR health data. | Multimedia information retrieval | Addressing problems with data alignment and fusion | Continuously develop and advance a dynamic data model personalized for aiding in healthy living. |
| | Search medical literature | Data quality: There are often "holes" (missing data) as well as errors in data sets in big data. | visual question answering from biomedical knowledge-bases | Improved data sharing policy | PHR/EMRs would be able to expand and amplify existing biomedical knowledge-base. |
| | Annotate and search images and multimedia | | Advancing machine learning algorithms to capitalize on available datasets. | Applications that can use the iOT data to improve healthy lifestyles, and provide timely user alerts. | |
| | Append PHR with (limited number of) personal iOT sensors | With the advance of time there will undoubtedly be advances in technology resulting in higher resolution sensors. Such varied resolution will need to be retroactively resolved. | Advances in health iOT. | | |
| | Analysis of limited examples of EMR data sequestered in health systems after acceptance of data sharing policy | There are real challenges of technology and policy in securing protected health information (PHI) and personally identifiable information (PII) in data. | | Personalized, real-time data analytics for continuous monitoring of patients keeping in mind the pre-existing health conditions. | |
| | | There is a need to address the challenges of "personalized analytics" in the presence of small and insufficient data. | | | |
| | | Finally, there are questions of data ownership and attributing data providers – particularly, when the multimedia healthcare data | | Applications that take advantage of health knowledge-bases. | |

| | | is assimilated piecemeal from a large number of datasets. | | | |
|---|---|---|---|---|---|

# 2. State of the art

Multimedia research in healthcare is not new. There have been several efforts reported in the literature that analyze some biomedical data which can be broadly organized as below:

1.  Computer-vision and Machine Learning for specific disease patterns in small to moderate sized data.
    This has been the most commonly applied effort in multimedia analytics for healthcare and is typically applied in CADx and screening applications. A simple search on "computer aided diagnosis health" on Google Scholar reveals that the top hits are all specific to a particular modality (X-ray, MRI, CT) for a particular organ or region (chest, abdomen, breast, brain) for a specific disease or set of closely related diseases.

2.  Biomedical and clinical informatics applications.
    Current research focuses on measuring outcomes on available population datasets or measuring impact of applications for a particular clinical need. Data gathering, normalizing, coding, and standards are of particular interest. Collecting available data for future research use is also of interest.

3.  Search medical literature
    Several applications are applying NLP, information retrieval, query understanding to hypermedia biomedical data sets. APIs aim to aid in application development.

4.  Search and annotate images and multimedia
    As part of steps above, methods are being developed for searching non-text media, and techniques for automated but noisy multimedia annotation, and learning from non-annotated or partially annotated data.

5.  Analyze PHR with (limited number of) personal sensors.
    The market is beginning to get flooded with a variety of health sensors ranging from connected weighing scales, smart phones and watches, blood pressure systems, automated pill reminders, etc. All these devices send the data back to the device manufacturers. Only a limited number of manufacturers make the data available for research, however. But, there have been initial efforts here as well.

6.  Analysis of EMR data sequestered in health systems - policy agreements.
    There has been an awakening on the need for using vast amounts of health data to make reliable predictions and develop meaningful applications. However, the data is sequestered within health systems, and behind complicated data sharing

agreements, mostly out of fear of liability, accidental exposure of PHI, and PII, and unclear rules about ownership and valuation of the data.

# 3. Challenges unmet

While there is a strong impetus and interest in advancing the state of the art in multimedia analytics of healthcare data, there are significant challenges facing the progress of this research.

(i)     With the advance of time there will undoubtedly be advances in technology resulting in higher resolution sensors. Such varied resolution will need to be retroactively resolved.

(ii)    There is a lack of annotated data which is key for training machine learning algorithms. Even the machine learning algorithms are only beginning to approach the notion of *artificial intelligence in medicine*. While this topic has been falling in and out favor over decades, advances is computing power have enabled previously compute resource intensive methods to become popular and tackle low-hanging fruit in big health data analytics.

(iii)   Data quality: There are often "holes" (missing data) as well as errors in data sets that can be hard to discover in *big data* but have the potential to be significantly detrimental to research outcomes.

(iv)    Despite international organizational efforts in gathering large data sets and interest in collecting and disseminating existing biomedical datasets, there is limited access to PHR/EMR health data. Undoubtedly, data alignment and fusion research techniques are necessary to enable dissemination of existing biomedical data sets.

(v)     However, it is not just a question of technology, but also of healthcare research policy – there are real challenges of securing protected health information (PHI) and de-identifying clinical data to keep personally identifiable information (PII) obscure from prying eyes.

(vi)    There is a need to address the following challenges of "personalized analytics" in the presence of small and insufficient data:
- What is the feasibility of the availability of annotated data in such instances?
- Will the annotations done for one patient be still valid for another patient, considering the personalized nature of disease or health condition?
- How does one validate a decision support system that has small and imprecise historic data?

- How does one quantify the precision and accuracy of proposed approaches and compare the efficiency of different approaches proposed for personalized analytics?
- Will unsupervised pattern discovery approaches (such as work by Balasubramanian et al. (2016)) be better than training based learning approaches?

(vii)    Finally, there are questions of data ownership and attributing data providers – particularly, when the multimedia healthcare data is assimilated piecemeal from a large number of datasets.

# 4. Achievable Milestones

## 5 years

With current research interest in discovering semantics in multimedia data, and continued interest in multimedia information retrieval and visual question answering from biomedical knowledge-bases, it is conceivable that these areas will be of strong research interest and result in significant advances in the next 5 years. However, semantics and visual question answering will remain as challenges for a longer period while the next wave of advances are made to improve data quality.

## 10 years

The next 5 years in the advance of multimedia healthcare research will probably see a surge in health focused personalized data acquisition through iOT sensors, but will also see the pressing need for data alignment and fusion; improved data sharing policy; applications that can use the iOT data to alert the user; and applications that aid improved interaction between knowledge extracting algorithms, the biomedical knowledge-bases and biomedical domain experts. It would be possible to have personalized, real-time data analytics for continuous monitoring of patients keeping in mind the pre-existing health conditions.

## 15 years

A longer horizon view depends critically on advances in the initial steps outlined above. If those are made, then it would be possible to start incorporating live data from sensors (hospital, personal, smart homes or smart surroundings) to continuously develop and advance a dynamic data model development that is both general for routine use and personalized for aiding in healthy living. It is also, therefore, appropriate to imaging that PHR/EMRs would be able to expand and amplify existing biomedical knowledge-bases.

# Chapter 11. **Smart Infrastructure**

## Chapter Editors:

Klara Nahrstedt, University of Illinois Urbana-Champaign
Ramesh Jain, University of California Irvine
Shih-Fu Chang, Columbia University

## Additional Workshop Participants:

Shri Narayanan, Hari Sundaram, Jie Yang (listener) and Zhengyou Zhang

## 1. Introduction

A healthy and robust physical infrastructure, including roads, bridges, buildings, power grid, agriculture, water grid and others, in our cities and rural areas represents a critical component for our society and its citizens. However, the current status of US infrastructure is ranked as D+ according to the latest report on America's infrastructure [Mynatt2017] and there is a major need to invest and innovate at multiple levels to improve the current situation. Especially, since our society is seeing major challenges such as population growth, aging population, increased pedestrian and vehicular traffic congestion in cities, water usage and electricity demand increase. We are seeing major urbanization. For example, in 2008, 50% of population lived in urban areas. It is predicted that by 2050, 70% of population will live in urban areas [Green2017].

**Definition:** Smart infrastructure represents the deep embedding of sensing, computing, communication, decision making and actuation into the traditional physical and human infrastructures for the purpose of increasing efficiency, resiliency and safety. For example, embedding sensors along the road can assist in controlling traffic lights and optimize traffic flow. Installing air pollution sensors at crossings of streets, and providing pollution maps to population allows citizens to react as well as contribute to cleaner air by less driving, and taking public transportation. Developing smart power grid capabilities allows efficiencies in generation, transmission, storage, and distribution of electricity, and increasing resiliency during natural disruptions, as well as easier inclusion of new energy sources and next-generation electrified vehicles.

As shown in Figure 1 [ITU2015], smart infrastructure and its technologies (ICT-based hard infrastructure and services layers) impact not only changes in the physical infrastructure (non-ICT-based hard infrastructure and natural environment), but also in the human infrastructure, represented by city government officials, city-workers, citizens, vendors, and other stakeholders, as well as cyber infrastructure for information dissemination and social media citizen communication.

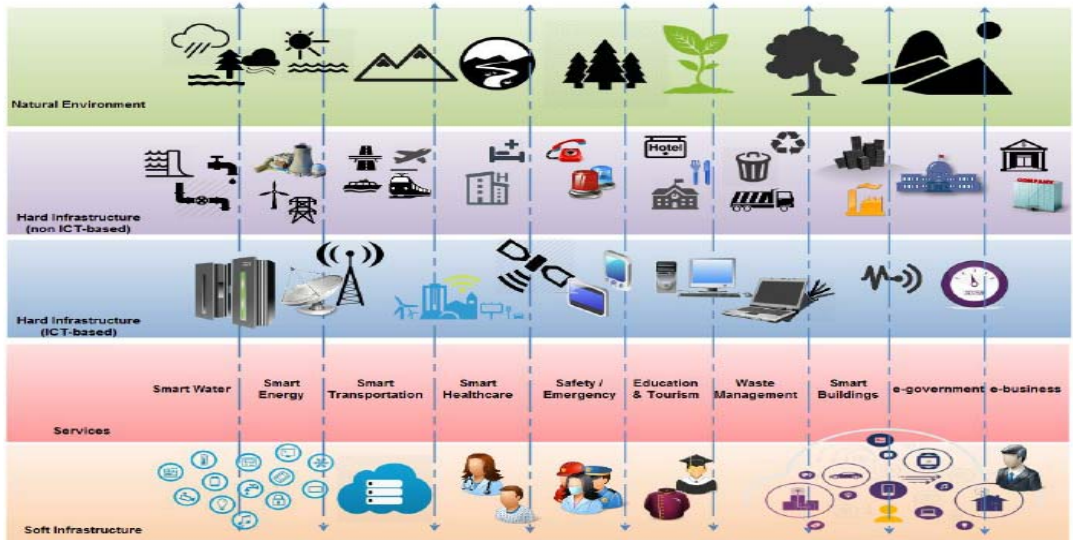

**Figure 1**: *Smart Infrastructure Layers for Smart Communities [ITU2015]*

**Scope**: In this chapter, we will discuss multimedia, multi-modal, multi-sensory-related challenges and research directions in smart infrastructure services to achieve robust and sustainable smart infrastructures for next generation smart cities and communities. We will analyze the challenges and research directions that come up within the technological workflow "closed loop" when multi-modal sensory and actuation devices, called also *Internet of Things (IoT)*, get embedded into various infrastructures such as transportation, food, water, energy, health care, buildings, safety and security. The technological workflow loop includes (1) sensory "things" development and deployment, (2) connection of "things", (3) digital data collection from "things", (4) processing, aggregation, analytics of correlation of data and metadata (context) according to human-physical models and urban domains/application services such as urban transportation, utilities (smart grid, water, gas, waste), urban living (smart buildings), public services (e-government, incident reporting), safety and security (police, first responders), sustainability, (5) comprehension, interpretation, reasoning about data and their findings, (6) creation of new services and actions to enable actionable information, and (7) actuation of "things" to gain new data and/or metadata or control processes in the physical world (e.g., traffic control or situation alert generation).

This workflow loop defines then the intelligent infrastructure as the integration of sensing and data analytics with municipal capabilities and services that enable evidence-based operations and decision making. The intelligent infrastructures can be identified via *descriptive, prescriptive, predictive and plan-full attributes*. Descriptive attributes provide an accurate and timely characterization of current state, prescriptive attributes recommend immediate and near-term actions, predictive attributes anticipate future assessments of available resources and capabilities, and plan-full attributes guide the complex decision making, actionable information and scenario planning [Mynatt2017]. In addition*, integrative attributes* will need to identify smart cities infrastructures to provide co-dependencies among infrastructures (e.g., food, energy, water) and resolution of conflicts (e.g., if one uses more energy to run pumps and avoid flooding of certain areas or one saves on energy, but allows certain areas to flood).

**Drivers**: The drivers of smart infrastructure technologies must be jointly the citizens of cities and rural areas, the governments of various cities, counties, states, and federal government, industrial vendors and academic community. Without the joint collaboration of all stakeholders, the changes to traditional infrastructures will not happen since cities themselves do not have sufficient tax incomes to push additional infrastructure innovations, vendors without business opportunities do not have incentives to lead and/or assist in innovating infrastructures, and academics without possible testbeds and collaboration with cities and companies do not have the realistic environments to test their ideas and innovative solutions. For example, Coen of Oostrom, CEO of OVG Real Estate in Netherlands has the mission to make the buildings sustainable and zero-energy by making use of new and smart technologies. For example, OVG considers power plants with bio-cogeneration and cooling from river water, adjustment of oxygen to the number of occupants, giving opportunity to create a healthy environment. Smart buildings use light and climate adjustments in their rooms, controlled by users' smartphones. OVG's system, called The Edge, uses 20,000 sensors and offers real-time information how the building is being used. This offers possibilities to make use and maintenance much more efficient. OVG showed that 35% of carbon emission in the world come from buildings [Oostrom2016], hence understanding the usage of building via processing of 'big data' is crucial to gain the intelligence for effective planning. OVG showed that business can drive the change in smart infrastructures and make changes in buildings to satisfy 96% sustainability according to the UK BREEAM (BRE Environmental Assessment Method) rating system (similar to LEED rating system of buildings).

**Unique Contributions**: Multimedia and multi-modal community can make crucial contributions within smart infrastructures and smart communities, because this community understands the embedding multi-sensory devices, the collection and correlation of multimedia data and metadata, utilization of context, multi-modal processing and analytics issues, multi-modal communication and time-sensitive streaming problems, dynamic actuation and adaptation according to human and physical constraints. Hence, in all parts of the smart infrastructure technological workflow loop, multimedia/multimodal community can contribute greatly:

- development of new multimedia sensors, including new cameras, sound devices, mobility sensors of people and cars, and other sensors;
- innovation in integrated real-time and safe distributed algorithms and protocols to enable timely and trusted connectivity among multi-modal and heterogeneous sensors, dissemination, streaming and delivery of synchronized multi-modal data and metadata to designated end-points;
- development of new signal processing and information-based algorithms to collect, compress and process reliably sensory data;
- exploration of new aggregation and analytics methods, including deep learning, to deal with very large multi-modal and correlated datasets;
- development of novel algorithms which will reason about predictions, coming from multimodal analytics, and provide actionable information to users such as power grid operators, city workers, citizens or building managers of smart infrastructures;
- development of new multimedia services that will allow for easy configuration, management and sustainability of existing and new infrastructures;
- innovations towards smart actuation devices that will change and adapt infrastructure's behavior based on the situation; and
- development of open testbeds that can easily integrate, maintain, upgrade heterogeneous types of sensors and analytics tools and APIs.

# 1 Summary Table and Research Roadmap

| Scenario | Current State-of-the-Art | Key Challenges and Unmet Needs | 5 Years Road Map | 10 Years Road Map | 15 Years Road Map |
|---|---|---|---|---|---|
| Urban Mobility | Sensing at intersections; Google map with traffic density; Usage of mobile phone; sparse deployment of pollution sensors; vehicle data of city service vehicle fleets; local room sensors in buildings. | -Monitor traffic flows and quality of air/water in real-time; -Inter-dependency between pollution, traffic flow (pedestrians and vehicles), and climate; -Matching renewable energy sources to next-generation electrified transportation; -Effective deployment of smart building operating systems with integrated multi-modal sensors and actuators that preserve privacy, and provide actionable intelligence for occupants, their | -Deployment of new arrays of sensors for roads, air, water, and mobility; - Alternative energy sources and more efficient energy distribution and storage systems; -Combination of information from physical sensors and online citizen communication; -Advanced analysis of all sensors to provide diverse urban mobility maps in real-time, secure and private manner; - Advanced analytics of all sensors within buildings to provide occupants mobility maps in real-time, | -Correlation, inference, reasoning about all the combined sensor data, with additional signals coming from other than vehicles, roads, utilities; - Development and deployment of open scalable testbeds for innovation of new technologies, services, and business models; - Development of integrated actionable intelligence between different domains such as food, energy, water (FEW). - Development of integrated and conflict free multi-modal technologies for diverse transportation sectors (e.g., rail, automobiles, trucks, ships) | -Develop new multi-modal multi-sector services communicating seamlessly with other infrastructures to contribute and use information to/from other sectors anytime, anywhere in safe, secure, private manner. |
| | | mobility patterns and usage. | secure and private manner. | | |
| Sustainability | Often no backward compatibility for many multimedia technologies used in urban infrastructures. | Urban infrastructures age differently (water, energy, and transportation) and see different investments in multi-modal technologies. This causes a major challenge how to integrate these technologies used in different infrastructures. | -Invest in multi-modal information compatibility and/or formats that do not age and change every 6 months; -Develop edge infrastructures that project aging sensors and actuators from security attacks. | -Develop long-term information formats and technologies that live at least 10 years similar to physical infrastructures; -Develop evolvable and adaptable multi-sensor technologies, algorithms and infrastructures for longer term (physical materials are developed to last for long time and this should be a similar goal for cyber-systems). | -Develop new multi-modal cyber-physical solutions that will be easily deployable, evolvable, and sustainable in terms of their forms, functionality, privacy and security. |

# 2  State of the Art

Major efforts regarding smart infrastructures within smart cities happen around the world. We will present selected set of smart cities and their infrastructures to motivate the various challenges and road map goals for the multimedia community.

- Singapore has a major initiative going on to create smart nation. Smart Nation initiative [Singapore2017] uses networks, data and info-comm technologies to enable citizens lead meaningful and fulfilled lives. Some interesting infrastructure innovations are in: (a) public transportation where mobile multimedia services for commuters keep people informed and enable them to make informed decisions on their journey plans (e.g., when their busses arrive, how busy the busses are and when they need to leave their homes or their offices), the commuter mobile services also allow the public transport authorities to better plan the overall bus system, (b) smart buildings and smart homes where multi-modal data analytics services analyze collected data from building sensors and home appliances and allow residents to detect activities, save energy, water, food, and other resources (e.g., motion sensors can sense elderly activity, occupancy sensors can save energy in office buildings, conference rooms, smart meters allow access to real-time water and electricity usage pattern);

- Songdo, South Korea is a new city built from scratch with advanced infrastructures utilizing many multimedia technologies [Green2013]. Most notable are (a) telepresence across the whole city where you can communicate with anybody from any building anytime; the video-chatting technology (similar to Skype) is integrated into televisions and subscription based classes; (c) tele-health to talk to the doctor remotely, and (c) multi-sensory system controlling underground system of pipes for automated waste management where garbage is sucked directly from people's apartments, consequently there are no street-corner trash cans or garbage trucks.
- Masdar City, United Arab Emirates,[Masdar2012] is a new completely renewable energy city, carbon neutral. They created urban heat oasis by the design of buildings with solar powers. Another innovate technology is deployed in their public mass transit and personal rapid transit (PRT). The motor vehicles are banned and only public transport and PRT enable transportation in the city. This design of transportation infrastructure allows for narrow and shaded streets that help funner cooler breezes across the city.
- Chicago, USA [Chicago2017] is an existing city which engaged to deploy an air pollution sensing infrastructure via the Array of Things (AoT) technology. The AoT are multi-modal devices with multiple different sensors that are installed across Chicago, monitor different aspects of pollution on streets of Chicago, and enable analysis of the pollution data to create pollution maps. The citizens then can access pollution maps and decide when to go out, and assist in keeping the air cleaner by driving less, improving their homes to pollute less. The plan is to install 500 sensor nodes measuring factors such as barometric pressure, light, carbon monoxide, ambient sound and temperature. The future goal is to create sensors to monitor other urban factors of interest such as flooding, precipitation, wind and pollutants.

# 3   Challenges

We need multi-modal and multi-dimensional technologies that bring together inter-dependent technologies from different infrastructures together. We need **fluid cities** [Venkatraman2014], not just smart cities, where different infrastructures with their multi-modal information talk to each other!

## 3.1.1   Sustainability of Multi-sensory Devices, Multi-Modal Information and Multi-Dimensional technologies

Many of the multi-modal and multi-sensory devices that are embedded in various smart infrastructures **age differently** in terms of upgrades, robustness, and functionality over time. This is especially challenging if one does incremental deployment of multimedia devices. Then the sensors/devices even in the same media group (e.g., cameras) can vary in their quality, their operating system support and they networking support. How do we ensure that the software and hardware of different ages still talk to each other after few years? Note that many sectors of smart cities (e.g., power grid, waste management) do not have the funds to do frequent upgrades of their sensory hardware infrastructure. In power grid, smart meters must last many years, otherwise the cost of electricity would go up rapidly.

Many companies do not provide backward **information compatibility** as new multimedia formats are coming up. This is especially true for multimedia devices such as video cameras, sound cards, graphics cards. For example, over the last 10-15 years, we went through 5 different generations of 3D cameras (several generations of Bumblebee cameras, and at least two generations of Microsoft Kinect cameras.

How should one build teleconferencing systems, on-demand content systems when multimedia devices are changes every year?

Many smart infrastructures **mature at different speeds and invest different amount of funds**. For example, smart grid has a road map, funds smart meters across generation, transmission and distribution grids to capture as many sensory signals as possible. On the other hand water management is moving forward very slowly with development of new sensors, cyber-data-management of captured data, and doing any advanced analytics. Many of the multimedia and multi-modal algorithmic, protocol, system, and service advances we know from entertainment industry (e.g., Flickr, YouTube, Netflix) are not present in smart cyber-physical infrastructures of smart cities because of (a) lack of funding, (b) lack of trust in new cyber-solutions; (c) lack of highly educated multi-modal computer scientists and computer engineers who would work in infrastructure-specific industries such as power grid, water grid, food industry. The closest deployment of computer scientists and engineers with multi-modal background is in automobile industry and overall transportation.

Multimedia software and hardware in smart infrastructures must be written with sustainability over 10-20 years in mind.

### 3.1.2  Security, Safety, and Privacy

The challenges in sustainability in terms of different software/operating system/functional services and their ages in sensory and other computing platforms, the information compatibility, and the infrastructures' different speed of deployment of multi-modal technologies causes major security, safety and privacy issues. If we deploy surveillance cameras, or noise sensors, and/or other sensing capabilities, how do we ensure that (a) they cannot be exploited as botnets to cause distributed denial of service attacks, (b) they are verified and software is written with the same care as electrical engineers design a power socket and certify via the National Electric Code (if somebody writes a Kinect software, who verifies what the software is actually doing?), (c) they are preserving privacy, and other safety concerns. How do we ensure that the multimedia data analytics is correct and not bias? How do we ensure that the learning algorithms are not adversarial and/or bias to serve one type of population and discriminate against other population in smart cities?

The multimedia software and hardware in smart infrastructures must be safe, trusted, secure and ensure privacy if desired.

### 3.1.3  Understanding Interoperability, Inter-dependencies

We talked about fluid cities where different infrastructures talk to each other. For example, in case of autonomous vehicles, the automobiles will communicate with roads, power grid (in case of Electric Vehicles), telecommunication infrastructure, users via their personal mobile phones and other infrastructures. Each infrastructure will have their information structures to capture, curate, manage, and analyze sensory data and yield decisions and actuations. How does one synchronize all the messages that are exchanged among infrastructure central cyber-points? Who is in charge to control the overall communication? How does one deal with different time-scale, different business decisions to deploy multi-modal devices and formats?

The multimedia software and hardware in smart infrastructures must be agile and adaptive, evolvable, capture dependencies and based on situational awareness adjust to different workflows in different sectors/infrastructures. A very good example, how different companies and content providers agreed on a streaming standard is DASH protocol. Hence, potential standardization will be required among multimedia devices, software, algorithms, and protocols to achieve interoperability. Also, plug-and-play, container-based platforms to plug new multimedia sensors, functions, learning algorithms will be needed.

Multimedia software and hardware and algorithms within them must have a representation and an understanding of inter-dependencies.

### 3.1.4  Real-time Multi-Modal Data Analytics, Knowledge, and Reasoning

We have done great advances in multimedia processing, and even context-aware multimedia processing. However, more needs to be done. Many of the smart infrastructures will have to deal with large scale of data and metadata of very diverse sensors, not just video and audio. How do we extract the same amount of information from accelerometer, gyrometer, compass as a surveillance camera extracts from video? Due to privacy demands, camera and video in smart infrastructures might be much less preferred, hence we will have to look at other multi-modal information to **infer mobility, density and contention points** in urban mobility and transportation. What is the minimal number of non-video sensors to infer the same information as from a video? What are the acceptable "anonymization" processes required before each of the sensory data can be included in various services?

To support high-level reasoning at the city scale, we need to integrate information extracted from divers data sources to build comprehensive knowledge representation ("data-to-knowledge") that can help understand the state of various entities of the smart city (people, environment, and infrastructures), discover their relations, and support reasoning of hypotheses and planning of new actions.

In case of video and images that can be collected by smart infrastructures in smart cities, e.g., smart buildings' entrances and other public places in cities, we will need to consider deep learning techniques to predict and prescribe what to do next as actions. However, how do we speed up deep learning algorithms, how do we avoid biases, how do we ensure appropriate training data? Can it happen that different smart infrastructures infer different things and contradict themselves giving users different options? How do we reconcile contradictions coming from different multi-modal information and different infrastructures? Do we need a concept of common context?

Another major issue will be to infer an actionable information from the various multi-modal signals. Do we have the reasoning and decision-making algorithms to extract actions and reason about them? We can predict all kinds of information and situations, but which ones are useful and actionable?

Multimedia software and hardware and algorithms must encompass deeper learning and reasoning to yield actionable information to operators of smart infrastructures and smart cities.

### 3.1.5  New Sensors

We need to develop also new multimedia sensors and new IoT with their embedded software for the next generation smart infrastructures. This includes cameras which can see multiple views as we see coming with Video360 cameras, cameras which produce not only RGB, but also polarized light, and work in other situations that smart infrastructures operate in (e.g., in tunnels for waste management or water management). We need to have sensors which are light, easily upgrade-able, robust, trusted, and can work in harsh condition if it is wet or dry environment, winter or summer.

# 4   Milestones

Here are few milestones that we can see for smart infrastructures:

**5 Years**

- New multimedia sensors and new IoT devices will need to be developed. The hardware is moving very fast in this direction and the multimedia community needs to follow with algorithms, software and overall system solutions.
- Sustainability must become a first order of a problem for smart infrastructures when deploying multi-modal solutions.
- Improved real-time multi-modal analysis will need to be available to smart infrastructure operators to yield human-led actions.
- Security, safety and privacy solutions need to become part of individual solutions within each smart infrastructure.

**10 Years**

- We will have a clear understanding of interdependencies among multiple sectors among smart infrastructures so that different modalities, devices, from different infrastructures are able to talk to each other, analyze each other's data in secure, safe and private manner.
- We will have first results about reasoning of multi-modal information.
- We will have adaptive and first generation of sustainable sensors, actuators, learning and analytics algorithms, multi-network protocols, sustainable information formats that will evolve.
- We will see development of new miniaturized sensors embedded in cloth, materials, and urban and rural infrastructures.

**15 Years**

- We will have first set of multi-modal, multi-sector services for fluid cities where smart infrastructures and their multi-modal services will be useful to each other in secure and private manner.

# Chapter 12. **Multi-Modal Research for Social Good**

**Chapter Editors:**

Alex Hauptmann, Carnegie Mellon University
Julia Hirschberg, Columbia University

**Workshop Participants:**

Dick Bulterman, Hari Sundaram, and Adam Wolfe

## 1. **Introduction**

In this chapter, we focus on the ways that multimodal research can foster social good.  The aim is to outline a vision, research and directions necessary for multimedia technologies to enhance our communities and human society in general. While multimodal research presents many challenges of genre collection and genre integration, it also presents major promise for tackling critical issues affecting the welfare of a society.  Areas such as healthcare, education, transportation, home care, security, and journalism are all areas in which multimodal efforts can help improve our society. The last few years have seen a tremendous surge in multimedia technology and applications, but much of this has been in the service of narrow goals such as autonomous vehicles, image/video classification and retrieval, analysis of human expression, or specific medical tools by improving underlying systems of networks, analytics, displays and interfaces.  Here we will examine several areas in which multimodal advances can work to improve society and the planet over the next 15 years.

Below are a few example areas where multimedia will contribute dramatically to social good in the coming years
- Home healthcare, rehabilitation and assistance
- Public safety and security
- News and social media analysis
- Facilitating societal dialog and cross-cultural understanding
- Improving education

The following table summarizes state of the art, key challenges and outlines a roadmap for future work on this topic.

| State of the Art | Key Research Challenges | Summary Research Roadmap & Timeline |
| --- | --- | --- |
|  |  |  |

|  |  | 5 years | 10 years | 15. years |
|---|---|---|---|---|
| **General:** | How can we integrate information from multiple modalities and represent the information?  How can we obtain necessary training data?  How can we protect individuals' privacy and safety?  How can we lower the cost of multimodal devices?  How can we improve human control of robotic devices and the quality of human interaction with them? |  |  |  |
| **Home healthcare, rehabilitation and assistance**: Smart devices using computer vision, ASR, and sensing to support virtual doctor visits, monitoring in the home; robot surgical aids and hospital care; daily fitness and health can be monitored by individuals | How can we identify key aspects of home and hospital healthcare that can be handled or facilitated easily by machines?  How can we improve the autonomy of the disabled? | More MM robots will improve the quality of life of the elderly at lesser cost. More tools supporting the autonomy of the blind, deaf and aged will be developed. Effective monitoring of health care in homes and nursing homes will be available. | Smart devices in the home will aid daily living.  In-home monitoring will detect medical and mental problems. MM robot services will be created to care for the elderly. Physical therapy by autonomous guide robots will be available. | Rehab and in-home therapy will reduce the need for hospital and rehab center stays. Assistance devices will extend the capabilities of the infirm. Robots will be able to gain and maintain trust with users and engage with them in human-like ways. |
| **Public safety and security:** Multimodal tools for detecting deception and illegal behavior deployed in cities, borders and airports using audio and visual recognition; predictive policing aids in alerting police to and solving crimes; social media can be mined to predict potential unsafe activities | How can we provide tools for MM analysis to public safety organizations and first responders? | More monitoring of traffic and crime in urban areas will be available. New interfaces with police body cams will be developed. Tools to coordinate response to crime and emergencies will be created | Better identification of audio/video information will aid crime prevention and prosecution as well as clearer evidence of police malfeasance. More cities will use drones for surveillance and security. MM approaches to deception detection in airports will be developed. | Autonomous aerial vehicles will patrol traffic and urban areas. Multi-sensor monitoring will be an integral part of public life. |
| **Transportation and Mobility**: Cameras and sensors monitor traffic, control traffic lights, and collect information on vehicles; self-driving vehicles heavily researched | How can academic research be incorporated into improving industry efforts to build safe self-driving vehicles? | "Safety-assisted" cars will be available. Routes will be identified that are safe for self-driving cars. | Vehicles will be more autonomous under normal driving conditions. | Self-driving vehicles will be used in most circumstances. Traffic patterns will be optimized over large regions. |
| **News and Social Media Analysis:** Research on identifying "fake news" in text and video | How do we distinguish truth from falsehood given contradictory MM evidence? | Tools for identifying "first post" of information and authenticity of images will incorporate MM input. | Automated tools for fake new flagging will integrate information from images, video, speech and text to provide better information for users. | Full automated fact-checking will be available together with confidence ratings. Most public events will be recorded. |

| Improving education: Robots and other MM applications can help to personalize education allowing students to learn on their own; MOOCs are widely deployed as aids to self-learning | How can we improve education via MM systems without creating student silos? | More apps will be developed for self-education. More devices will be created for younger students | Autonomous intelligent teaching systems for group education will be developed. | Teaching will become more system- than human-teacher-guided. Systems will integrate customized learning using user data plus personal history and environmental factors to assess knowledge retention and comprehension. |
|---|---|---|---|---|

# 2. State of the art

## 2.1 Home healthcare, rehabilitation and assistance

Generally we have seen major advances in computer vision, speech recognition, connected sensing device analysis through smart watches, phones, cameras and customized medical tools. These are being used to support virtual doctor visits, home installed sensors such room, chair, or bed occupancy sensors, motion sensors, intelligent thermostats, fans, air conditioners, refrigerators, for example. Monomodal diagnosis such as interpretation of radiographs for tumors, but not yet linked with evaluation of patient profiles in patient's electronic medical records. Many aspects of social good for healthcare are covered in the chapter on "Multimodality for healthcare", but we will merely illustrate the state of the art through some examples.

Currently, unimodal robots have begun to enjoy widespread use in homes, mainly as vacuums and pool cleaners. After years of development, the Swedish Electrolux Trilobite, a vacuum cleaning robot, became the first commercial home robot in 2001 (CYBERNETICZOO, 2017). A year later, iRobot introduced Roomba, at a tenth the cost of the Trilobite; it ran a behavior based controller. Since then, sixteen million Roombas have been sold around the world (IROBOT, 2017). Today these robots are able to build a complete 3D model of the house they clean and thus clean it more effectively.

Robots also are being used as surgical aids and in support of many hospital operations. The first such robot was Arthrobot, used for arthroscopic surgery in the early 1980s (Robodoc) a spin-out from IBM, which developed robotic systems for orthopedic surgeries, such as hip and knee replacements. More recently, Intuitive Surgical (Intuitive,Trevis) introduced the da Vinci system, a novel technology initially marketed to support heart bypass surgery and then used for treatment of prostate cancer.  While robotic surgical assistance has boomed in recent years, other hospital applications such as robotic delivery of hospital meal and medical record deliveries, have been less successful (Evans and Krishnamurthy 2005, Aethon).   However, other service industries such as hotels and warehouses (e.g.Amazon Robotics) have show that these services can indeed be automated.  However, in the medical domain in particular these services will clearly require more effective human-robot collaboration.

There are many examples of multimodal progress in rehabilitation and fitness. An increasing number of people track their daily activities in a variety of modalities such as GPS, motion sensors, heart rate sensors, galvanic skin response and video/imagery. Devices such as smart phones and smart watches are used to track daily activities such as steps, stairs, various activities, calories burned, and calories ingested using sensors for tracking daily activities to identify patterns that can affect health [Islam et al., 2015].  We can expect to hear much more about long-term impacts of these efforts over the next few years.
.
Research on coaching and monitoring for better health has shown success for example in monitoring hand washing to prevent transmission of infections across patients in hospital settings [Halyard, 2017; CENTRAK 2017]. Research projects have shown the feasibility of verifying the proper use of home infusion pumps by patients [Cai et al, 2015] , as well as monitoring and improving asthma inhaler technique to improve disease control and reduce expensive hospitalizations [Wang and Hauptmann, 2014]. There is increasing interest in the use of  smart multimodal systems to engage users interactively and with the use of social comparisons to improve health outcomes, such as through avatars and synthetic interviews. [MEDRESPOND 2017]

## 2.3 Public safety and security

Multimodal tools for detecting deceptive and possibly illegal behavior are already deployed in many cities, borders and airports already by city governments and law enforcement (Arikuma and Mochizuki 2016, Big Op-Ed). These tools collect audio and video information in order to analyze anomalous behavior indicated by gait and body gesture, facial expression, and speech characteristics to identify potential criminals and terrorists, as well as to identify known "persons of interest" by face recognition.  In the next 15 years these devices will increasingly employ information from cameras and microphones either stationary or attached to drones for surveillance.  While public spaces are generally considered to be legally open to such surveillance, issues of bias and violation of civil liberties may arise in the uses to which the information collected may be put.  Too, such information may be used to improve policing by limiting activity to times when it is truly needed and by removing human bias.  They may also prove helpful in managing crime scenes or in search-and-rescue operations.

In fact, to date, the multimodal tools deployed tend to be more useful for helping to solve crimes than to prevent them due to the low quality of person and event identification from audio and video and the lack of tools for searching massive video streams. The New York Police

Department's CompStat was the first tool pointing toward predictive policing (Perry et al 2013) and many police departments now use it (CompStat). Machine learning significantly enhances the ability to predict where and when crimes are more likely to happen and who may have committed them. However, predictive policing tools do raise the specter of innocent people being unjustifiably targeted, so research and resources should be directed toward ensuring this effect.  Multimodal techniques can be used to develop intelligent simulations for training law-enforcement personnel to collaborate even internationally. Horizon 2020 program, currently supports such attempts in projects such as LawTrain (LAW-TRAIN) The next step will be to move from simulation to actual investigations by providing tools that support such collaborations.  As multimodal search improves, better identification of audio/video information will provide better assistance in crime prevention and prosecution – as well as clearer evidence of police malpractice. Some cities have already added drones for surveillance purposes, and police use of drones to maintain security of ports, airports, coastal areas, waterways, industrial facilities is likely to increase, thus raising more concerns about privacy, safety, and other issues.

Tools also exist for scanning social media to search for certain events such as protests, demonstrations, or possible terrorist activities. Multimodal approaches can compare information from multiple sites and genres to improve search accuracy. Law enforcement agencies are increasingly interested in trying to detect plans for disruptive events from social media, and also to monitor activity at large gatherings of people to analyze security. Multimodal research on crowd simulations can help to determine how crowds can be controlled. A major challenge of course is to limit the potential for law enforcement agencies to use such tools to violate people's privacy.

The US Transportation Security Administration, Coast Guard, and the many other security agencies will probably increase their reliance on multimodal techniques to enable improve efficiency and efficacy (Tambe 2011). Research findings in vision, speech analysis, and gait analysis can aid interviewers, interrogators, and security guards in detecting possible deception and criminal behavior. For example, the TSA currently has an ambitious project to redo airport security nationwide (Neffenger 2016). The DARMS system is designed to improve efficiency and efficacy of airport security by relying on personal information to tailor security based on a person's risk categorization and the flights being taken. The future vision for this project is a tunnel that checks people's security while they walk through it. Once again, developers of this technology should be careful to avoid building in bias (e.g. about a person's risk level category) through use of datasets that reflect prior bias.

## 2.4 Transportation and Mobility

A survey by [Bonnes et al, 2016] provides a good overview of current smart transportation infrastructure and improved mobility through multimedia.  Cameras and sensors that control traffic lights, read license plates and observe vehicles can be found throughout the country (e.g. [Baran et al, 2016]). Smart parking lots that make drivers aware of available parking spaces are starting to appear [Shaikh et al., 2015]. Speeding cars can be automatically observed and photographed and tickets mailed to the owner of the car, as determined though the car's license plate.  Of course, the most attention has been given to self-driving cars, and, recently, even autonomous aerial vehicles. While a fully unassisted self-driving car in unconstrained traffic conditions is still a future goal, current advances of multimedia technology now enable warning when a car strays from its lane, adaptive cruise control, alerts about driver fatigue through steering and eye blink analysis, and pro-active collision avoidance braking.  All these serve to make the roads safer for all and reduce travel stress or delays.

## 2.5 News and Social Media Analysis

More subtle aspects in which multimodal research can contribute to social good are multimodal aids to those seeking to separate fact from fiction in the wealth of information provided in news and social media today.  One problem which has evolved with the rise of social media and which can affect our ability to address issues in multiple areas is the increasing prevalence of "fake news".  Such false but often widely disseminated and believed information can greatly hamper our ability to deal with real terrorist attacks and other criminal behavior, military confrontations, natural disasters, medical and emergencies by conflating false information with the true information we need to reason and plan effectively.  Governments, the military, corporations, disaster relief teams as well as journalists will profit enormously if we develop reliable methods to distinguish true news from false.

Despite the prevalence of fake news in social media and the wide recognition that it represents a serious problem ("How do we know what to believe?"), distinguishing truth from false data has proven an incredibly difficult task, compounded by the fact that, while some fake news is created deliberately with a specific goal in mind, other fake news is created or passed along innocently, by those with insufficient information, as rumors.  We believe that multimodal approaches to distinguishing fake news from truth have a high probability of success since they will provide multiple channels by which information can be evaluated.  The challenge of course will be integrating findings from each of these multimodal channels into a highly probable decision about the information in question.  To this end, we believe that the development of tools to integrate information from four modalities -- images, video, speech and text – would greatly improve the ability of those who need to know and act upon true information in multiple situations to distinguish fake from true information.

Tools to detect false information from single modalities have been the subject of much research already.  DARPA has recently initiated a large, multi-year research program on Media Forensics for "developing technologies for the automated assessment of the integrity of an image or video and integrating these in an end-to-end media forensics platform" [DARPA MEDIA FORENSICS, 2017]. For example, Hany Farid has developed techniques for detecting faked image content by identifying abnormal statistical features, low-level camera cues, as well as classifiers to distinguish photo-realistic computer graphics content from real photos [Gibney, 2017]. Google's reverse image search has effectively developed scalable image forensics methods to solve the provenance problem (finding copy sources of a an image) and there are also approaches to the photo genealogy problem (reconstructing the history of visual manipulations from sources to final products) (Kennedy & Chang, 2008). Much research has been done on deception detection from spoken cues as well as facial expressions, body gestures, and lexical features that perform significantly better in many cases than humans (DePaulo et al 2003; Newman et al 2003; Hirschberg et al 2005; Bachenko et al 2008; Frank et al 2008; Hancock et al 2008; Levitan et al 2016). These methods can be applied to audio, video, image, and text data to provide additional information about the deceptive intent of speakers. Classifiers developed to identify true from false text has already been integrated with acoustic/prosodic information to improve accuracy in human deception detection. Differences in online text sources over time can also be leveraged to identify differences and contradictions among sources, identifying outliers as well as copies made from a single source. Analysis of texts below the sentence level, can determine whether particular pieces of information from one source are corroborated or not by other sources. It should also be useful to measure how widespread the information is reported, how quickly it has spread, and whether reputable sources from diverse locations and political persuasions are reporting the news as fact (Kedzie et al 2015). Multimedia also contributes already to help human rights groups identify accurate or hoax video through applications such as reverse image search [https://citizenevidence.org/] and other tools [Piraces, 2017; Liang et. al, 2016.]

## 2.6 Improving education

Multimodal approaches already have an important role in education. As in healthcare, these applications also must involve considerable effort in determining how best to integrate multimodal approaches to education with face-to-face learning. However, multimodal research can enhance the quality of education at all levels, especially by through personalization. Robots have been popular educational devices since the 1980s when the MIT Media Lab developed the Lego Mindstorms kits. Intelligent Tutoring Systems for science, math, language, and other disciplines match students with robot machine tutors. Today, more sophisticated and versatile kits for use in K-12 schools are available from a number of companies that create robots with new sensing technologies that are programmable in a variety of languages. Ozobot is a robot that teaches children to code and reason deductively while configuring it to dance or play based on color-coded patterns (Ozobot). Cubelets help teach children logical thinking through assembling robot blocks to think, act, or sense, depending upon the function of the different blocks (Cubelets). Wonder Workshop's Dash and Dot spans a range of programming capabilities. Children eight years old and older can create simple actions using a visual programming language, Blockly, or build iOS and Android applications using C or Java (Meet Dash). PLEO rb is a robot pet that helps children learn biology by teaching the robot to react to different aspects of the environment (Pleo rb).

A few years ago, it seemed that Massive Open Online Courses (MOOCs) would revolutionize education, however this has largely not been the case, as MOOCs mostly have consisted of online videos [Udall, 2017]. The challenge for these systems is determining whether or not they truly improve student performance.  Schools and universities are therefore slow to adopt such technologies due to this lack of proof as well as their lack of funds.  However, the availability of online virtual reality teachers is likely to grow, as students take education more and more into their own hands.  Downloadable software and online systems such as Carnegie Speech or Duolingo provide foreign language training using Automatic Speech Recognition (ASR) and NLP techniques to recognize language errors and help users correct them (Van Lehn et al). Tutoring systems such as the Carnegie Cognitive Tutor (Carnegie Learning) have been used in US high schools to help students learn mathematics. Other ITS have been developed for training in geography, circuits, medical diagnosis, computer literacy and programming, genetics, and chemistry. Virtual tutors use software to mimic the role of a good human tutor by, for example, providing hints when a student gets stuck on a math problem.  In higher education a system called SHERLOCK (Lesgold et al 1988) is being used to teach Air Force technicians to diagnose electrical systems problems in aircraft. And the University of Southern California's Information Sciences Institute has developed more advanced avatar-based training modules to train military personnel being sent to international posts in appropriate behavior when dealing with people from different cultural backgrounds (Virtual Humans).

With growing quantities of educational materials freely available on the web in the nation's increasingly internet-savvy culture, students are turning to the internet for targeted educational material (e.g. video lectures, Wikipedia). The challenge will be to design a more realistic framework that builds on the reality of distributed, uncoordinated, multimedia educational resources, and that provides technical solutions to transform this reality into organized, integrated and re-useable multi-modal resources as envisioned by the pioneers who first spearheaded efforts like SCORM [Horta et al. *2004*].  The targeted scalable multi-source integration of distributed instructional materials has many technical challenges: 1) size of data sets and wide-spread diversity among data types and sources, 2) task semantic consistency, 3) complexity of relations and dependency-structure of models, and 4) large swaths of missing, uncertain or otherwise sparsely-sampled data.

## 3. Challenges

We group the challenges into two types: fundamental and domain specific. Fundamental challenges are shared across many areas within multimedia for social good, while domain specific challenges are mostly applicable to specific sub-fields.

**Foundational Challenges**
- The future challenges in all multimodal tasks will be how to integrate the information gleaned from each individual task.  For example, deceitful speech and text cues from the speech channel and visual duplicate cues linking image objects to external sources may validate the faked news hypothesis, while contradictory semantics or sentiments extracted from associated media (image, caption, speech) may reveal instances of sarcasm or parody causing misinformation in the dissemination process. As with any large data analysis task, the key will be to find gold training data. This can in fact be collected by first finding matching visual objects and text concepts (entities, attributes, relations, or actions), to identify related news or posts reporting the same event over time. Where there is divergence or controversy regarding some aspects of these events which later was resolved it is possible to identify useful training data.

- Joint representations across media types are difficult to create in general way and the world knowledge required for logical inferences is woefully incomplete.
- The ability to deal with noisy or incomplete data in different environments will require new approaches to fill in or disregard the incomplete or incorrect parts of the data.
- Understanding should not just deal with the content of media but also the impact on different viewers or recipients. New ideas will be required to analyze both the content and its potential impact.
- Situations where there is very little or no training data are currently a major stumbling block, especially as rare events that should still be detected automatically are difficult to learn given current machine learning techniques.
- Trust, ethics, and accountability are key issues to study as multimedia takes a greater role in our daily lives. Currently there remain critical multimedia shortcomings when we are concerned with social good. For example, interpretable systems that can explain their decisions and have self-knowledge could vastly increase trust in the systems.
- With the ubiquity of observational and multimodal systems integrated into daily life, privacy and safety continue to be an afterthought, where it should be a key concern to research and a major challenge for implementation in the real world.
- Cost will be another challenge to widespread adoption of personal and service robots.

**Domain-specific challenges**

- News and Social Media Analysis. Multimodal Research for Analysis and Pattern Detection in Public Media Detection of "Fake News" and beyond. What is truth? How do systems understand contradictions in text, images and video?
  A specific challenge is facilitating societal dialog and cross-cultural understanding, perhaps in the form of scalable efforts in supporting democracy through a virtual multi-modal dialogic discourse.
- Providing tools for multi-media analysis to public safety organizations and first responders. Integrating multimodal advances beyond theoretical papers, for actual use by police, firefighters, and EMS services, where improved communication, situation awareness, visual documentation and forensics can improve both public safety and national security.
- Home healthcare, rehabilitation and assistance. In hospitals as in nursing homes, such mundane activities as moving patients from beds to wheelchairs or gurneys which are now handled laboriously by humans can be delegated to multimodal machines which can not only lift but can use sensors to relay patient discomfort in the process and react appropriately, much as humans do now. Multimodal systems can potentially create greater autonomy for the blind or hearing impaired. Many such opportunities are opening up in the medical domain with the opportunity to create significant improvements in prevention, diagnosis, and treatment.

- How to control robots is now fairly well understood, but how to use that control to enable robots to interact with humans and other aspects of the world in human-like ways is still a major area of research. While today deep learning in robotics is limited by the lack of large labeled corpora of multimodal data, safer and more reliable hardware, less expensive sensors, improvements in speech recognition and in spoken dialogue systems are now making it possible to create more human-like robots.
- It appears the transportation and mobility issues are increasingly driven by business opportunities, so we expect to see companies taking the lead in this area supplanting much of what is currently still academic research.
- Multimedia for better education. This educational multimodal mining is a major challenge. Improving education through multi-modal systems is a key challenge area, where robust solutions have proven to be very hard to achieve.

# 4. **Achievable Milestones**

We describe some milestones that may be achievable within the next 5, 10, and 15 years.

**5 year Milestones:**

**Healthcare** More multimodal robots currently improve the quality of life of the elderly in assisted care and nursing homes as robot pet cats.    In coming years more powerful and cheaper devices will be able to support the development of new software for home robots, which will support current advances in speech understanding and image recognition together with low cost 3D sensors to enhance robots' interactions with people in their homes.  More than half a dozen startups around the world are developing AI-based robots for the home, now concentrating mainly on social interaction.  There will also be increasing development of ools and systems that provide greater autonomy for the blind, hearing impaired, and elderly. These will help  handicapped people become more autonomous by providing. navigational instruments for the blind and supporting an aging population through reminders and alerting of caregivers.  Effective monitoring of health conditions in nursing home as well as in independent living situations will also be developed.

Better low-cost sensors and improvements in vision and speech recognition and spoken dialogue systems are also becoming a reality. Better hearing aids and visual assistive devices will counter the effects of hearing and vision loss, improving safety and social connection. Families will also be increasingly able to maintain close communicative ties using multimodal tools for audio and video communication and monitoring of loved ones living alone.

**News and Social Media**: Tools and systems for determining the first post of a piece of news or information and the authenticity of a location in an image are a reality. Integrated multi-modal systems to assist users in scanning social media for verification or debunking of 'news' items will also be improved.

**Public Safety**

We expect more significant monitoring of traffic and urban areas through video cameras and sensors, with automated alerts about accidents, vandalism and other disturbances. Body cameras on police officers will be standard with initial smart tools to interactively help protect privacy during unintended recording and finding critical events. New tools for coordination will improve response time and utilization of resources in emergencies.

**Transportation and Mobility.** Partially 'safety assisting' cars will become a standard more of transportation. Specific routes, with installed guiding technology and limited access will be available for bus routes and highways. Technologies such as collision anticipation, proactive evasion actions will become widely available.

**Education:** More educational systems will be available as apps for specific subjects. The devices will observe the students visually and assess their engagement with the program. A virtual teacher will be available from a palette of educational approaches and interventions to extend time on task and improve success rates. Numerous smart, sensor-based, interactive computer tools will be available especially for children and K-5 grades.

**10 year Milestones:**
**Healthcare:** Within the next 10 years, smart devices in the home will help with daily living activities when needed, such as cooking and, if robot manipulation capabilities improve sufficiently, dressing and toileting. In-home health monitoring and health information access will be able to detect changes in mood or behavior and alert caregivers. Multimodal home/service robots will be available for more sophisticated home care of the elderly to enable them to stay longer in their own homes with enhanced quality of life.
For physical therapy we expect to see autonomous guide robots providing assistance. Hardware improvements in safety and reliability as well as better understanding of human-human interaction will enable new uses and capabilities of multimodal home/service robots.

**News and Social Media**: Automated tools for fake news flaggingo will integrate information from four modalities -- images, video, speech and text.  These will greatly improve the ability to distinguish fake from true information in multiple situations. Systems will be tracing multimedia information propagation and how it is changing along the way.

**Public Safety and Security**.
As multimodal search improves, better identification of audio/video information will provide better assistance in crime prevention and prosecution – as well as clearer evidence of police malpractice. Most cities will use drones for surveillance purposes, and to maintain security of ports, airports, coastal areas, waterways, industrial facilities, despite increasing concerns about privacy.  Police, firefighters, and EMS services will become much more efficient through improved situation awareness
In airports and at borders, security will increasingly rely on unobtrusive multimodal approaches to improve efficiency and increase security.  Multimodal techniques in vision, speech analysis, and gait analysis will be widely used for detecting potential deceptive and criminal behavior.

**Transportation and Mobility.** Vehicles and personal robots will become significantly more independent, although not fully autonomous in difficult situations such as crowds, noise, or with environmental factors such as snow, rain, fog. Human driving? will no longer be viewed as a desirable activity that provides personal freedom.

**Education:** With small amounts of human intervention, autonomous intelligent teaching system guided study groups will prove to be comparably effective and more efficient than human classroom teaching. Most learning will take place through interaction with personal devices that accumulate knowledge about the user during the learning process.

**15 years:**
**Healthcare**: Personalized rehabilitation and in-home therapy will reduce the need for hospital or care facility stays. Physical assistance devices such as intelligent walkers, wheelchairs, and exo-skeletons will extend the range of activities of an infirm individual.

Robots and softbots will have the ability to gain and maintain the trust of their users. Humans will accept interaction and ubiquity of automated multimodal assistance dialog systems. Systems will be engaging with users in human-appropriate ways.

**Public Safety.** Autonomous aerial vehicles will patrol urban areas and major traffic arteries. Unobtrusive multi-sensory observation will be an integral part of public life, through home monitoring, wearable sensors and public surveillance.

**Transportation and Mobility**: Self-driving vehicles will be used in virtually all circumstance. Personal mobility devices will complement and interact seamlessly with long distance transportation methods. Traffic patterns and routes will be globally optimized over large regions.

**News and Social Media**: Fully automated fact-checking systems will provide context, background and provenance of news stories, together with confidence ratings in their veracity. Automated intelligent systems will emerge to make sense of the digital hubbub, and allow users to identify the major voices amongst the crowds. Most events will be recorded in an 'objective' multimedia documentation of what has occurred.

**Education**: Teaching will become more intelligent-system-guided than human-teacher-driven with better outcomes. Educational systems will have access to different types of relevant knowledge beyond the subject matter. Systems will integrate customized learning using real-time brain sensing data together with environmental sensors and personal history to assess knowledge, retention and comprehension at most educational levels through college.

# Section 4  Training, Infrastructure and Funding

# Chapter 13. **Training the Next Generation**

**Chapter Editors:**

Louis-Philippe Morency, Carnegie Mellon University
Sharon Oviatt, Monash University
Carlos Busso, University of Texas at Dallas
Joyce Chai, Michigan State University

## 1. Introduction

Training the next generation of workers is a fundamental challenge for our modern society. This training enables not only the young generation to learn skills and knowledge for their future career, but it also enables current workers to transition to new jobs and opportunities. Given the significant impact multimodal and multimedia researchers will have on our society, it is important to train the next generation of scientists, programmers, researchers and professors who will be required to achieve these goals.

## 2. State of the art

There have been a considerable number of courses and textbooks developed to train students in the specialized field of multimedia indexing and retrieval. The focus of these pedagogical resources has focused on building systems and technologies to index and retrieve specific instances from a large collection of videos. Examples of systems using such technologies include Google's YouTube and Microsoft's Bing video search. However, given the breadth of the new research in multimodal and multimedia systems, these courses only address a small part of all the ongoing and future research directions.

An additional current problem is that most introductory textbooks on human-computer interaction still focus on keyboard-based graphical interfaces, with little or no content on new media (speech, gestures, writing, multi-touch, nonverbal behavior, etc.) or their multimodal-multimedia combinations. Introductory HCI textbooks' coverage of ubiquitous multisensor interfaces also is limited, as is coverage of more complex but increasingly common multimodal-multisensor interfaces. These topics now define the *dominant interface paradigm in computing*, and are especially prevalent on mobile systems such as cell phones.

In terms of textbooks intended to provide general training in the broad fields of multimodal and multimedia systems, there has been a striking lack of resources until very recently. One note-worthy exception is *The Paradigm Shift to Multimodality in Contemporary Computer Interfaces* [Oviatt and Cohen, 2015], which provides an introduction to the basics of multimodal interfaces and systems that is intended for advanced undergraduates. It includes an introduction to cognitive science foundations, signal and language processing, fusion-based architectures, commercial system examples, and other topics. For training graduate students and professionals in the field, another newer textbook and reference is available, *The Handbook of Multimodal-Multisensor Interfaces* [Oviatt et al. 2017]. This three-volume textbook includes extensive definitions of terminology, source materials and instructions for conducting hands-on projects, videos that illustrate systems, and focus questions to assist students with learning the material.  In the area of multimedia, a graduate-level reference book is available, *Frontiers of Multimedia Research* [Chang, 2018]. It includes surveys of emerging topics in multimedia that were presented by eleven leading researchers at the 2015 inaugural SIGMM Rising Stars Workshop.

In related fields, such as the study of multisensory cognitive science and neuroscience, research and training on multimodal brain circuits, multimodal perception, and multimodal behavioral patterns has been extremely active. This multimodal processing viewpoint has now superseded the historical focus on unimodal perceptual processing that dominated psychology for over a century. For a seminal reference book on this topic, which is available for graduate training, see *The Handbook of Multisensory Processing* [Calvert et al. 2004]. A second edition now is also available, *The New Handbook of Multisensory Processing* [Stein 2012]. Apart from textbooks, few courses exist for teaching multimodal interaction and computation.

# Challenges unmet

## Textbooks

One of the key challenge for many professors and advisors in multimodal research is to find a reference book. Most books in the area are edited books, where the chapters are written by different authors. These books are not appropriate for training students, since each chapter has its own specialized topic and bias. It is not always possible for students to read and understand  influential primary papers on multimodal topics involving different disciplinary perspectives, without first receiving training on terms and basic concepts. Research papers typically are written for a more professional audience, with a different purpose in mind besides training. Textbooks are required to present the material in a meaningful manner, taking into consideration students' academic background and training goals.

# 3. Milestones

This topic requires multidisciplinary training, which could begin in high school. The topic requires groups with complementary expertise, so Institute-level organizations would be ideal, and they could be supported internationally and include international scientist and trainee exchanges. Their missions could include cutting-edge research, running summer schools for students at different levels, and establishing corporate partnerships to pursue particularly important application directions.

Here are several potential mechanisms to train the next generation of researchers:

- Establish dedicated multi-disciplinary training program (similar to IGERT) or science and technology centers.
- Support summer schools or summer camps (similar to JHU's summer school on human language research) to train students and junior researchers in multimodal-multisensor interfaces and systems. Parallel sessions could be run on key topics, such as multimodal behavior in the wild, multimodal analytics, multimodal machine learning, cross-cultural perspectives on multimodal interaction and systems, etc.
- Create specialization within a degree program (e.g., certificate in MM within the computer science or information technology program)

- Instigate MM internship programs with industry and research labs for undergraduate and graduate students.
- Provide special fund for university labs (in MM) to host high-school students and teachers during summer. (e.g., funding related to K-12 STEM education. .
- Organize grand challenges on multimodal-multimedia topics that are structured to support student engagement and sustained learning

The ACM ICMI and ACM Multimedia Conference series have been major training grounds for students interested in multimodal interfaces and systems and multimedia research. They have  made it high priority to provide funding for student presenters and PhD doctoral session attendees. They have also provided hands-on training through participation in the many grand challenge workshops and competitions for  many years. However, the grand challenge events require more support (funding, infrastructure) to be sustainable. They currently tend to be organized in an ad hoc way and to be discontinued after a few years. They are labor intensive, and they lack the support required to become a sustained  training ground for accumulating knowledge and tools needed for education.

# 4. International dimension

One of the essential ingredients of multimodal and multimedia is multi-cultural. Where mono-disciplinary research often has clear problems set and tasks to solve, a multi-modal, multi-media and multi-cultural setting is often the playground for emerging insights. Machine-delivered coaching for health will be obvious in one culture and perceived as meddling with personal autonomy and privacy in another. Forms of man-machine interaction may be obvious in one culture and be completely misunderstood in other less visually oriented cultures. And, as a third example, the difference between words and pictures in one culture in terms of emotional impact may be quite different from the impact in other cultures (and largely varying per topic as well).

Also in the area of turning personal and private data into business is perceived different across different continents. When a considerable part of the new markets is with these data these differences should be part of the learning experience.

All three examples and the practical impact in internationally operating data-industries indicate the enormous benefit for an honors program of leaders of the future to be educated in a deliberate international and intercultural context. One such implementation could be in courses organized on the side of ACM Multimedia conferences, or as a summer school with partners from three continents.

# Chapter 14. **International Funding and Collaboration**

**Chapter Editors:**

Shih-Fu Chang, Columbia University
Joyce Chai, Michigan State University
Sharon Oviatt, Monash University
Arnold Smeulders, University of Amsterdam

## 1. Introduction

Multimedia and multimodal communities have enjoyed broad participation of researchers and practitioners from many regions in the world. Many conferences, like ACM Multimedia, since inception have been held with regular location rotation over different geographical areas. This shows the broad relevance of the research topics being addressed by the community to the actual needs of the local industry and society in different regions. From these, identification of common topics important to different countries and areas CAN help us discover opportunities for international cooperation and learn from best R&D programs and practices successfully deployed in different countries. Additionally, certain research agendas especially those related to cross-cultural interaction and understanding will greatly benefit from active participation of researchers and users of different backgrounds.

In this chapter, we first summarize a few existing international projects or initiatives closely related to the multimedia and/or multimodal field. We then present several new challenges and problems that call for further collaborative efforts from the international multimedia community. Finally, we briefly discuss a few ways to improve the process of international collaboration.

## 2. Examples of Current International Multimedia Initiatives

Under the EU Cordis program, the header of man-machine interaction, multimodal interaction is one of the themes. Man-machine interaction is important, but it usually focuses on a controlled environment for a restricted set of purposes. For the long-term vision of the multimedia community, we should aim for multimodal cross-cultural research in general with a much broader scope. This will involve understanding of multimodal behavior in the wild (e.g., not limited to the gestures for steering a computer) as well as understanding cross-cultural behavior (where the task of steering a computer is uniform across culture).

EU has also sponsored quite a few large projects on multimodal/multimedia research. In addition, many large projects on human-robot have significant components on MM. The following includes a few examples:

- http://www.chistera.eu/projects/camomile
- http://cordis.europa.eu/project/rcn/96289_en.html (old)
- http://www.mummer-project.eu/about/ (ongoing project, multimodal mall entertainment robot)
- http://www.tradr-project.eu/mission-2/
- https://robot-ears.eu/

Recently, a few large scale research programs organized by US NIST, DARPA and IARPA have seen successful active collaboration with multimedia researchers from outside the US. One notable program enjoying broad participation from across the world is the international video retrieval evaluation, TRECVID, that has been successful run since 2011 covering various multimodal tasks such as semantic concept detection, summarization, shot boundary detection, and multimedia event detection. Other example programs related to multimedia research also include the IARPA ALADDIN program focusing on multimedia event detection/recounting and the DARPA MediFor program focusing on development of robust forensics tools for detecting manipulations of images and videos and ensuring integrity of the media content.

# 3. New Challenges and Problems Calling for International Collaboration

- Multimodal analytics for diagnosing and monitoring medical and health-related intervention progress are critical areas that are beginning to emerge. At the moment, most of that work is still unimodal (e.g., analyzing speech signal features to detect Parkinson's). To become more accurate, including avoiding false positives and distinguishing a particular disease from the most closely related clinically confusable ones (based on expert doctor's reports), will require pursuing ultra-reliable fusion-based multimodal analysis-- for example of speech and other physical activity patterns (e.g., manual control during handwriting, facial control during emotional expression) in cases like Parkinson's detection. Beyond diagnostics, these analytics will be required to objectively evaluate intervention outcomes-- whether comparing drugs, physical therapy routines, surgery techniques, etc.)

- Multimodal analytics for detecting and supporting students' learning progress in school would be another example of a critical application area that most funders and countries would value.
- In times where it has become important which country is first and which are second, research could contribute to cross-cultural understanding. This is often multimodal in nature as what is said in words in one country is expressed by body language in another. For one, the Inuit better specify the various types of ice in lingual expressions, where Brazilian body language may well differentiate better among states of the soul. Cross-cultural research also urgent. Threats of terror have set the necessity is understanding body language: are good versus bad intentions readable across cultures? And, cross-cultural understanding fires back. It deepens the insight into the own culture: what habits do we have in food (as opposed to how other cultures illuminate us)? And, can we learn from the way people learn across cultures?

# 4. Mechanisms for Stimulating International Collaboration

- If anything would be appropriate for an internationally funded program of research it would be cross-cultural multimodal research. As argued above, these are the times to install these programs between national science foundations across continents.
- Share resources (e.g., data, infrastructure, software, evaluation metrics) across regions and countries. Establishing something like Robot Operation System (ROS) for the multimedia community will be beneficial. Such platforms provide open-source community resources for researchers/developers to share software tools for application development.  More information can be found at: http://www.ros.org/
- The multimodal versus multimedia communities currently are too disjoint-- and spread across ACM versus IEEE type activities. We should be seeking opportunities to cross-fertilize these communities more-- for example, by including them in summer school training programs.

# Chapter 15. **Data and Computing Infrastructure**

### Chapter Editor:

Alex Hauptmann, Carnegie Mellon University

*This section represents a summary of informal replies by workshop participants to questions on future data requirements, access to computing resources and other critical infrastructure needs of the multimedia/multimodal research community.*

**Q1. Data is always a problem, arguments have been made that defined datasets and evaluations associated with them are stifling creativity; on the other hand, proponents say publicly available data sets enable and accelerate quantifiable research progress. In light of this, how would future multimedia research benefit from public datasets? What types of data are needed that are not available now?**

- Well-structured datasets appropriate for making inferential assumptions about cause and effect are sorely lacking. They are extremely expensive to collect, and even more expensive to ground-truth code in a high-quality manner. There is a special need for **longitudinal multimodal datasets,** as others also have noted, which are especially labor intensive because they require collecting data on (1) multiple time-synchronized input streams, and (2) over an extended time period. At the moment, there is also a strong interest in collecting data in natural field setting (e.g., classrooms, clinics) and also on users' cell phones while mobile and engaged in their daily lives. This type of work, which is of extraordinary importance, will require major investments in new infrastructure for high-quality mobile data collection, and resources for teams of scientists to both collect and annotate the data.

  An additional hurdle for these datasets is the growing concern about privacy and protecting individuals from undesirable use of the data, such as for inappropriate spamming or even personal attacks on social media.

  Especially for research on person-centered interaction, although there is a rich amount of publicly available multimedia data (e.g., images and videos) in aggregate, the data centered on a particular person is rather sparse. There is very little longitudinal data available about a person. Without such data, it is difficult to perform more meaningful person-centered multimedia research. Moreover, due to privacy reasons, how could we obtain a sufficient amount of data to do research while protecting the privacy of individuals?

- There is also a lack of platforms that collect, analyze, and track longitudinal multimedia data. While there are numerous companies collecting this data, from fitness trackers, IoT devices, smart watches to Facebook and Google virtually none of this is available for research studies. A possible model for such a platform that can collect longitudinal person-centered research data on an "opt-in for research" basis may be the Apache Software Foundation approach, which builds is free and open-source cross-platform software, released under generous license with development and maintenance through an open community of developers.

- Many novel questions cannot be addressed by the existing datasets, such as longitudinal understanding of influences to the behavior of an individual. One specific example would be education, where there are no long term multimedia dataset available to study progress in learning. Beyond participating in shared tasks (e.g., on benchmark datasets), researchers should also be encouraged to identify new research problems and collect new datasets (which will become publicly available later) to address those problems. Some dedicated program similar to NSF CRI will be needed to fund such efforts.

- An alternative perspective is that we cannot expect to have large multimodal databases that are appropriate for every task or problem that we want to address. Thus it will be up to the community to develop algorithms and modules where multiple corpora are combined for training a bigger system, where missing features are properly handled, and unlabeled/unstructured data are still useful. The key is to have the infrastructure to share databases. Today, we only have individual efforts, where the corpora are shared by each group using their own websites. The problem is that the community may not be aware of valuable resources that are already available to them.

- Another major problem is the lack of reliable ground-truth annotation for multimedia/multimodal data. Different research teams may have their own setup for acquiring ground-truth annotation from crowd-sourcing or pseudo-annotations by using pre-selection and filtering algorithms data most likely to be correctly labeled on the web, and then manually verifying correctness of what will inevitably be a biased but "annotated" dataset. It is worth considering whether it makes sense for the MM community to create some joint annotation service (e.g., identify and maintain a pool of annotators who are experienced for MM, share annotation tools). This may help improve the quality of annotation. A previous model for this could be seen in the Linguistic Data Consortium (LDC) at the University of Pennsylvania, which has become the central repository for linguistic corpora with annotations.

**Q2. Do you see other critical infrastructure needs in the 5-10 year future? How important will GPUs be to research over the next 5-10 years? Why?**

- Within this time frame, massive datasets will be collected that are both multi-stream multimodal and longitudinal in nature. The need to automate data collection, annotation, and retrieval of key information-- and also the need to provide for adequate storage, will become critical to avoid undermining research progress.

  Specific infrastructure needs will relate to access to the data expected to be in the range of petabytes, and computational processing capacity suitable for multimodal analysis. NSF is currently funding research at a number of high-performance computing (HPC) facilities, which develop architectures to provide support for current and future research needs. However, the traditional high-performance computing infrastructure in the form of very fast computers or very large loosely coupled clusters are often not very suitable to multimedia research, since there are very high I/O requirements, large local cache needs as well as very fast (primarily GPU) processing. In light of the projected dramatic increase in multimedia use over the internet, one suggestion is to give multimedia/multimodal research a larger role in shaping the future of computing. Collaborations between MM researchers, HPC designers, collaborating with relevant industries to taking advantage of their insights, current best practices and anticipated needs. A more centrally organized collaboration with industry may also make access to industry resources such as computing clusters available to researcher that would otherwise lack the contacts and leverage to gain access to such resources.